\documentclass{article}

\usepackage[dandb, final]{neurips_2025}

\usepackage{amsmath}
\usepackage{mdframed} 
\usepackage{longtable}
\usepackage{multirow}
\usepackage{graphicx}
\usepackage[utf8]{inputenc} 
\usepackage[T1]{fontenc}    
\usepackage{url}            
\usepackage{booktabs}       
\usepackage{amsfonts}       
\usepackage{nicefrac}       
\usepackage{microtype}      
\usepackage{xcolor}         
\definecolor{iris}{rgb}{0.35, 0.31, 0.81}
\definecolor{amaranth}{rgb}{0.9, 0.17, 0.31}
\usepackage[colorlinks=true, linkcolor=amaranth, urlcolor=amaranth, citecolor=blue, anchorcolor=blue]{hyperref}
\usepackage{subcaption}
\usepackage[table]{xcolor}  
\usepackage{wrapfig}
\usepackage[most]{tcolorbox} 
\usepackage{multicol}
\usepackage{xcolor}
\usepackage{listings}
\usepackage{rotating}
\usepackage{tabularx}   

\tcbuselibrary{skins, breakable}

\lstdefinestyle{pythonstyle}{
  language=Python,
  basicstyle=\ttfamily\small,
  keywordstyle=\color{blue}\bfseries,
  commentstyle=\color{green!50!black}\itshape,
  stringstyle=\color{orange!90!black},
  numberstyle=\tiny\color{gray},
  numbers=left,
  stepnumber=1,
  numbersep=5pt,
  showstringspaces=false,
  breaklines=true,
  frame=none,
  tabsize=4,
  captionpos=b
}

\lstdefinelanguage{Spice}{
  morekeywords={subckt,ends,tsmc18dN,tsmc18dP},
  sensitive=false,
  morecomment=[l]{//},
  morestring=[b]"
}
\title{Toward Engineering AGI: Benchmarking the Engineering Design Capabilities of LLMs}

\author{
\small
\begin{tabularx}{\textwidth}{X}
Xingang Guo$^{1*\dagger}$, Yaxin Li$^{1*}$, Xiangyi Kong$^{1*}$, Yilan Jiang$^{1*}$,  
Xiayu Zhao$^{1*}$, Zhihua Gong$^{1*}$, Yufan Zhang$^{1*}$,
Daixuan Li$^1$, Tianle Sang$^1$, Beixiao Zhu$^1$, Gregory Jun$^1$, Yingbing Huang$^1$, Yiqi Liu$^1$, Yuqi Xue$^1$, Rahul Dev Kundu$^1$,
Qi Jian Lim$^1$, Yizhou Zhao$^2$, Luke Alexander Granger$^1$, Mohamed Badr Younis$^1$, Darioush Keivan$^1$, Nippun Sabharwal$^1$,
Shreyanka Sinha$^1$, Prakhar Agarwal$^1$, Kojo Vandyck$^1$, Hanlin Mai$^1$, Zichen Wang$^1$, Aditya Venkatesh$^1$, Ayush Barik$^1$,
Jiankun Yang$^1$, Chongying Yue$^1$, Jingjie He$^1$, Libin Wang$^1$, Licheng Xu$^1$, Hao Chen$^1$, Jinwen Wang$^1$, Liujun Xu$^1$,
Rushabh Shetty$^1$, Ziheng Guo$^1$, Dahui Song$^1$, Manvi Jha$^1$, Weijie Liang$^1$, Weiman Yan$^1$, Bryan Zhang$^1$,
Sahil Bhandary Karnoor$^1$, Jialiang Zhang$^1$, Rutva Pandya$^1$, Xinyi Gong$^1$, Mithesh Ballae Ganesh$^1$, Feize Shi$^1$,
Ruiling Xu$^1$, Yifan Zhang$^1$, Yanfeng Ouyang$^1$, Lianhui Qin$^3$, Elyse Rosenbaum$^1$, Corey Snyder$^3$,
Peter Seiler$^4$, Geir Dullerud$^1$, Xiaojia Shelly Zhang$^1$, Zuofu Cheng$^1$, Pavan Kumar Hanumolu$^1$, Jian Huang$^1$,
Mayank Kulkarni$^5$, Mahdi Namazifar$^5$, Huan Zhang$^1$, Bin Hu$^{1\dagger}$ \\
\text{$^{1}$University of Illinois at Urbana-Champaign,
$^{2}$University of Pennsylvania,} \\
\text{$^{3}$University of California San Diego,
$^{4}$University of Michigan,
$^{5}$Amazon AGI} \\
$^{*}$Core Contributors \quad
$^{\dagger}$Corresponding Author \\
\textbf{Project page: \href{https://agi4engineering.github.io/Eng-Design/}{https://agi4engineering.github.io/Eng-Design/}}\\
\textbf{Dataset page: \href{https://huggingface.co/datasets/opt1zer/EngDesign}{https://huggingface.co/datasets/opt1zer/EngDesign}}
\end{tabularx}}

\begin{document}
\maketitle
\begin{abstract}
Modern engineering, spanning electrical, mechanical, aerospace, civil, and computer disciplines, stands as a cornerstone of human civilization and the foundation of our society. However, engineering design poses a fundamentally different challenge for large language models (LLMs) compared with traditional textbook-style problem solving or factual question answering. Although existing benchmarks have driven progress in areas such as language understanding, code synthesis, and scientific problem solving, real-world engineering design demands the synthesis of domain knowledge, navigation of complex trade-offs, and management of the tedious processes that consume much of practicing engineers' time. 
Despite these shared challenges across engineering disciplines, no benchmark currently captures the unique demands of engineering design work. In this work, we introduce \textbf{\textsc{EngDesign}}, an \textbf{Eng}ineering \textbf{Design} benchmark that evaluates LLMs' abilities to perform practical design tasks across nine engineering domains. Unlike existing benchmarks that focus on factual recall or question answering, \textsc{EngDesign} uniquely emphasizes LLMs' ability to synthesize domain knowledge, reason under constraints, and generate functional, objective-oriented engineering designs. Each task in \textsc{EngDesign} represents a real-world engineering design problem, accompanied by a detailed task description specifying design goals, constraints, and performance requirements. \textsc{EngDesign} pioneers a \emph{simulation-based evaluation paradigm} that moves beyond textbook knowledge to assess genuine engineering design capabilities and shifts evaluation from static answer checking to dynamic, simulation-driven functional verification, marking a crucial step toward realizing the vision of engineering Artificial General Intelligence (AGI).
\end{abstract}

\section{Introduction}

Modern engineering, spanning electrical, mechanical, aerospace, civil, and computer disciplines \citep{chen2004electrical, grote2021springer, chen2002civil, blockley2012engineering}, stands as a cornerstone of human civilization and the foundation of our society. From the electrical grids powering our cities to the aerospace systems launching us beyond Earth, from analog integrated circuits forming the foundation of electronics to the structural designs supporting our skylines, engineering disciplines have woven themselves into the fabric of human existence \citep{ten2024modern, paul2023formal, gray2009analysis}.  At its core, engineering design is the systematic and creative process that engineers use to solve problems and create functional products, systems, or processes, transforming requirements and constraints into tangible solutions that meet human needs. Recent advances in large language models (LLMs) have led to remarkable performance on conventional question-answering (QA) benchmarks, with strong results across a wide range of tasks—from textbook-level scientific problem solving \citep{rein2024gpqa, wang2024mmlu, du2025supergpqa, zou2024dynamath}, to code synthesis \citep{chen2021evaluating, jain2024livecodebench, chan2024mle}, and even to answering textbook-level questions across various engineering domains \citep{kevian2024capabilities, li2024eee, skelic2025circuit}.
Today, industry leaders aspire to build on such progress in LLMs to create general-purpose AI engineers capable of bringing to life humanity’s boldest ambitions-from interstellar starships to Dyson spheres that harvest stellar energy-leveraging these foundation models’ vast knowledge across engineering domains and their potential to transcend traditional disciplinary boundaries that constrain human engineers \citep{businesswire2025p1ai}.

However, practical engineering design confronts LLMs with a grand challenge profoundly different and vastly more complex than conventional factual recall or textbook-level engineering problem solving. Specifically, real-world engineering design demands the synthesis of domain knowledge, navigation of complex trade-offs, management of the tedious processes that consume much of practicing engineers' time, and rigorous validation through domain-specific simulators to ensure that designs meet functional requirements and safety constraints. Yet despite progress in textbook-level engineering problem solving, no existing benchmark captures the distinctive demands of practical design work--with its multifaceted, open-ended challenges across multiple engineering disciplines.

In this work, we introduce \textsc{EngDesign}, the first benchmark for holistically evaluating LLMs on real-world, multi-domain engineering design challenges. 
Unlike traditional Question-Answer (QA) formats, our benchmark requires models to function as practicing engineers, producing functional solutions, such as dynamical system controllers, material structure designs, analog integrated circuits, or GPU architectures that meet rigorous performance requirements.  \textsc{EngDesign} 
uniquely emphasizes the synthesis of domain knowledge, constraint-based reasoning, and the generation of functional, objective-oriented designs,
spanning practical design problems across nine engineering domains: Operating System Design, Computer Architecture Design, Control System Design, Mechanical Systems, Structural Design, Digital Hardware Design, Analog Integrated Circuit Design, Robotics, and Signal Processing.
Each task in  \textsc{EngDesign} is accompanied by an executable evaluation pipeline that validates designs against task requirements using domain-specific simulation tools such as
SPICE simulations \citep{roberts1996spice}, structural finite element analysis \citep{bhavikatti2005finite}, MATLAB Control System Toolbox \citep{chiang1984matlab}, and other scientific software tailored to specific engineering domains. 
Our approach establishes a revolutionary evaluation paradigm: assessment using engineering simulation tools. 
This simulation-based approach fundamentally shifts evaluation from linguistic pattern matching to functional verification, ensuring that solutions are assessed based on their engineering merit rather than textual plausibility.

\begin{figure}[t!]
\includegraphics[width=1.0\textwidth]{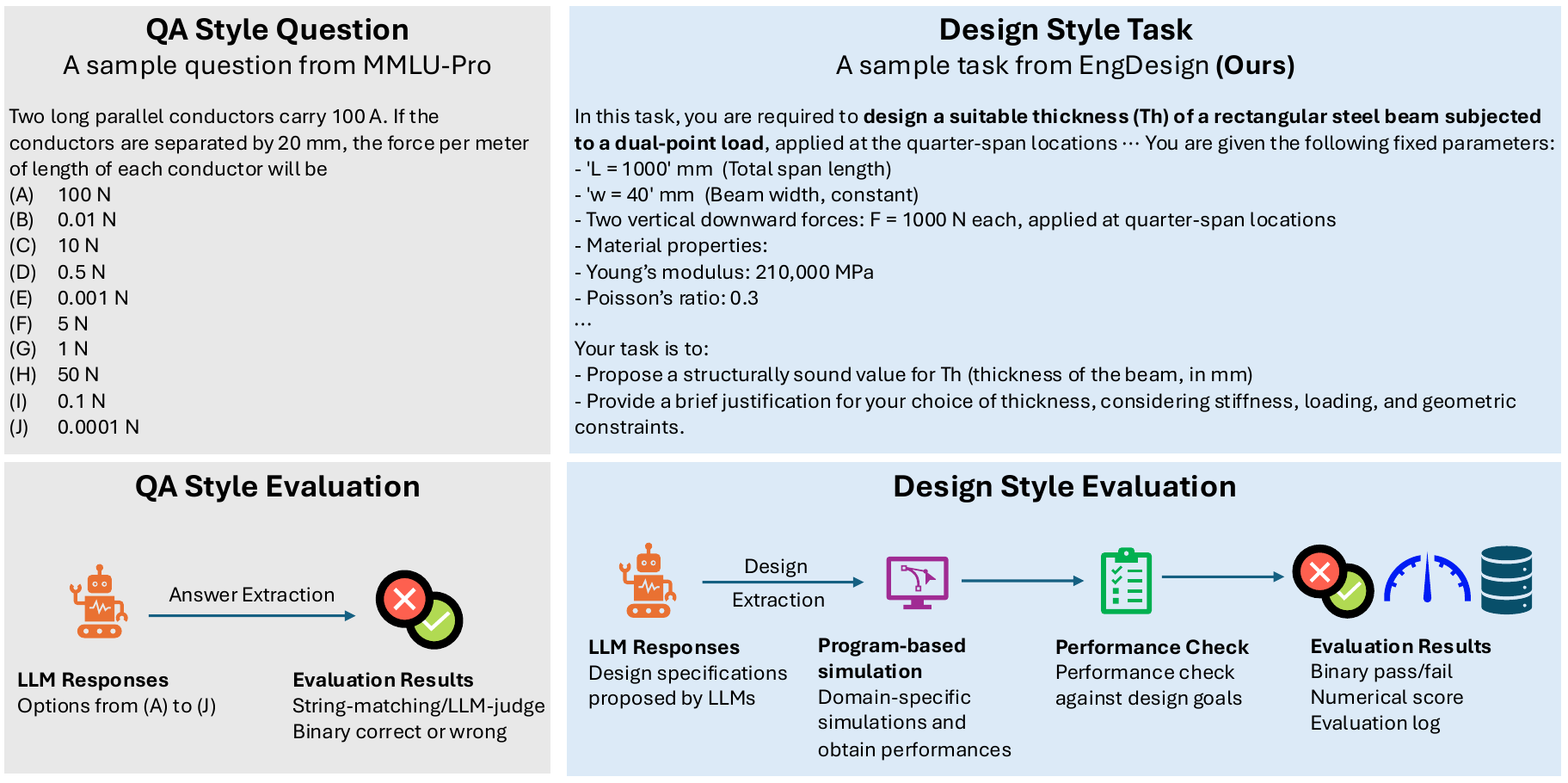}
\caption{Comparison between conventional QA-style benchmarks (left) and the design-style benchmark \textsc{EngDesign} (right). Conventional QA benchmarks evaluate LLMs through static answer extraction and string-matching, while \textsc{EngDesign} involves open-ended design tasks with potentially non-unique solutions. LLMs must propose candidate design specifications, which are evaluated via program-based simulations and performance validation pipelines.}
\label{figure:engdesign_pipeline}
\end{figure}

\begin{figure}[htbp]
  \centering
  \begin{subfigure}{\linewidth}
    \centering
    \includegraphics[width=\linewidth]{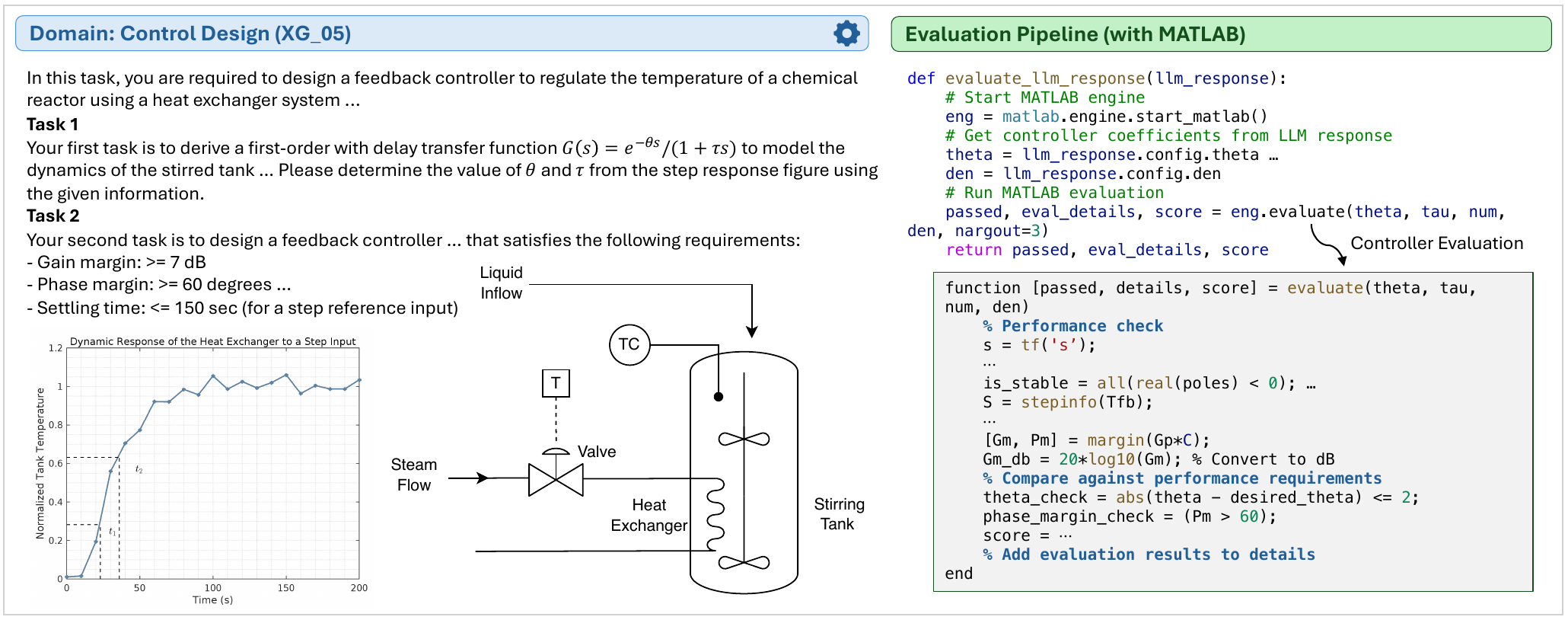}
  \end{subfigure}

  \begin{subfigure}{\linewidth}
    \centering
    \includegraphics[width=\linewidth]{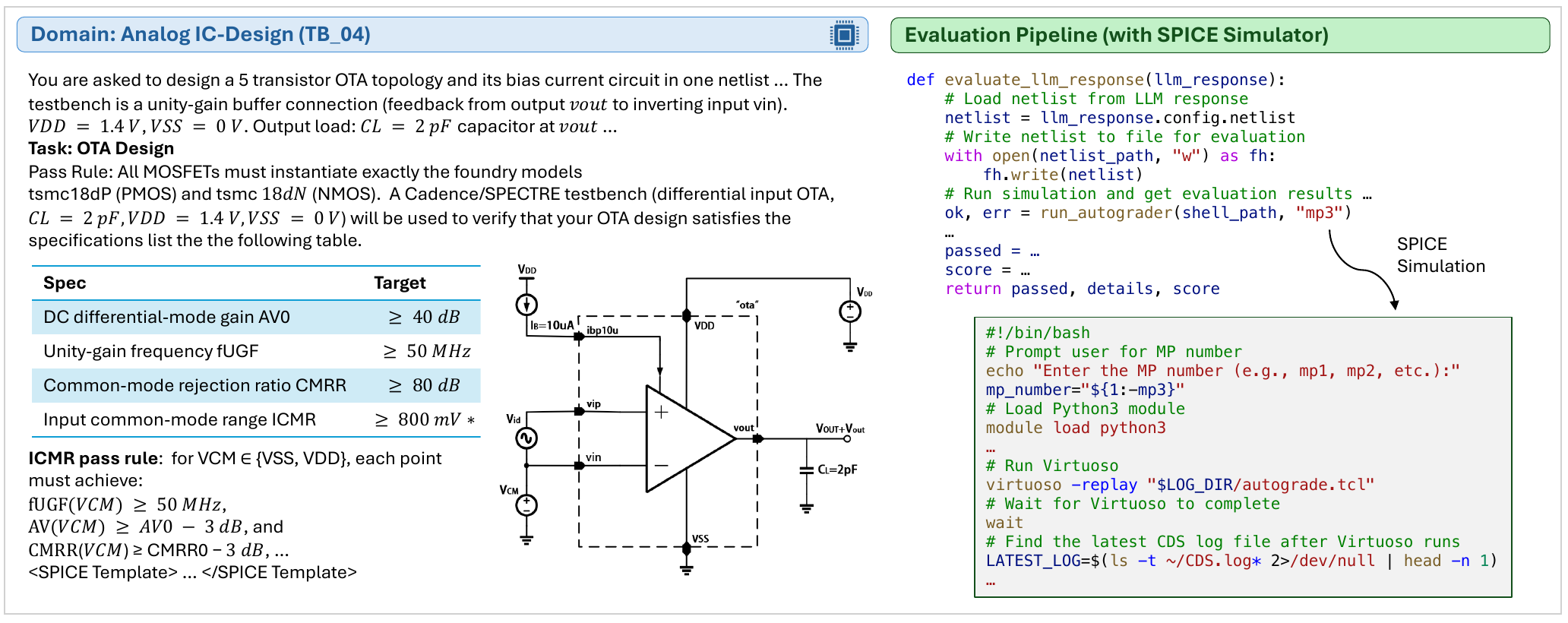}
  \end{subfigure}

  \begin{subfigure}{\linewidth}
    \centering
    \includegraphics[width=\linewidth]{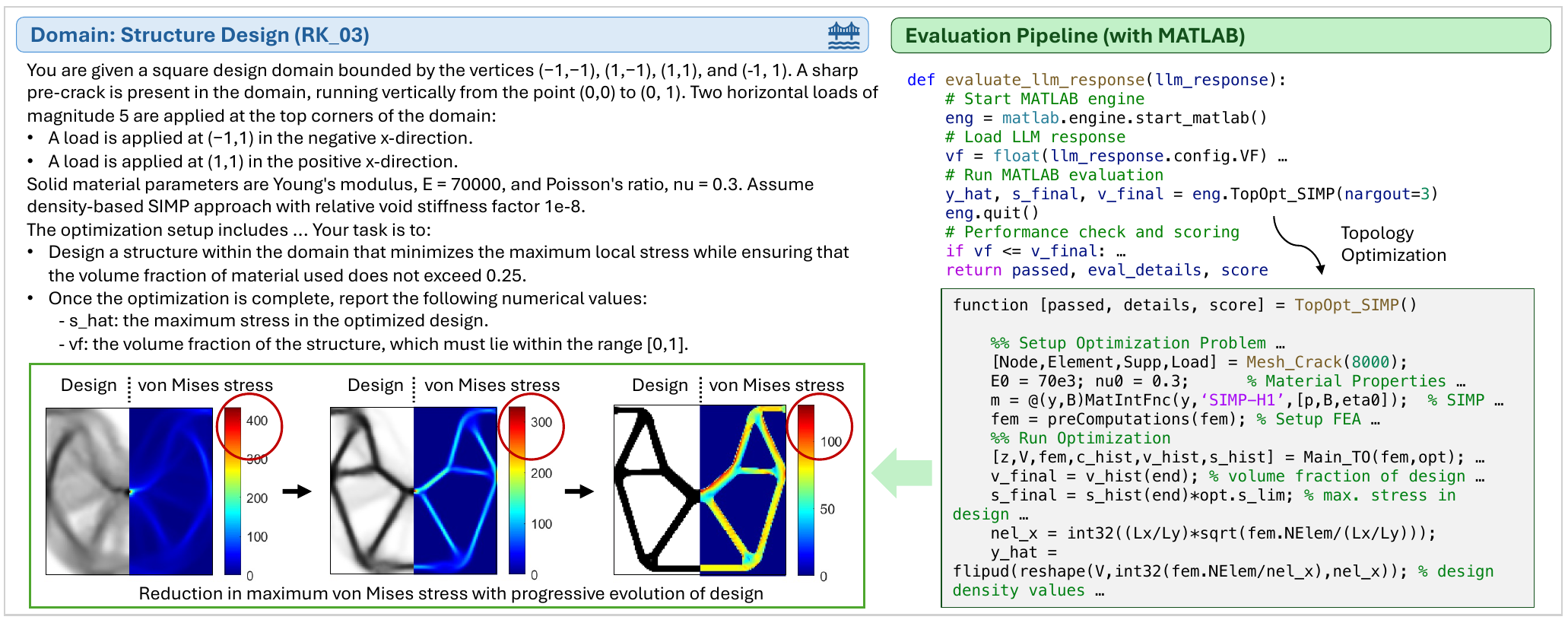}
  \end{subfigure}

  \begin{subfigure}{\linewidth}
    \centering
    \includegraphics[width=\linewidth]{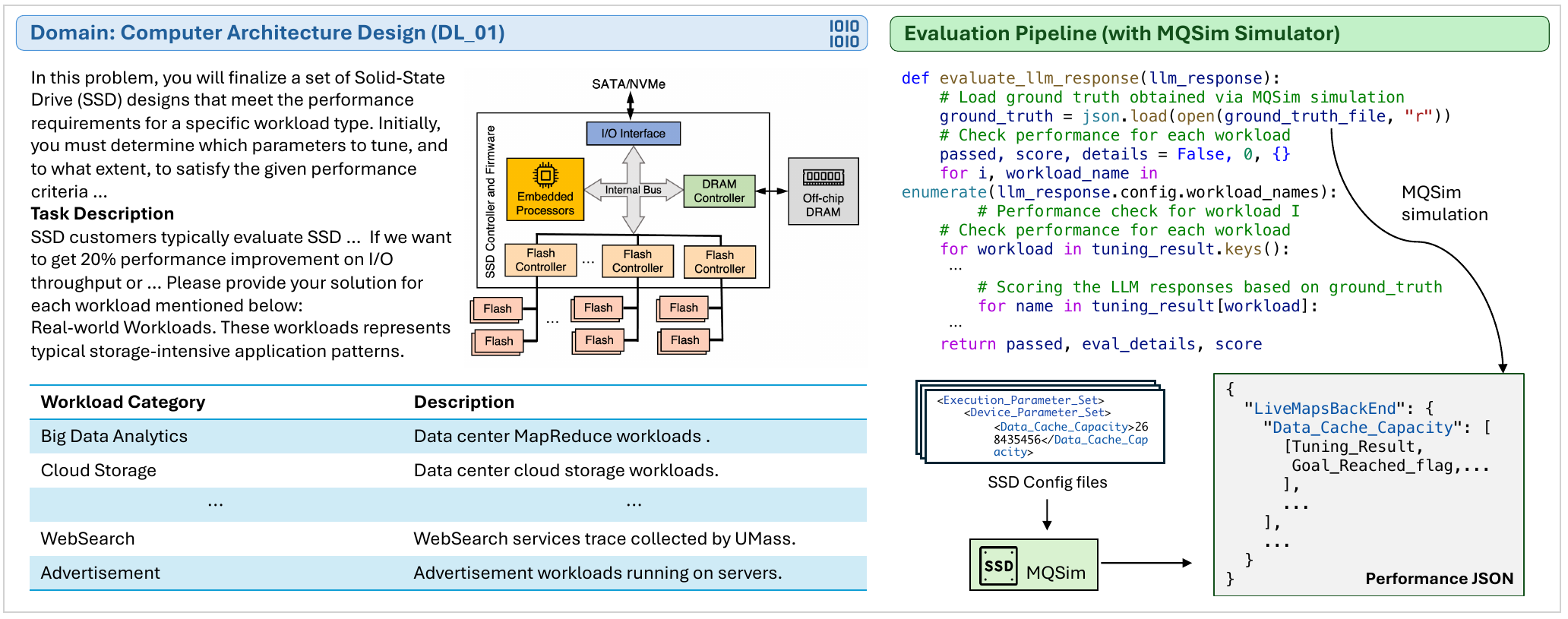}
  \end{subfigure}

  \caption{Selected demonstration tasks from \textsc{EngDesign}.}
  \label{fig:demos}
\end{figure}

\textsc{EngDesign} advances LLM evaluation methodology by moving beyond conventional question-answering benchmarks to address the open-ended, constraint-driven nature of real-world engineering design. Our key contributions include:
\begin{itemize}
\item \textbf{The first multi-domain engineering design benchmark.} Unlike QA benchmarks that test narrow factual or procedural knowledge, \textsc{EngDesign} evaluates LLMs on complex engineering design challenges spanning diverse engineering disciplines (e.g., mechanical, electrical, civil), requiring synthesis of domain knowledge, constraint satisfaction, and strong reasoning capabilities.
\item \textbf{Executable simulation-based evaluation.} We replace static correctness checks with dynamic, domain-specific evaluation pipelines. Each task includes human-designed task-specific evaluation scripts that rigorously verify functional feasibility (e.g., via simulation, constraint validation, or performance testing), ensuring objective, reproducible scoring grounded in real engineering standards.  
\item \textbf{Partial-credit grading for incomplete solutions.} Conventional benchmarks often use binary scoring, but \textsc{EngDesign}’s rubrics quantify incremental progress (e.g., 20/100 for delivering a stable controller that meets partial performance requirements). This granularity reveals nuanced capability gaps and rewards iterative refinement.  
\item \textbf{Empirical validation of design competencies.} Through large-scale evaluations across 10+ state-of-the-art LLMs including both general-purpose LLMs and reasoning models, we demonstrate \textsc{EngDesign}’s ability to expose critical limitations in AI systems—such as over-reliance on textual patterns or failure to handle trade-offs—that are invisible to traditional benchmarks.  
\end{itemize}

We believe that \textsc{EngDesign} marks a crucial milestone toward realizing engineering AGI by breaking new ground with its comprehensive coverage of engineering design problems across diverse domains, each requiring deep technical expertise and domain-specific simulation tools for rigorous evaluation. This benchmark not only measures what models know but what they can actually design, bridging the chasm between textbook-level knowledge understanding and practical engineering capability.

\section{\textsc{EngDesign}}
\label{sec:engdesign}

In this section, we introduce \textsc{EngDesign}, a multi-domain benchmark developed to evaluate the capabilities of LLMs in real-world engineering design tasks. Unlike conventional QA benchmarks, \textsc{EngDesign} adopts a rigorous simulation-based evaluation pipeline to assess model performance in practical, design-oriented scenarios. Table~\ref{stat-engdesign} summarizes the benchmark's key statistics: it comprises 101 design tasks spanning 9 engineering domains, with a total of 473 gradable items. While the task distribution may appear uneven, it naturally reflects the domain expertise of our contributors and the filtering rigor imposed by our multi-stage review process.
Notably, the average prompt length in \textsc{EngDesign} is 778.71 tokens, substantially higher than typical QA benchmarks\footnote{We adopt Byte Pair Encoding (BPE) tokenizer for OpenAI models, available at \href{https://github.com/openai/tiktoken}{Tiktoken}. Token length comparison between \textsc{EngDesign} and other QA benchmarks in Appendix \ref{app:token_num_com}.}, highlighting the contextual richness and complexity of realistic engineering design problems. Word clouds generated from \textsc{EngDesign} prompts are provided in Appendix~\ref{app:more_on_eng}.

Among the 101 tasks in \textsc{EngDesign}, 34 tasks require domain-specific scientific software such as MATLAB or Cadence for evaluation, while the remaining 67 tasks are fully open-sourced and evaluated using manually authored evaluation scripts. We consolidate these tasks into a subset called \textsc{EngDesign-Open} to support broader community adoption without licensing constraints. Additionally, 23 tasks of \textsc{EngDesign} incorporate images as part of the task input to LLMs.

\begin{table}
  \caption{Statistics of \textsc{EngDesign}. Token counts are computed vis \texttt{o200k\_base} in Tiktoken.}
  \label{stat-engdesign}
  \centering
  \begin{tabular}{l|ccccccc}
    \toprule                
    \multirow{2}{*}{\textbf{Engineering Domain}} & 
    \multicolumn{3}{c}{\textbf{\# tasks}} & 
    \multirow{2}{*}{\textbf{\# rubrics}} & 
    \multicolumn{3}{c}{\textbf{\# query tokens}} \\
    \cmidrule(lr){2-4}
    \cmidrule(lr){6-8}
     & \textbf{Open} & \textbf{Closed} & \textbf{Total}  &    & \textbf{Max} & \textbf{Min} & \textbf{Avg} \\
    \midrule
    Operating System Design &8 &0 &8 &66 &2938 &310 &1103.25 \\
    Computer Arch Design &5 &0 &5 &20 &4385 &2348 &3539.60\\
    Control Design  &7 &11 &18   &100      &1361   &209   &634.44   \\
    Mechanical Systems  &6 &1 &7  &32   &781   &225   &391.14   \\
    Structure Design &7 &6  &13 &25 &483 &186 &345.31\\
    Digital Hardware Design &13 &4 &17 &58 &1715 &206 &515.65  \\
    Analog IC Design &0 &5 &5 &23 &2136 &547 &1196.6\\
    Robotics  &10 &0 &10   &68      &1485   &192   &771.9   \\
    Signal Processing   &11 &7 &18      &81   & 2304  &151 &611.72   \\   
    \midrule
    Overall &67 &34 &101   &473   &4385   &151      & 778.71  \\
    \bottomrule
  \end{tabular}
\end{table}

\subsection{Task Structure}
Each task of \textsc{EngDesign} consists of the following four key components:

\textbf{1. Task Description.}  
This part is the query prompt fed into the LLMs, offering a clear and detailed definition of the engineering design problem, including design objectives, specifications, constraints.

\textbf{2. Evaluation Rubrics.}  Given the complexity of \textsc{EngDesign} tasks, each task is further decomposed into multiple gradable items evaluated individually during the performance check stage. The evaluation rubrics define the assessment criteria and scoring metrics, with a full score of 100.  This enables models to receive partial credit even if the design does not fully meet all specified requirements.

\textbf{3. Evaluation Pipeline.}  
Each task includes automated evaluation scripts that assess the LLMs' design. It returns a binary pass/fail indicator, numerical score, and evaluation logs for further analysis.

\textbf{4. Reference Design.}  
Each task provides a validated reference design that fully satisfies all specified requirements, ensuring the feasibility and realism of the design challenge.

Figure \ref{fig:demos} shows four demonstrated design examples from \textsc{EngDesign} including task descriptions and evaluation pipelines.

\subsection{Evaluation Pipeline}
\label{eval_pipe}

Unlike conventional benchmarks, which often have a single golden answer that allows evaluation through exact string matching or LLM-as-judge scoring, our engineering design tasks inherently lack a unique \textit{golden design}. For example, there are infinitely many valid controller designs that can regulate the temperature of a stirred tank modeled by a first-order system while satisfying both time-domain and frequency-domain performance requirements. 
To address this challenge, our benchmark introduces a \textbf{simulation-based evaluation pipeline}. Specifically, for each LLM-generated design, we first parse the key design components from the response automatically (such as code snippets, key parameters, etc.). These components are then fed into our evaluation pipeline, which runs domain-specific simulations to assess the design’s performance against the specified task requirements. Guided by a per-task rubric table, the evaluation pipeline outputs three key results: a binary pass/fail indicator, a numerical score between 0 and 100, and a detailed log that records the evaluation process for further analysis. Below we discuss each step in detail.

\textbf{Structured Responses from LLMs.}  
To constrain the output format for a wide range of LLMs and ensure compatibility with our evaluation pipeline, we adopt the popular open-source Python library \texttt{instructor} \citep{liu2024instructor}, which is built on top of Pydantic and facilitates structured LLM responses. By defining schema templates that specify expected fields, such as design parameters or code snippets, the \texttt{instructor} package enables LLMs to produce outputs in a pre-defined format. We provide an example at Figure \ref{fig:response_structure} in Appendix \ref{app:LLM_response}. Specifically, LLMs are instructed to construct their responses into two main parts: (1) a \texttt{reasoning} field, which contains the step-by-step reasoning process for solving the task, and (2) a \texttt{ConfigFile} class, which summarizes the final design results, including their design choices or code snippets. During evaluation, the fields defined in \texttt{ConfigFile} can be automatically parsed to trigger the simulation-based evaluation pipeline.

\textbf{Task-Specific Evaluation Pipeline.}  
A simulation-based evaluation pipeline was designed for each task. For example, control engineering tasks may evaluate closed-loop dynamics through metrics such as rise time, settling time, overshoot, and phase/gain margins via MATLAB. The evaluation pipeline outputs three key results: (1) a binary pass/fail indicator for meeting all performance requirements, (2) a numerical score (0–100) reflecting fine-grained performance analysis, and (3) an evaluation log capturing simulation outputs, performance metrics, or error messages.

\begin{figure}[t!]
\centering
\includegraphics[width=0.85\textwidth]{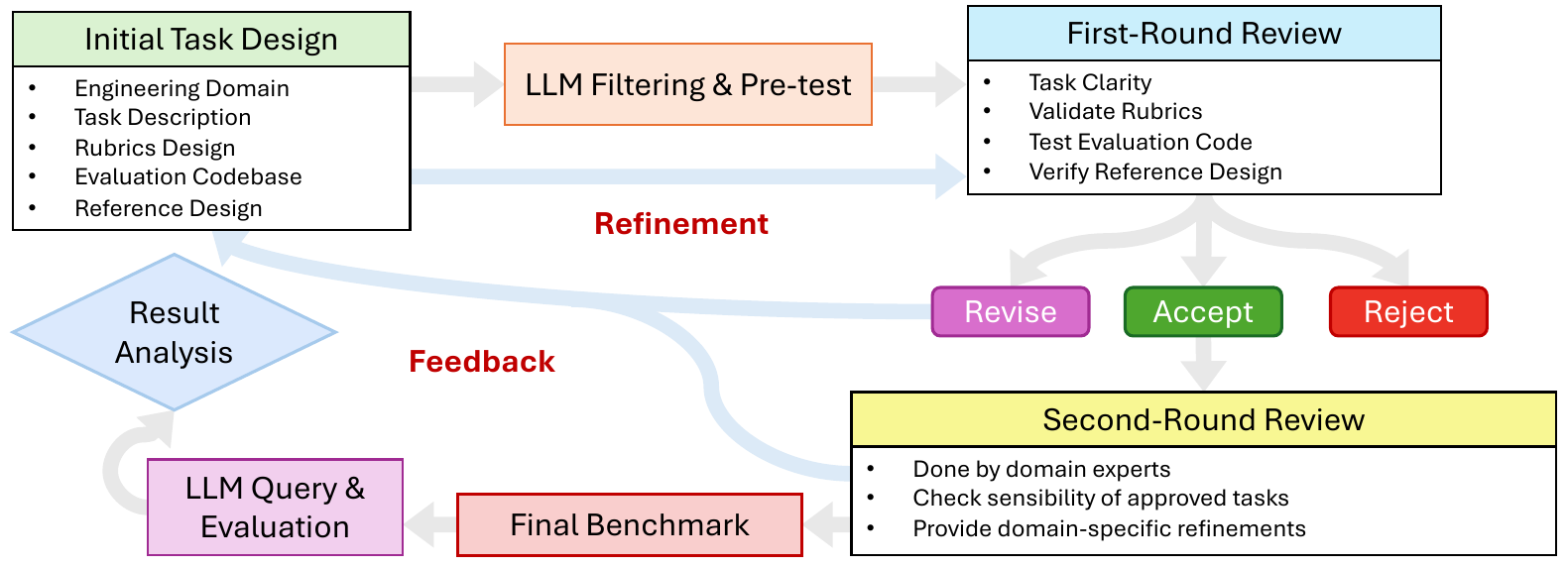}
\caption{An overview of the construction process of \textsc{EngDesign}, illustrating stages from initial task design, LLM filtering, and expert review to final benchmark integration, with iterative refinement and evaluation.}
\label{figure:overview}
\end{figure}

\subsection{Construction of \textsc{EngDesign}}
\label{sec:construction}

The construction of \textsc{EngDesign} follows a multi-stage process that integrates both automated validation and human expertise as shown in Figure~\ref{figure:overview}:

\begin{enumerate}
    \item \textbf{Initial Task Design.}  
    We recruited graduate students and researchers from various engineering disciplines to contribute initial task proposals, drawing on open-source resources and their domain knowledge. Contributors were guided to follow a standardized submission format, which includes a task description, evaluation rubric, executable evaluation codebase, and a reference solution.

    \item \textbf{LLM Filtering and Pre-Test.}  
    Submitted tasks undergo an initial filtering phase using a language model (o4-mini) to assess prompt sufficiency and the functionality of the evaluation code. Specifically, we prompt the LLM to classify the task’s engineering domain and determine whether the information provided is adequate to solve the problem. If the model flags missing or unclear details, we work with the original contributor to address those gaps.

    \item \textbf{First-Round Review.}  
    Tasks that pass the pre-test enter a first-round review involving close collaboration between reviewers and the original authors. The review ensures that:
    \begin{itemize}
        \item The task description is clear, self-contained, and complete.  
        \item The evaluation rubric is well-defined with partial credit, and aligns with the task goals.  
        \item The evaluation codebase is executable, robust, and handles typical edge cases.  
        \item The reference design meets all performance requirements, validating the task feasibility.
    \end{itemize}
    Based on this review, tasks fall into one of three categories:  
    \textbf{Accept}: the task meets all requirements and proceeds to expert review;  
    \textbf{Revise}: minor issues are identified and addressed in collaboration with the contributor; or 
    \textbf{Reject}: the task is excluded due to critical issues such as unfeasibility, triviality, or misalignment with design-oriented evaluation.

    \item \textbf{Second-Round Review with Domain Experts.}  
    Tasks accepted in the first round are reviewed by faculty experts in the corresponding engineering domain. These experts assess the technical soundness and relevance of each task to ensure it serves as a meaningful evaluation of LLMs in realistic design settings. 

    \item \textbf{Final Integration.}  
    Tasks that pass both review stages are standardized to match the benchmark’s formatting and structural guidelines, and are formally included in the \textsc{EngDesign}.
\end{enumerate}

\section{Experiments}
\label{sec:experiments}

\begin{table}
  \caption{Average pass rate (\%) results (each task evaluated over 3 trials). $^*$ Model is not multi-modal and was evaluated only on the text-only subset. We report text-only results for all models in Appendix \ref{app:add_eval}. The domain abbreviations are: AICD = Analog Integrated Circuit Design, Arch = Computer Architecture Design, Ctrl = Control Design, DHD = Digital Hardware Design, Mech = Mechanical Systems, OS = Operating System Design, Robo = Robotics, SigP = Signal Processing, Stru = Structure Design. The best results for each column are highlighted in bold.}
  \label{main-table}
  \centering
  \resizebox{\textwidth}{!}{
    \begin{tabular}{lcccccccccc}
      \toprule                
      \textbf{Model} & \textbf{Overall} & \textbf{AICD} & \textbf{Arch} & \textbf{Ctrl} & \textbf{DHD} & \textbf{Mech} & \textbf{OS} & \textbf{Robo} & \textbf{SigP} & \textbf{Stru} \\
      \midrule
      \multicolumn{11}{c}{\textbf{Chat Models}} \\
      \midrule
      GPT-4o & 15.68 & 0.00 & 20.00 & 18.52 & 10.26 & 0.00 & 4.17 & 26.67 & 17.65 & 25.64 \\
      DeepSeek-v3$^*$ & 17.92 & 0.00 & 0.00 & 27.27 & 25.64 & 0.00 & 0.00 & 16.67 & 12.82 & 38.89 \\
      Gemini-2.0-Flash & 14.16 & 0.00 & 0.00 & 15.15 & 16.67 & 4.76 & 0.00 & 20.00 & 5.13 & 36.11 \\
      Claude-3.7-Sonnet & 22.61 & 0.00 & \textbf{40.00} & 16.67 & 33.33 & 14.29 & 0.00 & 33.33 & 21.57 & 30.77 \\
      \midrule
      \multicolumn{11}{c}{\textbf{Reasoning Models}} \\
      \midrule
      o1 & 29.17 & 0.00 & \textbf{40.00} & 24.07 & 41.03 & 14.29 & \textbf{37.50} & 50.00 & 25.49 & 23.08 \\
      o3 & \textbf{34.38} & 0.00 & \textbf{40.00} & 35.19 & 20.51 & 23.81 & 25.00 & 63.33 & \textbf{41.18} & 30.77 \\
      o3-high & 33.57 & 0.00 & \textbf{40.00} & 27.78 & 20.51 & \textbf{28.57} & 30.43 & \textbf{68.97} & 35.29 & 41.03 \\
      o4-mini & 31.60 & 0.00 & 33.33 & 24.07 & 41.03 & 19.05 & 29.17 & 43.33 & 33.33 & 41.03 \\
      o4-mini-high & 34.04 & 0.00 & \textbf{40.00} & 27.78 & \textbf{47.22} & 19.05 & \textbf{37.50} & 46.67 & 35.29 & 35.90 \\
      Gemini-2.5-Pro & 29.54 & 0.00 & 0.00 & 33.33 & 43.59 & 0.00 & 9.52 & 56.67 & 12.82 & \textbf{50.00} \\
      DeepSeek-R1$^*$ & 25.53 & 0.00 & 36.36 & \textbf{36.36} & 38.46 & 4.76 & 5.26 & 26.67 & 20.51 & 41.67 \\
      Claude-3.7-Thinking & 20.07 & 0.00 & 33.33 & 18.52 & 17.95 & 9.52 & 0.00 & 40.00 & 19.61 & 28.21 \\
      \bottomrule
    \end{tabular}
  }
\end{table}

We evaluate a set of representative LLMs on the \textsc{EngDesign} and analyze their performance.

\textbf{Evaluated LLMs.} For chat models, we include GPT-4o, Claude-3.7-Sonnet, Gemini-2.0-Flash, and DeepSeek-v3. For reasoning models, we evaluate o1, o3, o3-high, o4-mini, o4-mini-high, Claude-3.7-Thinking, DeepSeek-R1, and Gemini-2.5-Pro.

\textbf{Evaluation Metrics.} We report three primary evaluation metrics: (1) \textbf{Average Pass Rate}, (2) \textbf{Average Score}, and (3) \textbf{Reasoning Robustness}. Each task is evaluated over three independent trials per model. To measure reasoning robustness, we compute the ratio between the number of tasks where all three trials passed and the number of tasks where at least one trial passed. This ratio, which ranges from 0 to 1, serves as a straightforward indicator of an LLM’s reasoning consistency. A value closer to 1 indicates higher robustness and more stable reasoning behavior across repeated runs.
Formal definitions of all evaluation metrics are provided in Appendix~\ref{app:eval_metric}.

\subsection{Main Results}

Table \ref{main-table} presents the average pass rates and average score across evaluated LLMs, across 9 distinct engineering design areas. The average score demonstrates a strong correlation with the pass rate, which we provided the detailed results in Appendix \ref{app:add_eval}. We make the following key observations.

\textbf{\textsc{EngDesign} presents a highly challenging benchmark, with even the best-performing models achieving only modest pass rate.}
As shown in Table \ref{main-table}, no model surpasses a 35\% overall pass rate. The top-performing model, o3, achieves 34.38\%, while widely-used chat models such as GPT-4o and Gemini-2.0-Flash attain around 15\%. Analog IC design tasks stands as the most difficult ones where all the models got 0\% pass rate. These results highlight the difficulty of the benchmark, which stems from its requirements for domain-specific expertise, multi-step reasoning, and design trade-off considerations, capabilities that go beyond the strengths of current LLMs.  

\begin{figure}[t!]
\includegraphics[width=1\linewidth]{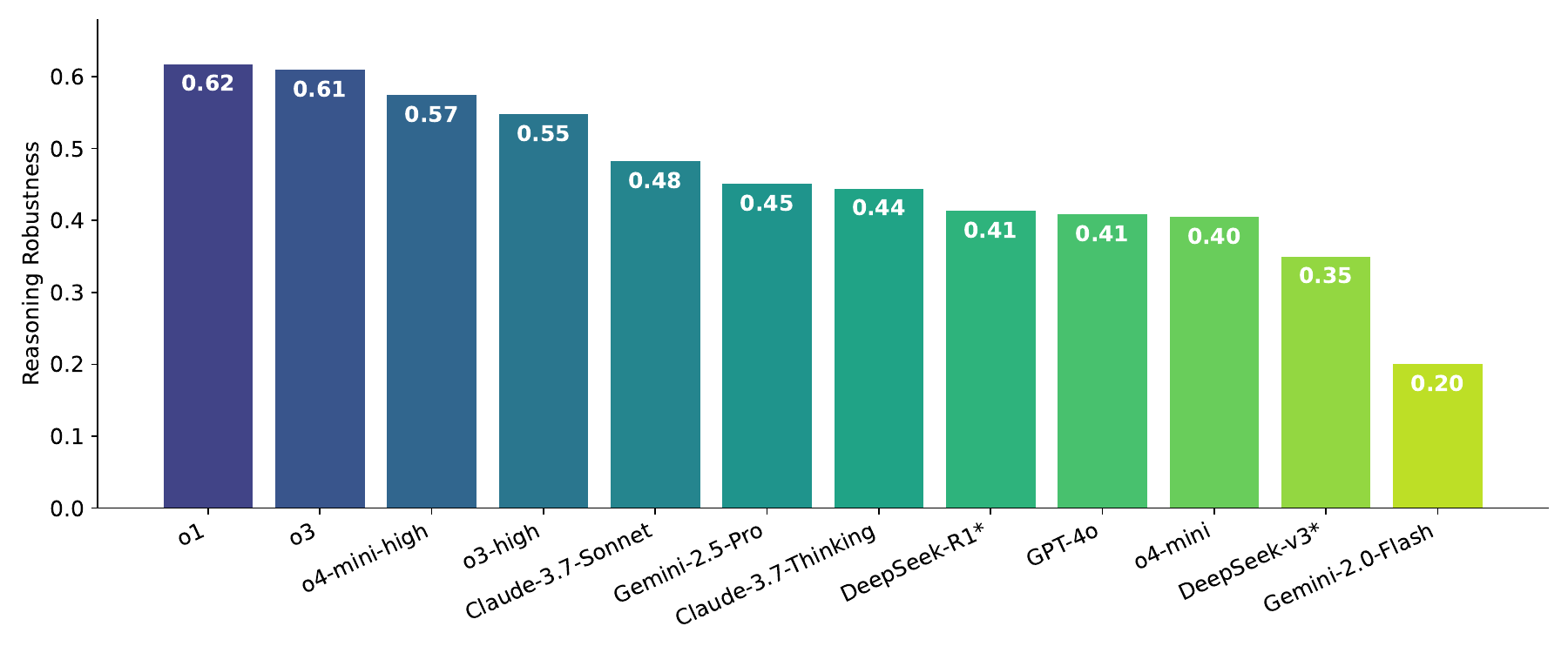}
\caption{Reasoning robustness distribution of evaluated LLMs on \textsc{EngDesign}.}
\label{fig:reasoning_robustness}
\end{figure}

\textbf{Reasoning models outperform general-purpose chat models.} Models designed with enhanced reasoning capabilities, including {o3}, {o4-mini}, and {DeepSeek-R1}, consistently outperform general-purpose chat models across most engineering domains. However, this trend does not hold for Claude models, where both {Claude-3.7-Sonnet} and {Claude-3.7-Thinking} exhibit similar pass rates despite their differences in reasoning emphasis.

\textbf{Reasoning Models are more robust in general.} Figure \ref{fig:reasoning_robustness} shows the reasoning robustness of all evaluated models. Reasoning-focused models such as {o1}, {o3}, and {o4-mini-high} achieve the highest robustness scores (0.62, 0.61, and 0.57, respectively), indicating strong consistency across repeated trials on tasks they are capable of solving. In contrast, chat-oriented models like Gemini-2.0-flash and DeepSeek-v3 exhibit substantially lower robustness (0.20 and 0.35), reflecting less stable reasoning behavior. Notably, the gap in reasoning robustness between top-performing reasoning models and weaker chat models is more pronounced than in average pass rate as shown in Table \ref{main-table}—up to a $3\times$ performance difference in robustness compared to a $2\times$ difference in pass rate. This highlights robustness as a distinguishing strength of reasoning models.

\begin{figure}[h]
\includegraphics[width=\linewidth]{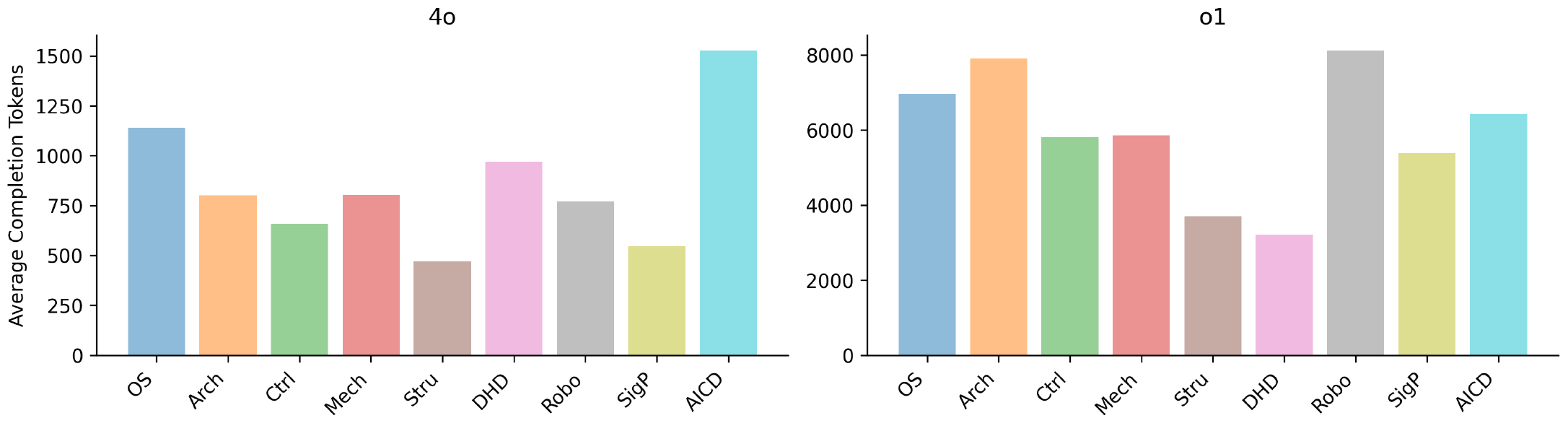}
\caption{Average token consumption of evaluated LLMs on \textsc{EngDesign}.}
\label{fig:token_num}
\end{figure}

\textbf{Token Consumption}. Reasoning models require substantially more compute due to longer inference time. We analyze the number of completion tokens generated across different models. As shown in Figure \ref{fig:token_num}, the reasoning model (o1) produces significantly more tokens than the non-reasoning model (GPT-4o) to achieve performance gains. Optimizing the latency–performance trade-off will be essential for making future models more accessible and deployable in real-world engineering design scenarios. Additional results are provided in Appendix \ref{app:token_consumption}.

\subsection{Iterative Design}
\label{sec:iter_design}

\begin{figure}[t]
\includegraphics[width=\linewidth]{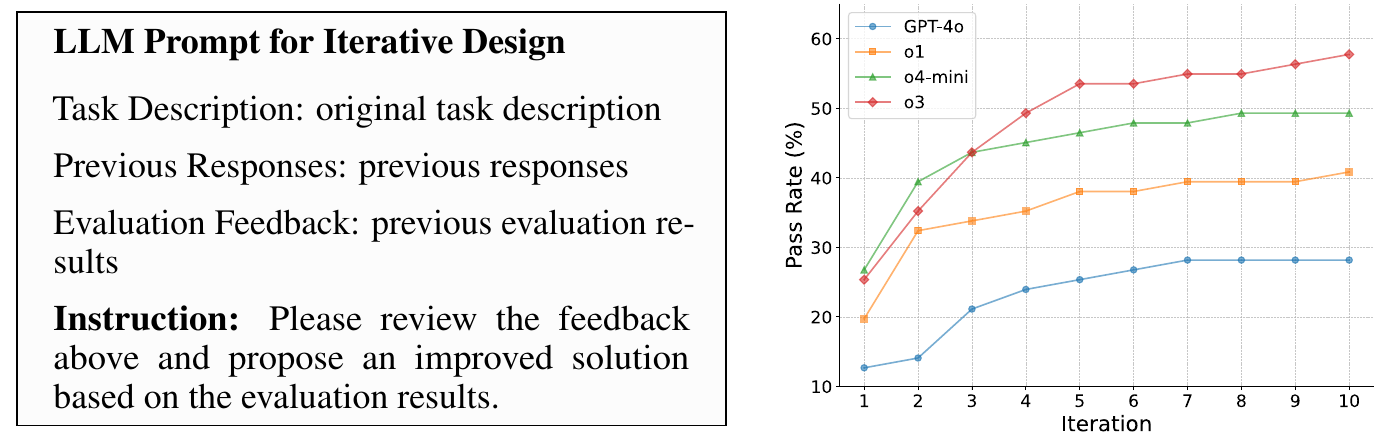}
\caption{\textbf{Left}: LLM prompt for iterative design. \textbf{Right}: Average pass rate across iterations, showing how performance evolves with feedback-driven updates.}
\label{fig:iterative}
\end{figure}

To emulate the workflow of human engineers, we implement an iterative design protocol that allows LLMs to refine their solutions based on feedback from previous attempts. In our implementation, the LLM is provided with its previous design output along with corresponding evaluation results, such as scores, performance metrics, and diagnostic logs, and is then prompted to generate an improved design in the subsequent iteration. The prompt we used for iterative design is presented in Figure \ref{fig:iterative}. 

As shown in Figure \ref{fig:iterative}, model performance consistently improves with additional iterations. Notably, {o3} achieves almost a 60\% pass rate after ten iterations.\footnote{Due to time and resource constraints, the iterative design protocol was applied to a selected subset (71 tasks) of \textsc{EngDesign} tasks using four representative models: GPT-4o, {o1}, {o3}, and {o4-mini}. As a result, reported pass rates in the iterative experiments are slightly lower than the ones report in Table \ref{main-table}. Nonetheless, this subset demonstrates the effectiveness of feedback-driven refinement and highlights the potential of LLMs to improve design quality through iterative design.}. However, we also observed that iterative design does not help in all cases. For example, in Analog IC design tasks, models still fail to meet the requirements even after ten iterations.

\subsection{Failure Analysis}
To better understand the limitations of LLMs in engineering design tasks, we define five primary error categories observed in \textsc{EngDesign}: 
\begin{enumerate}
    \item \textbf{Domain Knowledge Error (DKE).}  The model lacks essential engineering knowledge required to interpret or solve the task correctly.
    \item \textbf{Constraint Violation Error (CVE).}  The model generates designs that violate explicitly stated constraints, such as structural limits, performance bounds, or budgetary requirements.
    \item \textbf{Prior Knowledge Overreliance (PKO).}  The model applies memorized patterns or textbook solutions without adapting them to the specific requirements of the task, resulting in suboptimal or invalid outputs.
    \item \textbf{Hallucination (HAL).}  The model fabricates content, such as equations, parameters, or designs, that is unsupported by the task description or domain knowledge.
    \item \textbf{Computation Error (CE).}  The model performs arithmetic or symbolic computations incorrectly despite correct methodological reasoning.
    \item \textbf{Others.} Residual errors that do not fit the above categories
\end{enumerate}

\begin{table}[h]
  \caption{Error types statistics for three representative models.}
  \label{tab:additional_error}
  \centering
    \begin{tabular}{l|cccccc}
      \toprule                
      \textbf{Model} & \textbf{DKE} & \textbf{CVE} & \textbf{PKO} & \textbf{HAL} & \textbf{COM} & \textbf{Others}  \\
      \midrule
      o4‑mini	 &33.3\% &25.2\% &18.8\% &12.6\% &9.0\% &0.9\%  \\
      Gemini‑2.5‑Pro	 &31.9\% &31.9\% &15.9\% &12.4\% &6.2\% &1.7\% \\
      Claude‑3.7‑Sonnet	&30.7\% &36.0\% &10.5\% &13.2\% &7.0\% &2.6\% \\
      \bottomrule
    \end{tabular}
\end{table}

We analyzed the responses of three representative models (o4-mini, Gemini-2.5-Pro, and Claude-3.7-Sonnet) on \textsc{EngDesign} tasks where they failed evaluation, and manually annotated each failure with its corresponding error types. The results are summarized in Table~\ref{tab:additional_error}. Given the complexity of engineering design problems, many responses exhibited multiple failure modes; we therefore allowed multi-label annotations per task. 

Our analysis yields several key observations\footnote{Additional failure cases and qualitative analyses are provided in Appendix~\ref{app:failure_analysis}.}:

\textbf{1. Dominant failure modes:} Insufficient domain knowledge and constraint violations together account for roughly 55–67\% of all failures. This indicates that LLMs frequently struggle to apply domain-specific principles, satisfy task constraints, and generalize beyond memorized patterns.

\textbf{2. Second-tier issues:} Over-reliance on prior knowledge and hallucinations contribute 25–30\%, suggesting that incorrect reuse or misapplication of known facts remains a significant challenge.

\textbf{3. Low arithmetic fragility:} Pure computational errors are rare ($\le$9\%), implying that numerical reasoning is not the primary bottleneck for current frontier models.

\section{Related Work}

\textbf{Large Language Models.} Advances in large language models (LLMs) have been propelled by scaling, instruction tuning, and improved reasoning. Frontier models such as GPT-3 and GPT-4~\citep{brown2020language, achiam2023gpt}, Claude~\citep{anthropic2024claude37}, Gemini~\citep{team2023gemini}, and DeepSeek~\citep{liu2024deepseek} exhibit broad general capabilities. Recent work enhances reasoning through architectural and prompting innovations, including Chain-of-Thought~\citep{wei2022chain} and least-to-most prompting~\citep{zhou2022least}. These developments have led to reasoning-oriented models such as OpenAI’s o-series~\citep{openai2024o3o4mini}, Claude-Thinking~\citep{anthropic2024visiblethinking}, Gemini 2.5 Pro~\citep{google2025gemini25}, and DeepSeek-R1~\citep{guo2025deepseek}. Yet, their effectiveness in domain-specific contexts, particularly engineering design, remains largely unexplored. \textsc{EngDesign} fills this gap by evaluating LLMs within realistic, simulation-based engineering workflows.

\textbf{General Purpose Benchmarks for LLMs.}  
A variety of benchmarks assess LLM reasoning and problem-solving capabilities. MMLU~\citep{hendrycks2020measuring} and MMLU-Pro~\citep{wang2024mmlu} evaluate broad subject knowledge, while GAIA~\citep{mialon2023gaia} and HLE~\citep{phan2025humanity} test long-context and high-difficulty reasoning. HumanEval~\citep{chen2021evaluating} measures code generation accuracy, GPQA~\citep{rein2024gpqa} targets graduate-level science, and GSM8k~\citep{cobbe2021training} focuses on arithmetic reasoning. Recent work such as DynaMath~\citep{zou2024dynamath} evaluates VLMs' mathematical reasoning robustness.

\textbf{LLMs in Engineering Domains.}  
Recent studies have started to explore the applicability of LLMs to engineering contexts. For example, \citep{syed2024benchmarking, kevian2024capabilities, guo2025resbench, xu2025evaluating, xiong2025can, eslaminia2024fdm, ogo2024can} evaluate LLMs across engineering subfields using curated QA datasets. However, these efforts only focuses one specific domain or largely focus on factual recall rather than generative design capabilities. Several domain-specific studies examine LLM-assisted design workflows. For example, ControlAgent~\citep{guo2024controlagent} integrates control theory solvers for automated controller design, while AnalogCoder~\citep{lai2024analogcoder} and SPICED~\citep{chaudhuri2024spiced} target analog circuit and SoC design using prompt engineering and retrieval-based techniques. Retrieval-augmented and programmatic strategies have also been explored to enhance LLM performance~\citep{ghosh2024retrieval, alsaqer2024potential}. Other studies examine LLMs in mechanical design~\citep{lu2024constructing}, cross-domain creativity~\citep{jia2024enabling}, and computational engineering~\citep{hamann2024perspective, xu2024unlocking, majumder2024exploring, makatura2023can}. Compared to prior work, \textsc{EngDesign} spans multiple engineering domains and introduces a novel evaluation framework that combines generative outputs from LLMs with simulation-based evaluation to assess performance in realistic design settings.

\section{Conclusion}

In this work, we introduced \textsc{EngDesign}, a comprehensive benchmark for rigorously evaluating large language models (LLMs) in realistic, multi-domain engineering design scenarios. \textsc{EngDesign} emphasizes end-to-end design synthesis, constraint satisfaction, and simulation-based validation across diverse engineering disciplines. Our results show that \textsc{EngDesign} poses substantial challenges: even the most capable frontier models achieve only modest performance. We envision \textsc{EngDesign} as a foundation for future research at the intersection of language, reasoning, and engineering intelligence. By providing a standardized and reproducible evaluation testbed, we hope to accelerate progress toward the long-term goal of developing general-purpose AI engineers.

\section*{Acknowledgment}

Xingang Guo and Bin Hu are generously supported by the Amazon AICE award.

{
\small
\bibliographystyle{plainnat}
\bibliography{neurips_2025}
}

\medskip
\newpage
\section*{NeurIPS Paper Checklist}

The checklist is designed to encourage best practices for responsible machine learning research, addressing issues of reproducibility, transparency, research ethics, and societal impact. Do not remove the checklist: {\bf The papers not including the checklist will be desk rejected.} The checklist should follow the references and follow the (optional) supplemental material.  The checklist does NOT count towards the page
limit. 

Please read the checklist guidelines carefully for information on how to answer these questions. For each question in the checklist:
\begin{itemize}
    \item You should answer \answerYes{}, \answerNo{}, or \answerNA{}.
    \item \answerNA{} means either that the question is Not Applicable for that particular paper or the relevant information is Not Available.
    \item Please provide a short (1–2 sentence) justification right after your answer (even for NA). 
\end{itemize}

{\bf The checklist answers are an integral part of your paper submission.} They are visible to the reviewers, area chairs, senior area chairs, and ethics reviewers. You will be asked to also include it (after eventual revisions) with the final version of your paper, and its final version will be published with the paper.

The reviewers of your paper will be asked to use the checklist as one of the factors in their evaluation. While "\answerYes{}" is generally preferable to "\answerNo{}", it is perfectly acceptable to answer "\answerNo{}" provided a proper justification is given (e.g., "error bars are not reported because it would be too computationally expensive" or "we were unable to find the license for the dataset we used"). In general, answering "\answerNo{}" or "\answerNA{}" is not grounds for rejection. While the questions are phrased in a binary way, we acknowledge that the true answer is often more nuanced, so please just use your best judgment and write a justification to elaborate. All supporting evidence can appear either in the main paper or the supplemental material, provided in appendix. If you answer \answerYes{} to a question, in the justification please point to the section(s) where related material for the question can be found.

IMPORTANT, please:
\begin{itemize}
    \item {\bf Delete this instruction block, but keep the section heading ``NeurIPS Paper Checklist"},
    \item  {\bf Keep the checklist subsection headings, questions/answers and guidelines below.}
    \item {\bf Do not modify the questions and only use the provided macros for your answers}.
\end{itemize}


\begin{enumerate}

\item {\bf Claims}
    \item[] Question: Do the main claims made in the abstract and introduction accurately reflect the paper's contributions and scope?
    \item[] Answer: \answerYes{} 
    \item[] Justification: We introduced \textsc{EngDesign} benchmark in this work to benchmark the engineering design capabilities of current LLMs. The dataset details have been discussed in Section \ref{sec:engdesign}, and our experimental results are included in Section \ref{sec:experiments}.

\item {\bf Limitations}
    \item[] Question: Does the paper discuss the limitations of the work performed by the authors?
    \item[] Answer: \answerYes{} 
    \item[] Justification: We have discussed the limitations of our work in Appendix \ref{app:lim_impact}.

\item {\bf Theory assumptions and proofs}
    \item[] Question: For each theoretical result, does the paper provide the full set of assumptions and a complete (and correct) proof?
    \item[] Answer: \answerNA{} 
    \item[] Justification: Our paper does not include theoretical results.

    \item {\bf Experimental result reproducibility}
    \item[] Question: Does the paper fully disclose all the information needed to reproduce the main experimental results of the paper to the extent that it affects the main claims and/or conclusions of the paper (regardless of whether the code and data are provided or not)?
    \item[] Answer: \answerYes{} 
    \item[] Justification: We have open-sourced our dataset and evaluation pipeline to reproduce our work. 

\item {\bf Open access to data and code}
    \item[] Question: Does the paper provide open access to the data and code, with sufficient instructions to faithfully reproduce the main experimental results, as described in supplemental material?
    \item[] Answer: \answerYes{} 
    \item[] Justification: We have provided open access to the benchmark data including all the task prompt, domains, and topics. In addition, we have provided a github repo contains all the evaluations logs and evaluation pipeline for each task of \textsc{EngDesign}.

\item {\bf Experimental setting/details}
    \item[] Question: Does the paper specify all the training and test details (e.g., data splits, hyperparameters, how they were chosen, type of optimizer, etc.) necessary to understand the results?
    \item[] Answer: \answerYes{} 
    \item[] Justification: We can explained our experimental setup in detail in Appendix \ref{sec:exp_setup}.

\item {\bf Experiment statistical significance}
    \item[] Question: Does the paper report error bars suitably and correctly defined or other appropriate information about the statistical significance of the experiments?
    \item[] Answer: \answerYes{} 
    \item[] Justification: We report error bars in Table \ref{main-table-w-error} and Table \ref{main-table2-w-error}, which presents the detailed evaluation results. Specifically, during evaluation, we ran each task three independent trials to compute the mean and standard deviation results. 

\item {\bf Experiments compute resources}
    \item[] Question: For each experiment, does the paper provide sufficient information on the computer resources (type of compute workers, memory, time of execution) needed to reproduce the experiments?
    \item[] Answer: \answerYes{} 
    \item[] Justification: We have discussed the required compute resources in Appendix \ref{sec:exp_setup}.
    
\item {\bf Code of ethics}
    \item[] Question: Does the research conducted in the paper conform, in every respect, with the NeurIPS Code of Ethics \url{https://neurips.cc/public/EthicsGuidelines}?
    \item[] Answer: \answerYes{} 
    \item[] Justification: This work adheres to the NeurIPS Code of Ethics. All experiments were conducted responsibly, without harm to individuals or groups, and without the use of sensitive or personally identifiable data. We also carefully considered the potential societal impact of our methods and reported both the limitations and potential risks in the paper.

\item {\bf Broader impacts}
    \item[] Question: Does the paper discuss both potential positive societal impacts and negative societal impacts of the work performed?
    \item[] Answer: \answerYes{} 
    \item[] Justification: We have discussed the broader impacts of our work in Appendix \ref{app:lim_impact}.
    
\item {\bf Safeguards}
    \item[] Question: Does the paper describe safeguards that have been put in place for responsible release of data or models that have a high risk for misuse (e.g., pretrained language models, image generators, or scraped datasets)?
    \item[] Answer: \answerNA{} 
    \item[] Justification: Our work does not involve the release of any pretrained models, generative systems, or large-scale scraped datasets that would present a high risk of misuse. \textsc{EngDesign} is a benchmark comprising structured engineering design tasks, and all associated data is manually curated and domain-specific. It is intended solely for the evaluation of model capabilities in engineering design and does not raise the same risks associated with general-purpose generative models or unfiltered data releases.

\item {\bf Licenses for existing assets}
    \item[] Question: Are the creators or original owners of assets (e.g., code, data, models), used in the paper, properly credited and are the license and terms of use explicitly mentioned and properly respected?
    \item[] Answer: \answerYes{} 
    \item[] Justification: Tasks in \textsc{EngDesign} are curated by individual task contributors, and the corresponding references are cited properly (see Table \ref{task-level-description}).

\item {\bf New assets}
    \item[] Question: Are new assets introduced in the paper well documented and is the documentation provided alongside the assets?
    \item[] Answer: \answerYes{} 
    \item[] Justification: The paper introduces a new benchmark, \textsc{EngDesign}, which includes a suite of engineering design tasks along with structured evaluation scripts. Each task is documented with a clear task description, LLM response instruction, and evaluation codebase. Additionally, we provide metadata and instructions for running evaluations, reproducibility guidelines, and license information alongside the released assets. The assets are shared through publicly accessible repositories to support transparent and reproducible research.

\item {\bf Crowdsourcing and research with human subjects}
    \item[] Question: For crowdsourcing experiments and research with human subjects, does the paper include the full text of instructions given to participants and screenshots, if applicable, as well as details about compensation (if any)? 
    \item[] Answer: \answerNA{} 
    \item[] Justification: Our study does not involve crowdsourcing nor research with human subjects.

\item {\bf Institutional review board (IRB) approvals or equivalent for research with human subjects}
    \item[] Question: Does the paper describe potential risks incurred by study participants, whether such risks were disclosed to the subjects, and whether Institutional Review Board (IRB) approvals (or an equivalent approval/review based on the requirements of your country or institution) were obtained?
    \item[] Answer: \answerNA{} 
    \item[] Justification: Our study does not involve crowdsourcing nor research with human subjects.

\item {\bf Declaration of LLM usage}
    \item[] Question: Does the paper describe the usage of LLMs if it is an important, original, or non-standard component of the core methods in this research? Note that if the LLM is used only for writing, editing, or formatting purposes and does not impact the core methodology, scientific rigorousness, or originality of the research, declaration is not required.
    \item[] Answer: \answerNA{} 
    \item[] Justification: The core development of \textsc{EngDesign} in this work does not involve LLMs as any important, original, or non-standard components.

\end{enumerate}

\newpage
\appendix
\section{Limitations and Broader Impact}
\label{app:lim_impact}

\subsection*{Limitations}

While \textsc{EngDesign} provides a comprehensive benchmark for evaluating LLMs in engineering design tasks, some limitations remain:

\begin{enumerate}
    \item \textsc{EngDesign} currently covers 9 engineering design domains. Although these are diverse and representative, the benchmark does not yet encompass the full breadth of engineering disciplines or subfields.
    
    \item Our evaluation includes a selection of widely used and high-performing LLMs. However, it is not exhaustive. Many emerging or smaller-scale models are not included in our evaluation.
    
    \item We employ the \texttt{instructor} framework to enforce structured output from LLMs. While this tool has shown reliable performance throughout our evaluations, there are instances where LLMs fail to adhere to the expected output format. As \texttt{instructor} is still under active development, future improvements will be necessary to further enhance its robustness and compatibility across models.
\end{enumerate}

\subsection*{Broader Impact}

This work aims to advance our understanding of how LLMs perform in complex, real-world engineering design scenarios. However, we highlight the following considerations:

\begin{itemize}
    \item \textbf{Not a replacement for human engineers.} Our benchmark is intended to assess LLM capabilities, not to replace the expertise and judgment of professional engineers. Engineering design involves safety-critical decision-making, physical prototyping, and domain-specific knowledge that current LLMs cannot fully replicate.

    \item \textbf{Supporting real-world design workflows.} We hope that LLMs, when properly evaluated and deployed, can assist engineers by automating routine tasks, enhancing ideation, and accelerating early-stage design processes—particularly in settings with limited access to expert resources.

    \item \textbf{Need for rigorous safety checks.} Any LLM-generated design, particularly those intended for deployment in safety-critical domains such as aerospace, biomedical devices, civil infrastructure, or autonomous systems, must undergo thorough validation and safety assessment by qualified human experts before real-world application.

    \item \textbf{Ethical use and accessibility.} Care must be taken to ensure that advances in AI-assisted engineering design benefit a broad and diverse community. Efforts should be made to avoid misuse, bias amplification, or overreliance on unverified outputs.
\end{itemize}

\section{More on Related Work}
\label{app:related_work}

Recent years have seen a surge of domain-specific engineering benchmarks that evaluate AI systems on isolated reasoning or coding sub-tasks. However, most existing efforts fall short of assessing end-to-end design capability, that is, the ability to synthesize, implement, and validate complete engineering systems under realistic constraints. Below, we review representative benchmarks across key engineering domains and highlight how EngDesign differs by emphasizing holistic design and simulation-based validation.

\paragraph{Operating Systems}

OSVBench \citep{li2025osvbench} focuses on the specification and verification of simplified operating system components using formal methods. While it advances progress in formal verification, its tasks are limited to checking correctness of given code fragments. In contrast, \textsc{EngDesign-OS} tasks require participants to design and simulate new operating system subsystems (e.g., schedulers, memory allocators), moving beyond static specification toward dynamic design.

\paragraph{Computer Architecture}

Benchmarks such as FIXME \citep{wan2025fixme} and ChatCPU \citep{wang2024chatcpu}  center on verifying or patching existing CPU components through code reasoning. These tasks measure correctness and local optimization. \textsc{EngDesign-Arch}, by comparison, tasks models with designing and optimizing new micro-architectures and quantitatively validates performance through simulation, bridging reasoning and generative design.

\paragraph{Control Systems Design}

ControlBench \citep{kevian2024capabilities} and ControlEval \citep{guo2024controlagent} examine language models’ ability to answer control-theoretic questions or tune simple PID controllers. These benchmarks remain largely static, focusing on analytical understanding rather than synthesis. \textsc{EngDesign-Ctrl} extends the scope to diverse control system design problems, including state-feedback, robust control, and optimal control, and evaluates designs via dynamic simulations.

\paragraph{Mechanical Design}

Prior mechanical design studies, such as MechAgents \citep{ni2024mechagents}
developed a set of AI agents can solve specified elasticity problem, which focuses on one type of problems. \textsc{EngDesign-Mech} introduces multiple mechanism design tasks with physics-based simulation and objective trade-offs, providing a consistent evaluation framework across mechanical systems.

\paragraph{Structural Engineering}

Benchmarks like AEC-bench \citep{liang2025aecbench} that integrate a benchmark for architecture, engineering, and construction field domain questions and DrafterBench \citep{li2025drafterbench} that focuses on evaluting LLMs on technical drawing revision all focuses on speicify question and answer evaluation scenario. \textsc{EngDesign-Stru} instead requires sizing new structures under material and load trade-offs, capturing the creative reasoning central to structural design.

\paragraph{Digital Hardware Design}

VerilogEval \citep{liu2023verilogeval} focuses on writing and verifying RTL code snippets. These benchmarks measure correctness at the snippet level but not full design integration. In contrast, \textsc{EngDesign-DHD} targets complete module-level design with explicit trade-offs between latency, area, and power—mirroring real-world hardware co-design challenges.

\paragraph{Analog and Integrated Circuit Design}

Recent efforts such as AnalogCoder \citep{lai2025analogcoder} and AnalogCoder-Pro \citep{lai2025analogcoderpro} focuses on agent design on IC design domains. EngDesign-AICD advances this direction by provide a evaluation testbed and validates models' performance through SPICE-like simulation.

\paragraph{Robotics}
Benchmarks such as \citep{goebel2025can, yin2024safeagentbench} target high-level reasoning, perception, or planning. These are complementary but orthogonal to engineering design, as they do not involve building controllers or physical systems. \textsc{EngDesign-Robo} bridges this gap by coupling low-level control design with physics simulation, assessing dynamic feasibility and safety in robotic mechanisms.

\paragraph{Signal Processing} 
SensorBench \citep{quan2025sensorbench} focuses on sensor data. They emphasize data understanding, but not system-level design. \textsc{EngDesign-SigP} introduces tasks involving filter design, sensor configuration, and system-level optimization, encompassing a broader range of design decisions.

Across these domains, \textsc{EngDesign} distinguishes itself by requiring end-to-end design reasoning, cross-disciplinary synthesis, and simulation-based validation. Whereas prior benchmarks evaluate correctness or reasoning on partial tasks, EngDesign systematically measures an AI model’s ability to conceptualize, implement, and verify complex engineered systems—an essential step toward assessing the real-world utility of intelligent design agents.

\section{More on \textsc{EngDesign}}
\label{app:more_on_eng}

\subsection{Prompt Token Length Comparison}
\label{app:token_num_com}

Table \ref{token_count} reports the average number of tokens in the input prompts for various benchmarks, measured using the \texttt{o200k\_base} tokenizer. Notably, \textsc{EngDesign} exhibits significantly longer prompts (averaging 778.71 tokens) compared to other popular QA-style benchmarks such as MMLU‑Pro (61.76), HLE (250.03), and GSM8K (58.46). This reflects the greater contextual and structural complexity involved in realistic engineering design tasks, which often require extensive problem descriptions and domain-specific constraints.

\begin{table}[h]
  \caption{Average token counts across different benchmarks using the \texttt{o200k\_base} tokenizer.}
  \label{token_count}
  \centering
  \resizebox{\textwidth}{!}{%
    \begin{tabular}{lcccccc}
      \toprule                
      \textbf{Benchmark} & \textbf{MMLU‑Pro} & \textbf{HLE} & \textbf{SuperGPQA} & \textbf{DynaMATH} & \textbf{GSM8K} & \textbf{EngDesign}  \\
      \midrule
      Average Token \# & 61.76 & 250.03 & 57.89 & 27.83 & 58.46 & 778.71 \\
      \bottomrule
    \end{tabular}
  }
\end{table}

\subsection{More on construction of \textsc{EngDesign}}

As described in Section~\ref{sec:construction}, the development of \textsc{EngDesign} follows a rigorous multi-stage process, including initial task design and two rounds of review to ensure task quality and relevance. During the initial stage, over 40 task contributors submitted a total of 186 task proposals. Following thorough evaluation and revision across the two review phases, 101 tasks were accepted for inclusion in \textsc{EngDesign}, while the remaining 85 were discarded due to various issues such as insufficient clarity, infeasibility, or misalignment with the benchmark's design-oriented goals.

\subsection{Word Clouds of \textsc{EngDesign}}

\begin{figure}[h]
\centering
\includegraphics[width=0.95\textwidth]{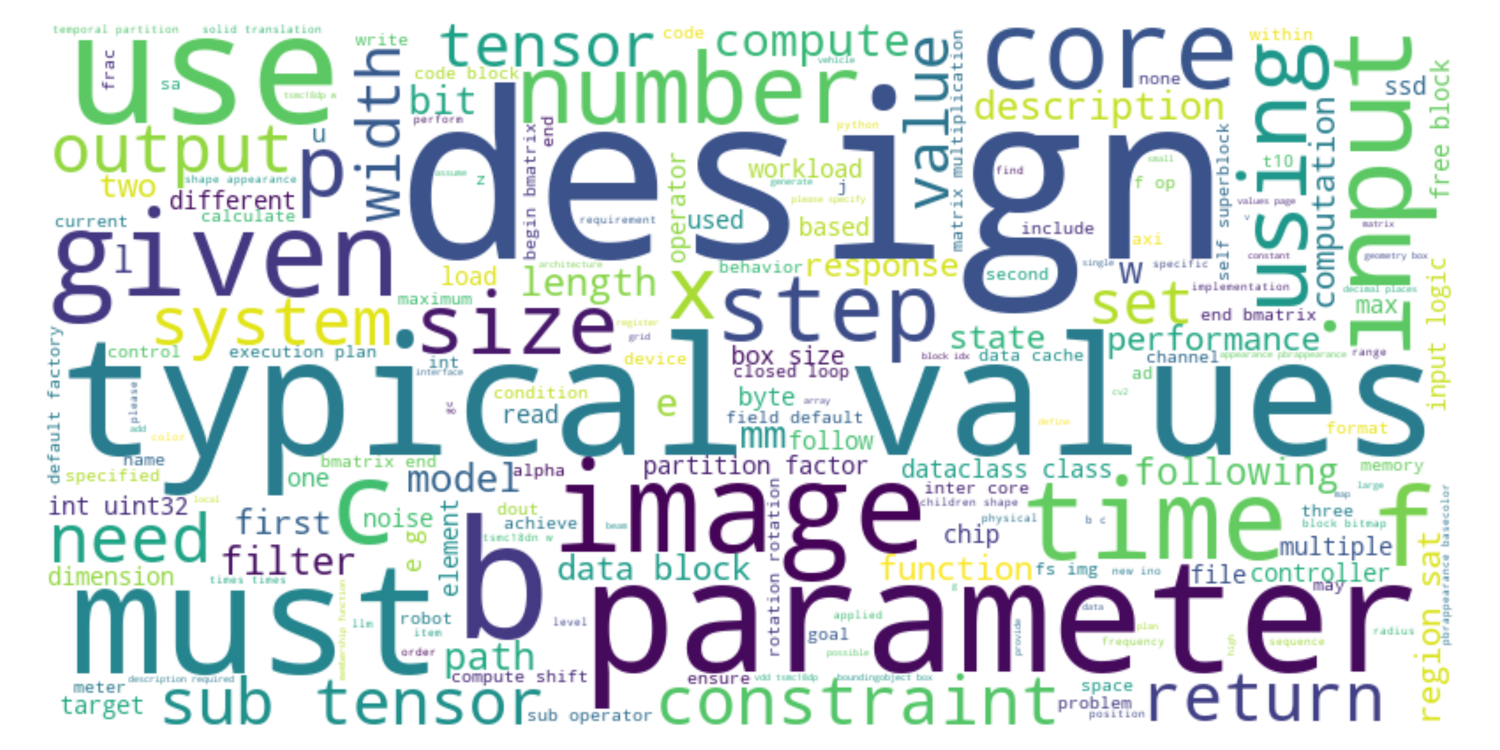}
\caption{Word Cloud of the overall \textsc{EngDesign} Prompts.}
\label{fig:engdesign_wordcloud}
\end{figure}

\begin{figure}
\centering
\begin{subfigure}[t]{0.33\textwidth}
    \centering
    \includegraphics[width=\textwidth]{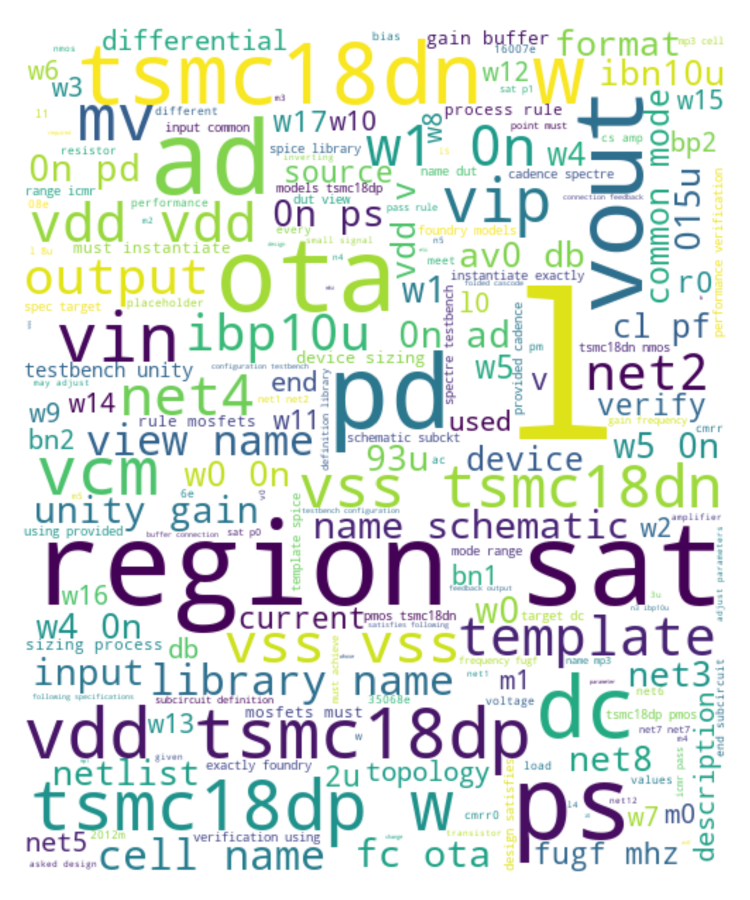}
    \caption{Analog integrated circuit design.}
\end{subfigure}%
\hfill
\begin{subfigure}[t]{0.33\textwidth}
    \centering
    \includegraphics[width=\textwidth]{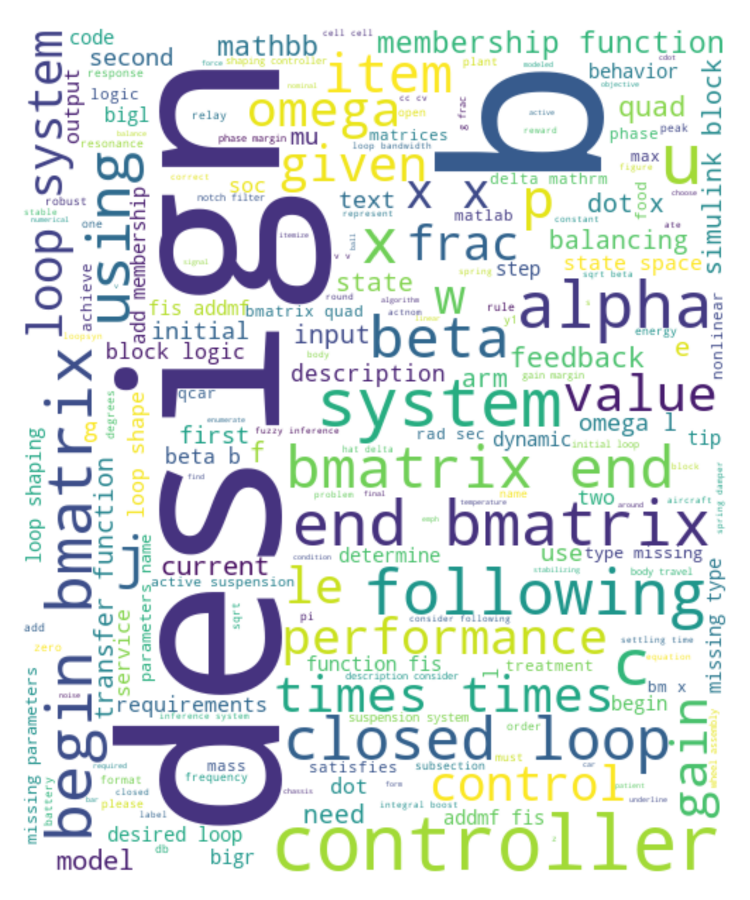}
    \caption{Control Design}
\end{subfigure}
\hfill
\begin{subfigure}[t]{0.33\textwidth}
    \centering
    \includegraphics[width=\textwidth]{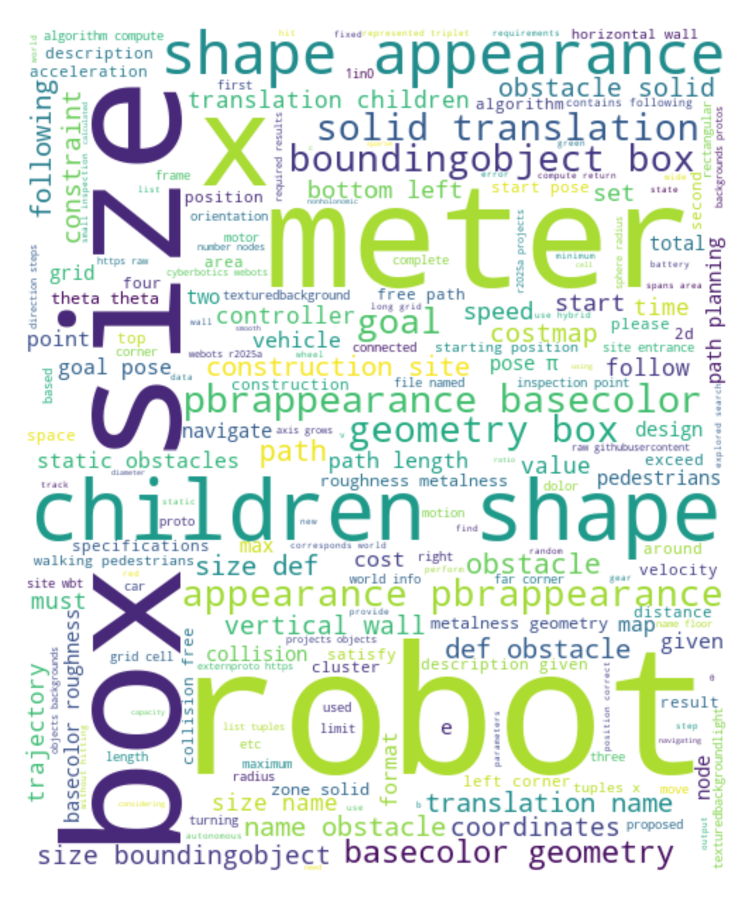}
    \caption{Robotics}
\end{subfigure}%

\vspace{0.5cm}

\begin{subfigure}[t]{0.33\textwidth}
    \centering
    \includegraphics[width=\textwidth]{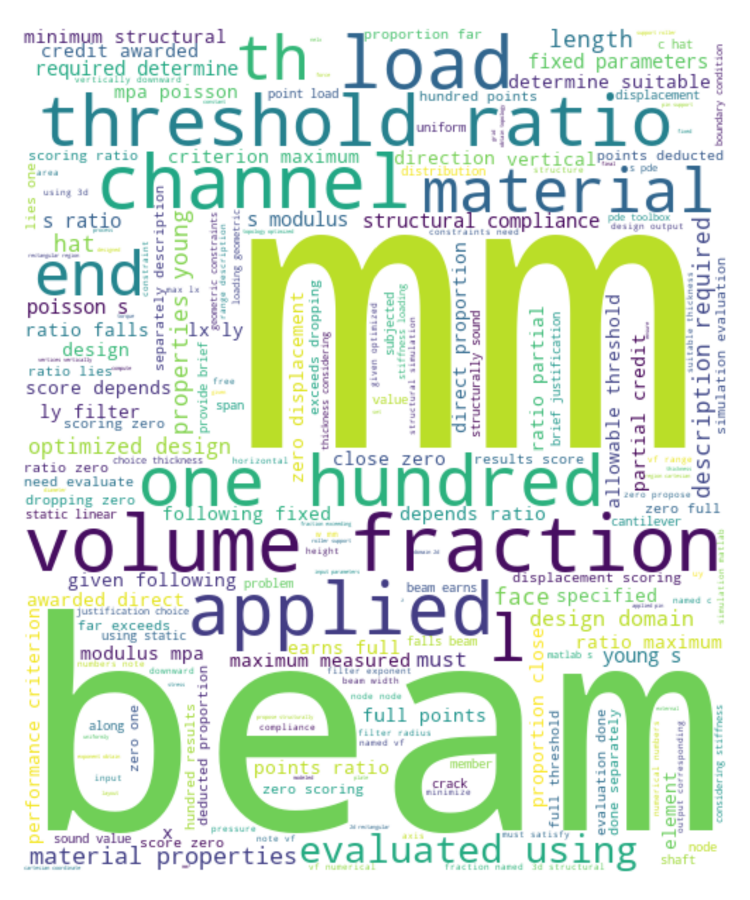}
    \caption{Structure Design}
\end{subfigure}
\hfill
\begin{subfigure}[t]{0.33\textwidth}
    \centering
    \includegraphics[width=\textwidth]{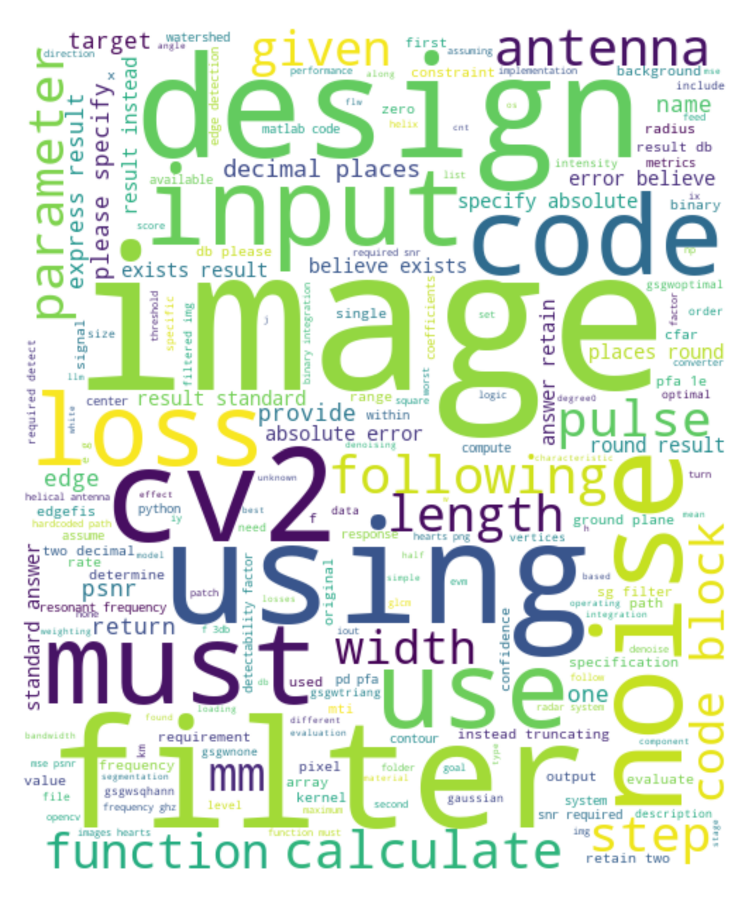}
    \caption{Signal Processing}
\end{subfigure}%
\hfill
\begin{subfigure}[t]{0.33\textwidth}
    \centering
    \includegraphics[width=\textwidth]{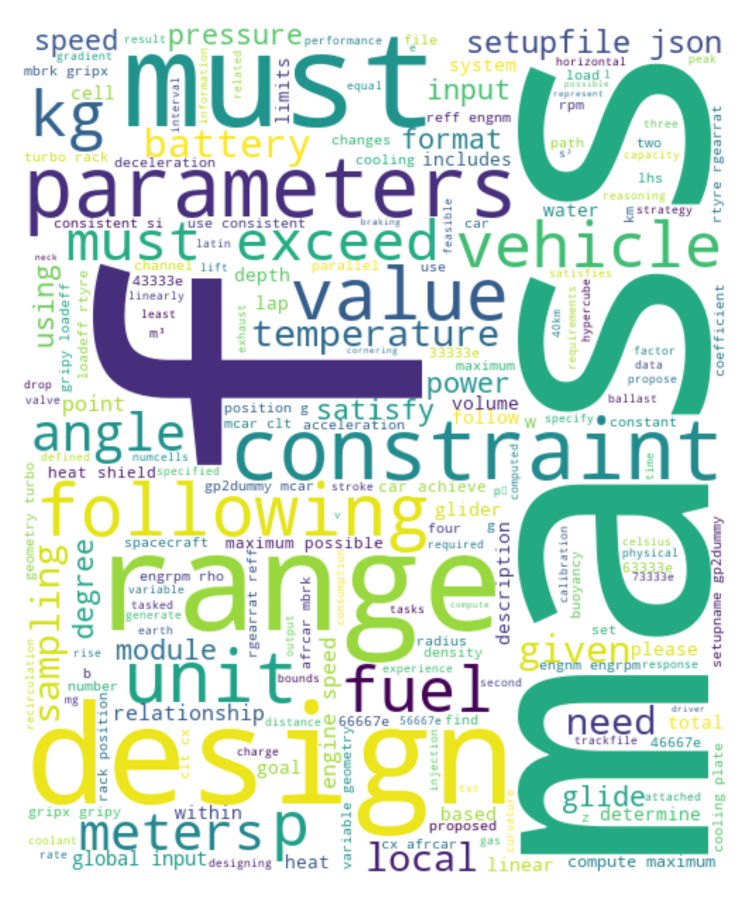}
    \caption{Mechanical Systems}
\end{subfigure}

\vspace{0.5cm}

\begin{subfigure}[t]{0.33\textwidth}
    \centering
    \includegraphics[width=\textwidth]{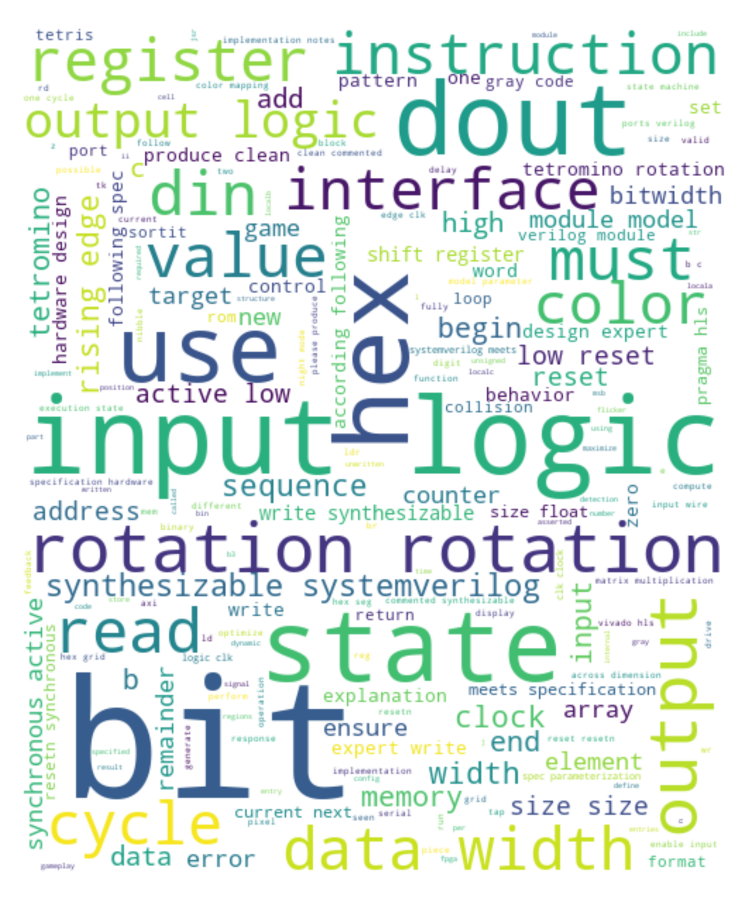}
    \caption{Digital Hardware Design}
\end{subfigure}%
\hfill
\begin{subfigure}[t]{0.33\textwidth}
    \centering
    \includegraphics[width=\textwidth]{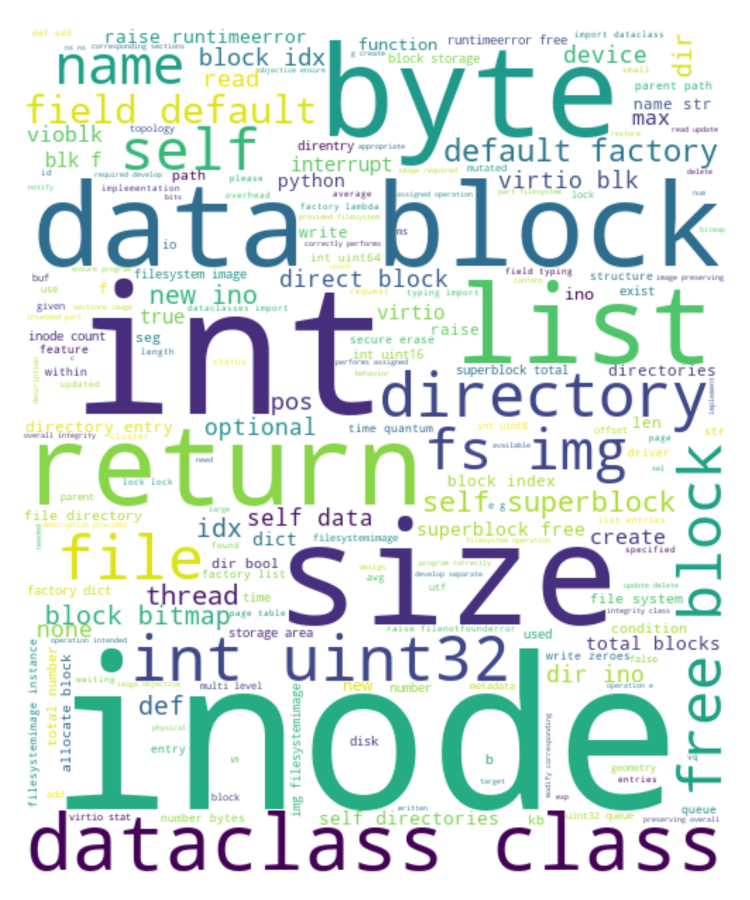}
    \caption{Operating System Design}
\end{subfigure}
\hfill
\begin{subfigure}[t]{0.33\textwidth}
    \centering
    \includegraphics[width=\textwidth]{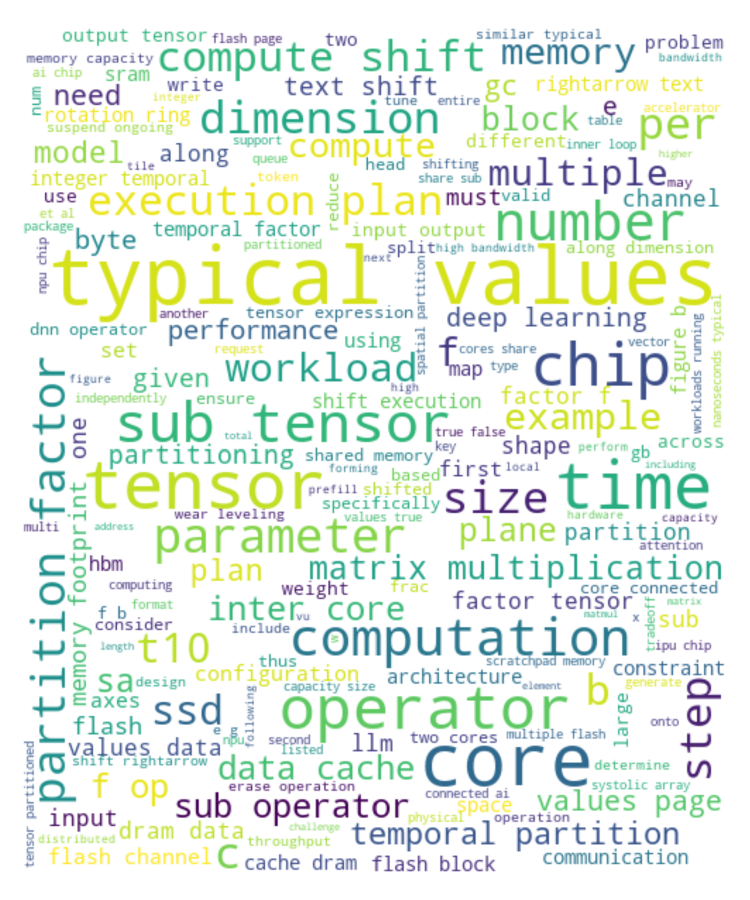}
    \caption{Computer Architecture Design}
\end{subfigure}

\caption{Word clouds of \textsc{EngDesign} for each engineering domain.}
\label{fig:appendix_wordclouds}
\end{figure}

Figure~\ref{fig:engdesign_wordcloud} and Figure~\ref{fig:appendix_wordclouds} present the word clouds of the most frequently occurring terms across all prompts and within each domain, respectively. These visualizations underscore the benchmark’s emphasis on design-specific vocabulary and highlight the diversity of engineering contexts it encompasses.

\section{More on Evaluation}

Figure \ref{figure:engdesign_pipeline} illustrates the benchmarking pipeline for evaluating LLMs on engineering design tasks. In this section, we provide more details on the evalutions of \textsc{EngDesign}.

\begin{figure}[t!]
\includegraphics[width=1.0\textwidth]{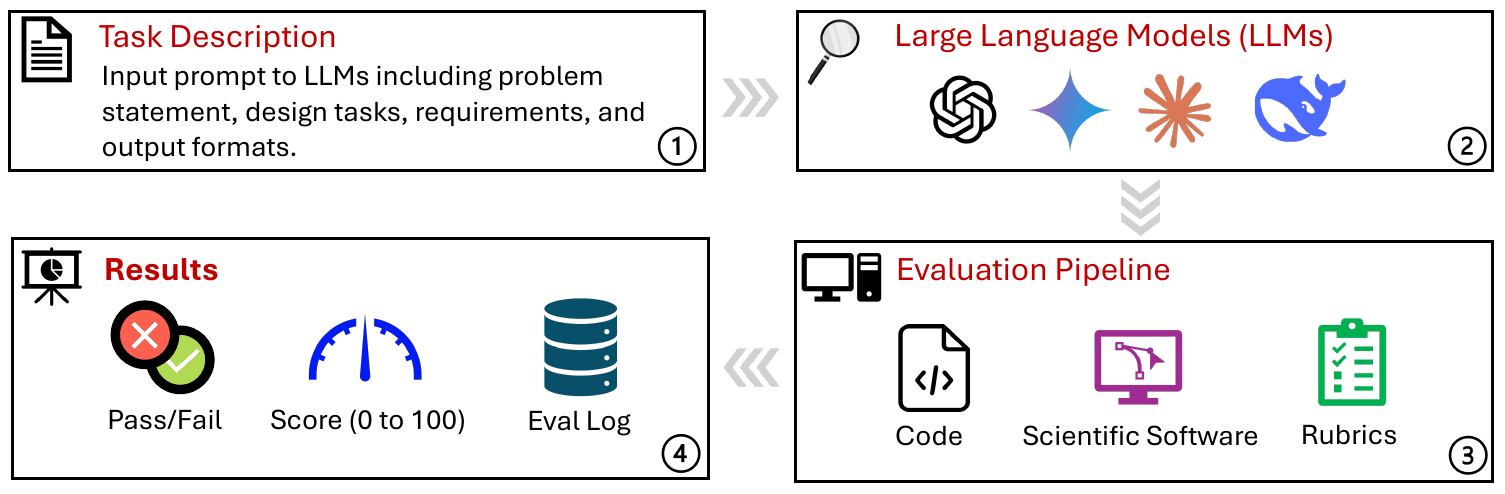}
\caption{Overview of the \textsc{EngDesign} Benchmarking Pipeline. The process begins with a task description curated by task contributors (Step 1). This prompt is then submitted to various LLMs (e.g., GPT, Claude, Gemini, DeepSeek) to generate candidate design proposals (Step 2). The generated outputs are evaluated using an automated codebase that includes code execution, integration with scientific software (e.g., MATLAB, Webots), and standardized scoring rubrics (Step 3). Finally, the results, including pass/fail status, quantitative scores, and detailed evaluation logs, are recorded for further analysis (Step 4).}
\label{figure:engdesign_pipeline}
\end{figure}

\subsection{Detailed Experimental Setup}
\label{sec:exp_setup}
We use the default configuration for each LLM during evaluation. Table~\ref{tab:llm_setup} provides detailed information on the model names, setup parameters, and special configurations used for each model.

\begin{table}[h]
\caption{Model names and configuration details for each evaluated LLM.}
\label{tab:llm_setup}
\centering
\begin{tabular}{l|l}
\toprule
\textbf{Model} & \textbf{Model Pointer \& Setup} \\
\midrule
GPT-4o & \texttt{gpt-4o-2024-08-06} \\
\midrule
o1 & \texttt{o1-2024-12-17} \\
\midrule
o3 & \texttt{o3-2025-04-16}, reasoning effort: \texttt{medium/high} \\
\midrule
o4-mini & \texttt{o4-mini-2025-04-16}, reasoning effort: \texttt{medium/high} \\
\midrule
Claude-3.7-Sonnet & \texttt{claude-3-7-sonnet-20250219} \\
\midrule
Claude-3.7-Thinking & \texttt{claude-3-7-sonnet-20250219}, thinking token budget: \texttt{16,000} \\
\midrule
Gemini-2.0-Flash & \texttt{gemini-1.5-pro} \\
\midrule
Gemini-2.5-Pro & \texttt{gemini-2.5-pro-preview-03-25} \\
\midrule
DeepSeek-v3 & \texttt{DeepSeek-V3} \\
\midrule
DeepSeek-R1 & \texttt{DeepSeek-R1} \\
\bottomrule
\end{tabular}
\end{table}

All evaluations in this work were conducted via API access to proprietary LLMs (e.g., OpenAI, Anthropic, Google, and DeepSeek). As a result, the computational burden on our end was minimal, with no need for local GPU infrastructure. The primary cost was associated with inference-time API usage, which varies depending on the model types and token numbers. Since we did not perform model training or fine-tuning, the overall computational footprint of our study remains low.

\subsection{LLM Response Structure}
\label{app:LLM_response}
\begin{figure}[h]
\centering
\includegraphics[width=0.75\textwidth]{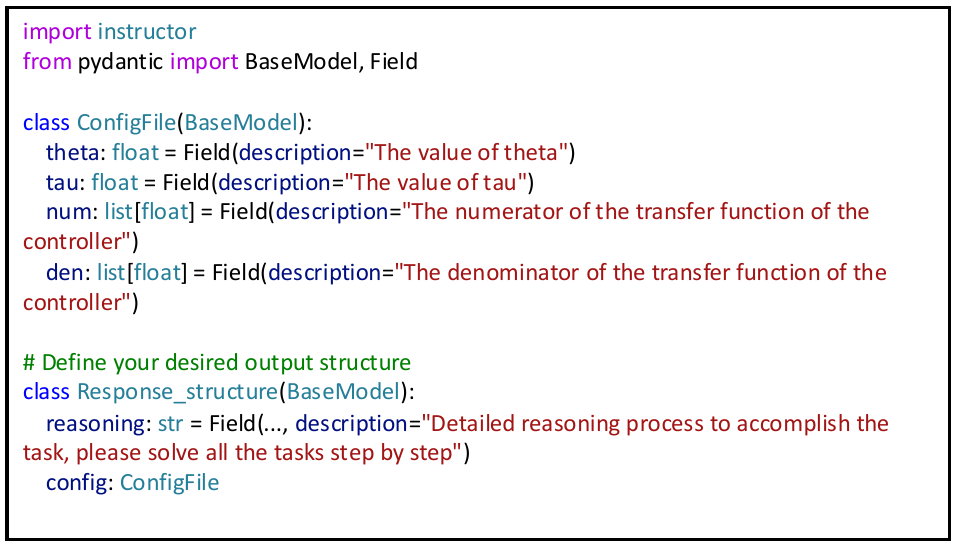}
\caption{An example of response structure class for \texttt{instructor} (Task XG\_05).}
\label{fig:response_structure}
\end{figure}

As discussed in Section~\ref{eval_pipe}, we use \texttt{instructor} to construct LLM responses by defining a response structure class, as illustrated in Figure~\ref{fig:response_structure}. For each task, the response structure class includes two keys: \texttt{reasoning} and \texttt{config}. The \texttt{reasoning} component prompts the LLM to perform a detailed step-by-step reasoning process to complete the task. The \texttt{config} component defines a task-specific \texttt{ConfigFile} class, which guides the LLM to produce key design components in a predefined format. This structured output facilitates automated evaluation by making it easy to parse LLMs' design choices or code snippets.

\subsection{More on Evaluation Metrics}
\label{app:eval_metric}

We evaluate model performance using three metrics: Average Pass Rate, Average Score, and Average Response Length. Each task is tested over $N$ ($N = 3$) independent trials per model. For each metric, we aggregate at the task level and then summarize across all tasks to ensure equal weighting.

\paragraph{Average Pass Rate}

For each task $i$, we define the pass rate as:
\[
\text{pass\_rate}_i = \frac{\text{number of successful trials}}{N}
\]
The overall pass rate is then computed as the mean of per-task pass rates:
\[
\mu_{\text{pass}} = \frac{1}{N} \sum_{i=1}^{N} \text{pass\_rate}_i
\]
We report $\mu_{\text{pass}}$ along with its standard deviation across tasks:
\[
\sigma_{\text{pass}} = \sqrt{\frac{1}{N-1} \sum_{i=1}^{N} (\text{pass\_rate}_i - \mu_{\text{pass}})^2}
\]

\paragraph{Average Score}

For each task $i$, the score is averaged over $N$ trials:
\[
\text{score}_i = \frac{1}{N} \sum_{j=1}^{3} x_{i,j}
\]
where $x_{i,j}$ is the score in trial $j$. The overall score is the mean of per-task scores:
\[
\mu_{\text{score}} = \frac{1}{N} \sum_{i=1}^{N} \text{score}_i
\]
with corresponding standard deviation:
\[
\sigma_{\text{score}} = \sqrt{\frac{1}{N-1} \sum_{i=1}^{N} (\text{score}_i - \mu_{\text{score}})^2}
\]

\paragraph{Reasoning Robustness}  Let \( T \) denote the set of all evaluated tasks. For a given model, each task \( t \in T \) is evaluated over $N$ independent trials. Define:

\begin{itemize}
    \item \( N_{\text{all}} \): the number of tasks for which {all $N$ trials passed}, and
    \item \( N_{\text{any}} \): the number of tasks for which {at least one trial passed}.
\end{itemize}

Then, the \emph{Reasoning Robustness} \( R \in [0, 1] \) is defined as:
\[
R = \frac{N_{\text{all}}}{N_{\text{any}}}
\]

This metric quantifies the model's consistency in reasoning. A value of \( R = 1 \) indicates perfect robustness. In other words, whenever a task is solvable by the model, it succeeds consistently across all trials. On the other hand, lower values indicate higher variance or instability in the model’s reasoning process across repeated attempts.

\subsection{Additional Evaluation Results}
\label{app:add_eval}

In this section, we present additional evaluation results.
Table \ref{main-table2} provide the average score results of evaluated LLMs on \textsc{EngDesign}. 

\begin{figure}[t!]
\centering
\includegraphics[width=1\textwidth]{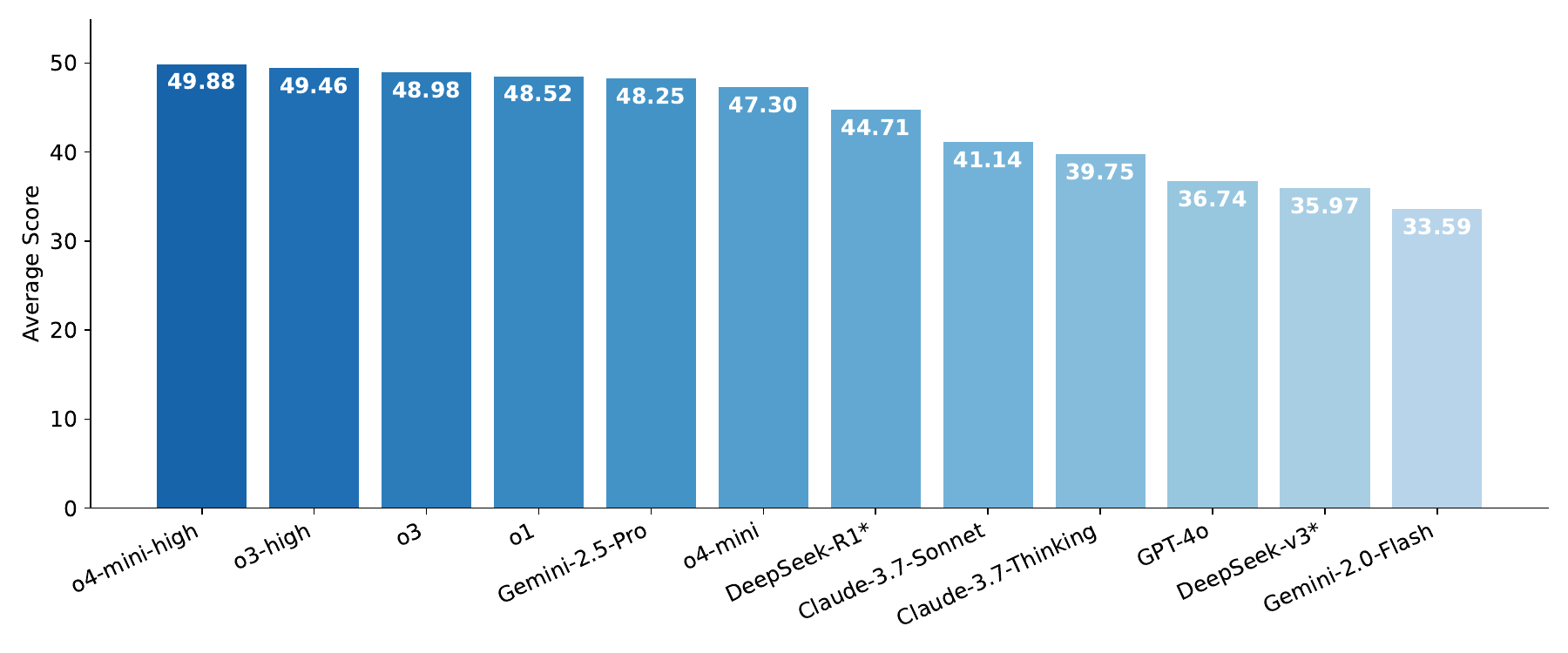}
\caption{Average Score of evaluted LLMs on \textsc{EngDesign}.}
\label{fig:score_dis}
\end{figure}

Figure \ref{fig:score_dis} presents the average score distribution across all evaluated models. The overall ranking is consistent with the pass rate results in Table \ref{main-table}, with reasoning-focused models such as {o4-mini-high}, {o3-high}, and {o3} achieving the highest scores. Notably, the score differences are less pronounced than the corresponding differences in pass rates, suggesting that while these models may not consistently meet strict pass criteria, they still generate partially correct or near-complete solutions that earn substantial credit. In addition, Table \ref{main-table-text-only} and Table \ref{main-table2-text-only} show the average pass and score for text-only tasks. Table \ref{main-table-open} and Table \ref{main-table2-open} present the evaluation results for \textsc{EngDesign-Open}.  Finally, we provide the results with error bar in Table \ref{main-table-w-error} and Table \ref{main-table2-w-error}.

\begin{table}
  \caption{Average score results (each task evaluated over 3 trials) under different domains.}
  \label{main-table2}
  \centering
  \resizebox{\textwidth}{!}{
    \begin{tabular}{lcccccccccc}
      \toprule                
      \textbf{Model} & \textbf{Overall} & \textbf{AICD} & \textbf{Arch} & \textbf{Ctrl} & \textbf{DHD} & \textbf{Mech} & \textbf{OS} & \textbf{Robo} & \textbf{SigP} & \textbf{Stru} \\
      \midrule
      \multicolumn{11}{c}{\textbf{Chat Models}} \\
      \midrule
      GPT-4o & 36.74 & 25.71 & 18.15 & 35.30 & 41.69 & 29.05 & 26.33 & 55.80 & 41.45 & 34.60 \\
      DeepSeek-v3* & 35.97 & 17.33 & 0.00 & 45.58 & 39.67 & 24.05 & 25.94 & 48.47 & 34.55 & 38.68 \\
      Gemini-2.0-Flash & 33.59 & 7.00 & 0.00 & 39.45 & 38.25 & 27.14 & 28.15 & 53.47 & 30.03 & 31.51 \\
      Claude-3.7-Sonnet & 41.14 & \textbf{31.00} & 23.79 & 43.94 & 46.03 & 30.24 & 29.55 & 58.97 & 44.90 & 37.11 \\      
      \midrule
      \multicolumn{11}{c}{\textbf{Reasoning Models}} \\
      \midrule
      o1 & 48.52 & 19.00 & \textbf{49.09} & 47.48 & 63.28 & 40.48 & 58.35 & 57.22 & 52.61 & 32.57 \\
      o3 & 48.98 & 20.33 & 35.82 & \textbf{59.37} & 28.82 & 40.24 & 54.52 & 70.13 & \textbf{57.44} & 44.77 \\
      o3-high & 49.46 & 20.33 & 32.80 & 53.19 & 33.36 & \textbf{45.24} & \textbf{65.59} & \textbf{73.67} & 53.51 & 47.48 \\
      o4-mini & 47.30 & 11.00 & 25.72 & 48.80 & 57.36 & 35.00 & 56.90 & 61.38 & 51.28 & 42.07 \\
      o4-mini-high & \textbf{49.88} & 12.67 & 36.97 & 54.87 & \textbf{64.28} & 36.43 & 63.96 & 56.72 & 48.45 & 44.15 \\
      Gemini-2.5-Pro & 48.25 & \textbf{31.00} & 0.00 & 53.70 & 58.05 & 29.05 & 44.05 & 62.43 & 39.63 & \textbf{55.02} \\
      DeepSeek-R1* & 44.71 & 22.00 & 0.00 & 56.27 & 54.38 & 26.19 & 37.26 & 53.82 & 48.08 & 40.31 \\
      Claude-3.7-Thinking & 39.75 & 22.67 & 26.02 & 43.87 & 29.59 & 37.62 & 28.25 & 59.38 & 45.58 & 40.39 \\
      \bottomrule
    \end{tabular}
  }
\end{table}

\begin{table}
  \caption{Average pass rate (\%) results (text-only tasks).}
  \label{main-table-text-only}
  \centering
  \resizebox{\textwidth}{!}{
    \begin{tabular}{lcccccccccc}
      \toprule                
      \textbf{Model} & \textbf{Overall} & \textbf{AICD} & \textbf{Arch} & \textbf{Ctrl} & \textbf{DHD} & \textbf{Mech} & \textbf{OS} & \textbf{Robo} & \textbf{SigP} & \textbf{Stru} \\
      \midrule
      \multicolumn{11}{c}{\textbf{Chat Models}} \\
      \midrule
      GPT-4o & 14.98 & 0.00 & 20.00 & 20.83 & 10.26 & 0.00 & 4.17 & 26.67 & 16.67 & 27.78 \\
      DeepSeek-v3$^*$ & 18.42 & 0.00 & 0.00 & \textbf{33.33} & 25.64 & 0.00 & 0.00 & 16.67 & 13.89 & 38.89 \\
      Gemini-2.0-Flash & 13.12 & 0.00 & 0.00 & 4.17 & 16.67 & 4.76 & 0.00 & 20.00 & 5.56 & 36.11 \\
      Claude-3.7-Sonnet & 21.97 & 0.00 & \textbf{40.00} & 25.00 & 33.33 & 14.29 & 0.00 & 33.33 & 13.89 & 33.33 \\
      \midrule
      \multicolumn{11}{c}{\textbf{Reasoning Models}} \\
      \midrule
      o1 & 28.95 & 0.00 & \textbf{40.00} & 25.00 & 41.03 & 14.29 & \textbf{37.50} & 50.00 & 22.22 & 25.00 \\
      o3 & 34.38 & 0.00 & \textbf{40.00} & 20.51 & 20.51 & 23.81 & 25.00 & 63.33 & \textbf{41.67} & 41.67 \\
      o3-high & \textbf{34.51} & 0.00 & \textbf{40.00} & \textbf{33.33} & 20.51 & \textbf{28.57} & 30.43 & \textbf{68.97} & 36.11 & 44.44 \\
      o4-mini & 31.58 & 0.00 & 33.33 & 20.83 & 41.03 & 19.05 & 29.17 & 43.33 & 30.56 & 44.44 \\
      o4-mini-high & 33.78 & 0.00 & \textbf{40.00} & 20.83 & \textbf{47.22} & 19.05 & \textbf{37.50} & 46.67 & 36.11 & 38.89 \\
      Gemini-2.5-Pro & 28.89 & 0.00 & 0.00 & 25.00 & 43.59 & 0.00 & 9.52 & 56.67 & 13.89 & \textbf{50.00} \\
      DeepSeek-R1$^*$ & 23.77 & 0.00 & 36.36 & 20.83 & 38.46 & 4.76 & 5.26 & 26.67 & 22.22 & 41.67 \\
      Claude-3.7-Thinking & 18.30 & 0.00 & 33.33 & 20.83 & 17.95 & 9.52 & 0.00 & 40.00 & 11.11 & 30.56 \\
      \bottomrule
    \end{tabular}
  }
\end{table}

\begin{table}
  \caption{Average score results (text-only results).}
  \label{main-table2-text-only}
  \centering
  \resizebox{\textwidth}{!}{
    \begin{tabular}{lcccccccccc}
      \toprule                
      \textbf{Model} & \textbf{Overall} & \textbf{AICD} & \textbf{Arch} & \textbf{Ctrl} & \textbf{DHD} & \textbf{Mech} & \textbf{OS} & \textbf{Robo} & \textbf{SigP} & \textbf{Stru} \\
      \midrule
      \multicolumn{11}{c}{\textbf{Chat Models}} \\
      \midrule
      GPT-4o & 37.57 & 25.71 & 0.00 & 43.75 & 41.69 & 29.05 & 26.33 & 55.80 & 36.59 & 34.98  \\
      DeepSeek-v3* & 37.20 & 17.33 & 0.00 & 60.17 & 39.67 & 24.05 & 25.94 & 48.47 & 34.93 & 38.68 \\
      Gemini-2.0-Flash & 33.87 & 7.00 & 0.00 & 42.58 & 38.25 & 27.14 & 28.15 & 53.47 & 30.86 & 31.51 \\
      Claude-3.7-Sonnet & 41.39 & 31.00 & 0.00 & 48.54 & 46.03 & 30.24 & 29.55 & 58.97 & 42.54 & 37.70  \\      
      \midrule
      \multicolumn{11}{c}{\textbf{Reasoning Models}} \\
      \midrule
      o1 & 48.94 & 19.00 & 0.00 & 53.92 & 63.28 & 40.48 & 58.35 & 57.22 & 54.59 & 32.79  \\
      o3 & 48.02 & 20.33 & 0.00 & \textbf{60.58} & 28.82 & 40.24 & 54.52 & 70.13 & \textbf{59.76} & 46.00 \\
      o3-high & \textbf{50.39} & 20.33 & 0.00 & 58.08 & 33.36 & \textbf{45.24} & \textbf{65.59} & \textbf{73.67} & 56.44 & 48.93  \\
      o4-mini & 48.18 & 11.00 & 0.00 & 50.04 & 57.36 & 35.00 & 56.90 & 61.38 & 52.46 & 43.08  \\
      o4-mini-high & 49.76 & 12.67 & 0.00 & 49.96 & \textbf{64.28} & 36.43 & 63.96 & 56.72 & 51.65 & 45.30  \\
      Gemini-2.5-Pro & 48.33 & \textbf{31.00} & 0.00 & 50.50 & 58.05 & 29.05 & 44.05 & 62.43 & 42.93 & \textbf{55.02} \\
      DeepSeek-R1* & 44.24 & 22.00 & 0.00 & 53.21 & 54.38 & 26.19 & 37.26 & 53.82 & 50.42 & 40.31 \\
      Claude-3.7-Thinking & 39.47 & 22.67 & 0.00 & 47.71 & 29.59 & 37.62 & 28.25 & 59.38 & 43.92 & 41.25  \\
      \bottomrule
    \end{tabular}
  }
\end{table}

\begin{table}
  \caption{Average pass rate (\%) results for \textsc{EngDesign-Open}.}
  \label{main-table-open}
  \centering
  \resizebox{\textwidth}{!}{
    \begin{tabular}{lcccccccccc}
      \toprule                
      \textbf{Model} & \textbf{Overall} & \textbf{AICD} & \textbf{Arch} & \textbf{Ctrl} & \textbf{DHD} & \textbf{Mech} & \textbf{OS} & \textbf{Robo} & \textbf{SigP} & \textbf{Stru} \\
      \midrule
      \multicolumn{11}{c}{\textbf{Chat Models}} \\
      \midrule
      GPT-4o & 14.74 & N/A & 20.00 & \textbf{16.67} & 11.11 & 0.00 & 4.17 & 29.63 & 20.00 & 0.00 \\
      DeepSeek-v3$^*$ & 8.94 & N/A & 0.00 & N/A & 14.81 & 0.00 & 0.00 & 18.52 & 8.33 & N/A \\
      Gemini-2.0-Flash & 10.34 & N/A & 0.00 & 0.00 & 12.50 & 5.56 & 0.00 & 22.22 & 8.33 & 0.00 \\
      Claude-3.7-Sonnet & 22.52 & N/A & \textbf{40.00} & 8.33 & 29.17 & 16.67 & 0.00 & 37.04 & 23.33 & 0.00 \\
      \midrule
      \multicolumn{11}{c}{\textbf{Reasoning Models}} \\
      \midrule
      o1 & 36.54 & N/A & \textbf{40.00} & 0.00 & 48.15 & 16.67 & 37.50 & 55.56 & 36.67 & 0.00 \\
      o3 & 35.90 & N/A & \textbf{40.00} & 0.00 & 7.41 & 27.78 & 25.00 & 70.37 & \textbf{60.00} & 0.00 \\
      o3-high & 36.77 & N/A & \textbf{40.00} & 0.00 & 11.11 & \textbf{33.33} & 30.43 & \textbf{74.07} & 50.00 & 0.00 \\
      o4-mini & 36.54 & N/A & 33.33 & 0.00 & 48.15 & 22.22 & 29.17 & 48.15 & 50.00 & 0.00 \\
      o4-mini-high & \textbf{40.52} & N/A & \textbf{40.00} & 0.00 & \textbf{58.33} & 22.22 & \textbf{37.50} & 51.85 & 50.00 & 0.00 \\
      Gemini-2.5-Pro & 30.00 & N/A & 0.00 & 0.00 & 44.44 & 0.00 & 9.52 & 62.96 & 20.83 & 0.00 \\
      DeepSeek-R1$^*$ & 22.03 & N/A & 0.00 & N/A & 44.44 & 5.56 & 5.26 & 29.63 & 16.67 & N/A \\
      Claude-3.7-Thinking & 19.74 & N/A & 33.33 & 0.00 & 7.41 & 11.11 & 0.00 & 44.44 & 30.00 & 0.00 \\
      \bottomrule
    \end{tabular}
  }
\end{table}

\begin{table}
  \caption{Average score results for \textsc{EngDesign-Open}.}
  \label{main-table2-open}
  \centering
  \resizebox{\textwidth}{!}{
    \begin{tabular}{lcccccccccc}
      \toprule                
      \textbf{Model} & \textbf{Overall} & \textbf{AICD} & \textbf{Arch} & \textbf{Ctrl} & \textbf{DHD} & \textbf{Mech} & \textbf{OS} & \textbf{Robo} & \textbf{SigP} & \textbf{Stru} \\
      \midrule
      \multicolumn{11}{c}{\textbf{Chat Models}} \\
      \midrule
      GPT-4o & 36.03 & N/A & 18.15 & 47.00 & 26.85 & 29.44 & 26.33 & 62.00 & 37.79 & 30.00 \\
      DeepSeek-v3* & 31.50 & N/A & 0.00 & N/A & 20.37 & 28.06 & 25.94 & 53.85 & 30.94 & N/A \\
      Gemini-2.0-Flash & 34.74 & N/A & 0.00 & 35.00 & 17.50 & 31.67 & 28.15 & 59.41 & 36.36 & 30.00 \\
      Claude-3.7-Sonnet & 38.71 & N/A & 23.79 & 46.00 & 29.17 & 35.28 & 29.55 & 65.52 & 36.41 & 30.00 \\      
      \midrule
      \multicolumn{11}{c}{\textbf{Reasoning Models}} \\
      \midrule
      o1 & 51.64 & N/A & \textbf{49.09} & 42.67 & 57.04 & 35.00 & 58.35 & 63.57 & 47.70 & 30.00 \\
      o3 & 47.48 & N/A & 35.82 & 41.33 & 7.41 & 41.39 & 54.52 & 77.93 & \textbf{64.18} & 30.00 \\
      o3-high & 49.61 & N/A & 32.80 & \textbf{47.17} & 13.33 & \textbf{45.00} & \textbf{65.59} & \textbf{79.13} & 57.54 & 30.00 \\
      o4-mini & 52.08 & N/A & 25.72 & \textbf{47.17} & 50.37 & 40.83 & 56.90 & 68.20 & 59.37 & 30.00 \\
      o4-mini-high & \textbf{54.02} & N/A & 36.97 & 46.33 & \textbf{59.58} & 42.50 & 63.96 & 63.02 & 54.46 & 30.00 \\
      Gemini-2.5-Pro & 47.88 & N/A & 0.00 & 46.00 & 47.78 & 30.56 & 44.05 & 69.37 & 46.13 & 30.00 \\
      DeepSeek-R1* & 45.66 & N/A & 0.00 & N/A & 48.15 & 30.56 & 37.26 & 59.80 & 50.62 & N/A \\
      Claude-3.7-Thinking & 34.74 & N/A & 26.02 & 37.67 & 7.41 & 35.00 & 28.25 & 65.98 & 39.07 & 30.00 \\
      \bottomrule
    \end{tabular}
  }
\end{table}

\begin{sidewaystable}
  \caption{Average pass rate (\%) (each task evaluated over 3 trials) under different domains with error bar.}
  \label{main-table-w-error}
  \centering
  \resizebox{\textwidth}{!}{
    \begin{tabular}{lcccccccccc}
      \toprule                
      \textbf{Model} & \textbf{Overall} & \textbf{AICD} & \textbf{Arch} & \textbf{Ctrl} & \textbf{DHD} & \textbf{Mech} & \textbf{OS} & \textbf{Robo} & \textbf{SigP} & \textbf{Stru} \\
      \midrule
      \multicolumn{11}{c}{\textbf{Chat Models}} \\
      \midrule
      GPT-4o & 15.68$\pm$ 0.79 & 0.00$\pm$ 0.00 & 20.00$\pm$ 0.00 & 18.52$\pm$ 2.62 & 10.26$\pm$ 3.63 & 0.00$\pm$ 0.00 & 4.17$\pm$ 5.89 & 26.67$\pm$ 4.71 & 17.65$\pm$ 0.00 & 25.64$\pm$ 3.63 \\
      DeepSeek-v3* & 17.92$\pm$ 1.56 & 0.00$\pm$ 0.00 & 0.00$\pm$ 0.00 & 27.27$\pm$ 7.42 & 25.64$\pm$ 3.63 & 0.00$\pm$ 0.00 & 0.00$\pm$ 0.00 & 16.67$\pm$ 4.71 & 12.82$\pm$ 3.63 & 38.89$\pm$ 7.86 \\
      Gemini-2.0-Flash & 14.16$\pm$ 0.31 & 0.00$\pm$ 0.00 & 0.00$\pm$ 0.00 & 15.15$\pm$ 11.34 & 16.67$\pm$ 1.14 & 4.76$\pm$ 6.73 & 0.00$\pm$ 0.00 & 20.00$\pm$ 8.16 & 5.13$\pm$ 7.25 & 36.11$\pm$ 3.93 \\
      Claude-3.7-Sonnet & 22.61$\pm$ 0.95 & 0.00$\pm$ 0.00 & \textbf{40.00$\pm$ 0.00} & 16.67$\pm$ 4.54 & 33.33$\pm$ 0.00 & 14.29$\pm$ 0.00 & 0.00$\pm$ 0.00 & 33.33$\pm$ 4.71 & 21.57$\pm$ 2.77 & 30.77$\pm$ 0.00 \\      
      \midrule
      \multicolumn{11}{c}{\textbf{Reasoning Models}} \\
      \midrule
      o1 & 29.17$\pm$ 2.25 & 0.00$\pm$ 0.00 & \textbf{40.00$\pm$ 0.00} & 24.07$\pm$ 2.62 & 41.03$\pm$ 3.63 & 14.29$\pm$ 0.00 & 37.50$\pm$ 0.00 & 50.00$\pm$ 14.14 & 25.49$\pm$ 2.77 & 23.08$\pm$ 6.28 \\
      o3 & \textbf{34.38$\pm$ 1.70} & 0.00$\pm$ 0.00 & \textbf{40.00$\pm$ 0.00} & 35.19$\pm$ 5.24 & 20.51$\pm$ 7.25 & 23.81$\pm$ 6.73 & 25.00$\pm$ 10.21 & 63.33$\pm$ 4.71 & \textbf{41.18$\pm$ 0.00} & 38.46$\pm$ 6.28 \\
      o3-high & 33.57$\pm$ 1.16 & 0.00$\pm$ 0.00 & \textbf{40.00$\pm$ 0.00} & 27.78$\pm$ 9.07 & 20.51$\pm$ 7.25 & \textbf{28.57$\pm$ 0.00} & 30.43$\pm$ 5.26 & \textbf{68.97$\pm$ 1.57} & 35.29$\pm$ 4.80 & 41.03$\pm$ 3.63 \\
      o4-mini & 31.60$\pm$ 0.49 & 0.00$\pm$ 0.00 & 33.33$\pm$ 9.43 & 24.07$\pm$ 6.93 & 41.03$\pm$ 7.25 & 19.05$\pm$ 6.73 & 29.17$\pm$ 5.89 & 43.33$\pm$ 4.71 & 33.33$\pm$ 2.77 & 41.03$\pm$ 3.63 \\
      o4-mini-high & 34.04$\pm$ 0.50 & 0.00$\pm$ 0.00 & \textbf{40.00$\pm$ 16.33} & 27.78$\pm$ 4.54 & \textbf{47.22$\pm$ 3.93} & 19.05$\pm$ 6.73 & \textbf{37.50$\pm$ 0.00} & 46.67$\pm$ 9.43 & 35.29$\pm$ 4.80 & 35.90$\pm$ 3.63 \\
      Gemini-2.5-Pro & 29.54$\pm$ 0.95 & 0.00$\pm$ 0.00 & 0.00$\pm$ 0.00 & 33.33$\pm$ 4.29 & 43.59$\pm$ 3.63 & 0.00$\pm$ 0.00 & 9.52$\pm$ 6.36 & 56.67$\pm$ 4.71 & 12.82$\pm$ 3.63 & \textbf{50.00$\pm$ 0.00} \\
      DeepSeek-R1* & 25.53$\pm$ 3.35 & 0.00$\pm$ 0.00 & 0.00$\pm$ 0.00 & \textbf{36.36$\pm$ 7.42} & 38.46$\pm$ 6.28 & 4.76$\pm$ 6.73 & 5.26$\pm$ 7.86 & 26.67$\pm$ 9.43 & 20.51$\pm$ 7.25 & 41.67$\pm$ 6.80 \\
      Claude-3.7-Thinking & 20.07$\pm$ 2.41 & 0.00$\pm$ 0.00 & 33.33$\pm$ 9.43 & 18.52$\pm$ 10.48 & 17.95$\pm$ 9.59 & 9.52$\pm$ 6.73 & 0.00$\pm$ 0.00 & 40.00$\pm$ 0.00 & 19.61$\pm$ 2.77 & 28.21$\pm$ 3.63 \\
      \bottomrule
    \end{tabular}
  }
\end{sidewaystable}

\begin{sidewaystable}
  \caption{Average score results (each task evaluated over 3 trials) under different domains with error bar.}
  \label{main-table2-w-error}
  \centering
  \resizebox{\textwidth}{!}{
    \begin{tabular}{lcccccccccc}
      \toprule                
      \textbf{Model} & \textbf{Overall} & \textbf{AICD} & \textbf{Arch} & \textbf{Ctrl} & \textbf{DHD} & \textbf{Mech} & \textbf{OS} & \textbf{Robo} & \textbf{SigP} & \textbf{Stru} \\
      \midrule
      \multicolumn{11}{c}{\textbf{Chat Models}} \\
      \midrule
      GPT-4o & 36.74$\pm$0.86 & 25.71$\pm$2.83 & 18.15$\pm$4.93 & 35.30$\pm$0.28 & 41.69$\pm$7.07 & 29.05$\pm$2.69 & 26.33$\pm$5.12 & 55.80$\pm$1.41 & 41.45$\pm$0.93 & 34.60$\pm$1.92 \\
      DeepSeek-v3* & 35.97$\pm$0.56 & 17.33$\pm$1.89 & 0.00$\pm$0.00 & 45.58$\pm$2.39 & 39.67$\pm$2.64 & 24.05$\pm$2.63 & 25.94$\pm$2.81 & 48.47$\pm$4.64 & 34.55$\pm$3.76 & 38.68$\pm$2.50 \\
      Gemini-2.0-Flash & 33.59$\pm$1.33 & 7.00$\pm$0.00 & 0.00$\pm$0.00 & 39.45$\pm$7.04 & 38.25$\pm$2.74 & 27.14$\pm$2.33 & 28.15$\pm$7.45 & 53.47$\pm$1.84 & 30.03$\pm$3.42 & 31.51$\pm$1.84 \\
      Claude-3.7-Sonnet & 41.14$\pm$0.63 & \textbf{31.00$\pm$4.32} & 23.79$\pm$2.17 & 43.94$\pm$0.67 & 46.03$\pm$2.87 & 30.24$\pm$2.63 & 29.55$\pm$3.06 & 58.97$\pm$1.16 & 44.90$\pm$1.22 & 37.11$\pm$1.79 \\      
      \midrule
      \multicolumn{11}{c}{\textbf{Reasoning Models}} \\
      \midrule
      o1 & 48.52$\pm$1.33 & 19.00$\pm$3.56 & \textbf{49.09$\pm$2.61} & 47.48$\pm$1.02 & 63.28$\pm$2.87 & 40.48$\pm$1.35 & 58.35$\pm$2.43 & 57.22$\pm$2.24 & 52.61$\pm$2.75 & 32.57$\pm$5.30 \\
      o3 & 48.98$\pm$1.93 & 20.33$\pm$7.32 & 35.82$\pm$4.58 & \textbf{59.37$\pm$3.52} & 28.82$\pm$7.31 & 40.24$\pm$5.54 & 54.52$\pm$10.78 & 70.13$\pm$0.82 & \textbf{57.44$\pm$1.35} & 44.77$\pm$5.92 \\
      o3-high & 49.46$\pm$1.95 & 20.33$\pm$3.77 & 32.80$\pm$2.08 & 53.19$\pm$3.52 & 33.36$\pm$6.05 & \textbf{45.24$\pm$4.86} & \textbf{65.59$\pm$6.28} & \textbf{73.67$\pm$3.69} & 53.51$\pm$6.36 & 47.48$\pm$1.74 \\
      o4-mini & 47.30$\pm$1.32 & 11.00$\pm$3.27 & 25.72$\pm$11.62 & 48.80$\pm$6.14 & 57.36$\pm$5.32 & 35.00$\pm$3.25 & 56.90$\pm$0.28 & 61.38$\pm$2.58 & 51.28$\pm$0.96 & 42.07$\pm$4.34 \\
      o4-mini-high & \textbf{49.88$\pm$1.08} & 12.67$\pm$2.36 & 36.97$\pm$12.24 & 54.87$\pm$3.35 & \textbf{64.28$\pm$5.04} & 36.43$\pm$1.54 & 63.96$\pm$4.00 & 56.72$\pm$3.94 & 48.45$\pm$4.09 & 44.15$\pm$0.14 \\
      Gemini-2.5-Pro & 48.25$\pm$3.24 & \textbf{31.00$\pm$5.72} & 0.00$\pm$0.00 & 53.70$\pm$3.75 & 58.05$\pm$4.94 & 29.05$\pm$2.69 & 44.05$\pm$2.18 & 62.43$\pm$1.25 & 39.63$\pm$6.50 & \textbf{55.02$\pm$1.71} \\
      DeepSeek-R1* & 44.71$\pm$3.78 & 22.00$\pm$10.68 & 0.00$\pm$0.00 & 56.27$\pm$5.10 & 54.38$\pm$8.12 & 26.19$\pm$0.67 & 37.26$\pm$6.91 & 53.82$\pm$3.31 & 48.08$\pm$5.26 & 40.31$\pm$7.77 \\
      Claude-3.7-Thinking & 39.75$\pm$2.14 & 22.67$\pm$4.92 & 26.02$\pm$11.93 & 43.87$\pm$6.09 & 29.59$\pm$7.28 & 37.62$\pm$6.45 & 28.25$\pm$2.06 & 59.38$\pm$1.59 & 45.58$\pm$2.92 & 40.39$\pm$1.98 \\
      \bottomrule
    \end{tabular}
  }
\end{sidewaystable}


\subsection{Token Consumption}
\label{app:token_consumption}
\begin{figure}
\centering
\begin{subfigure}[h]{0.45\textwidth}
    \centering
    \includegraphics[width=\textwidth]{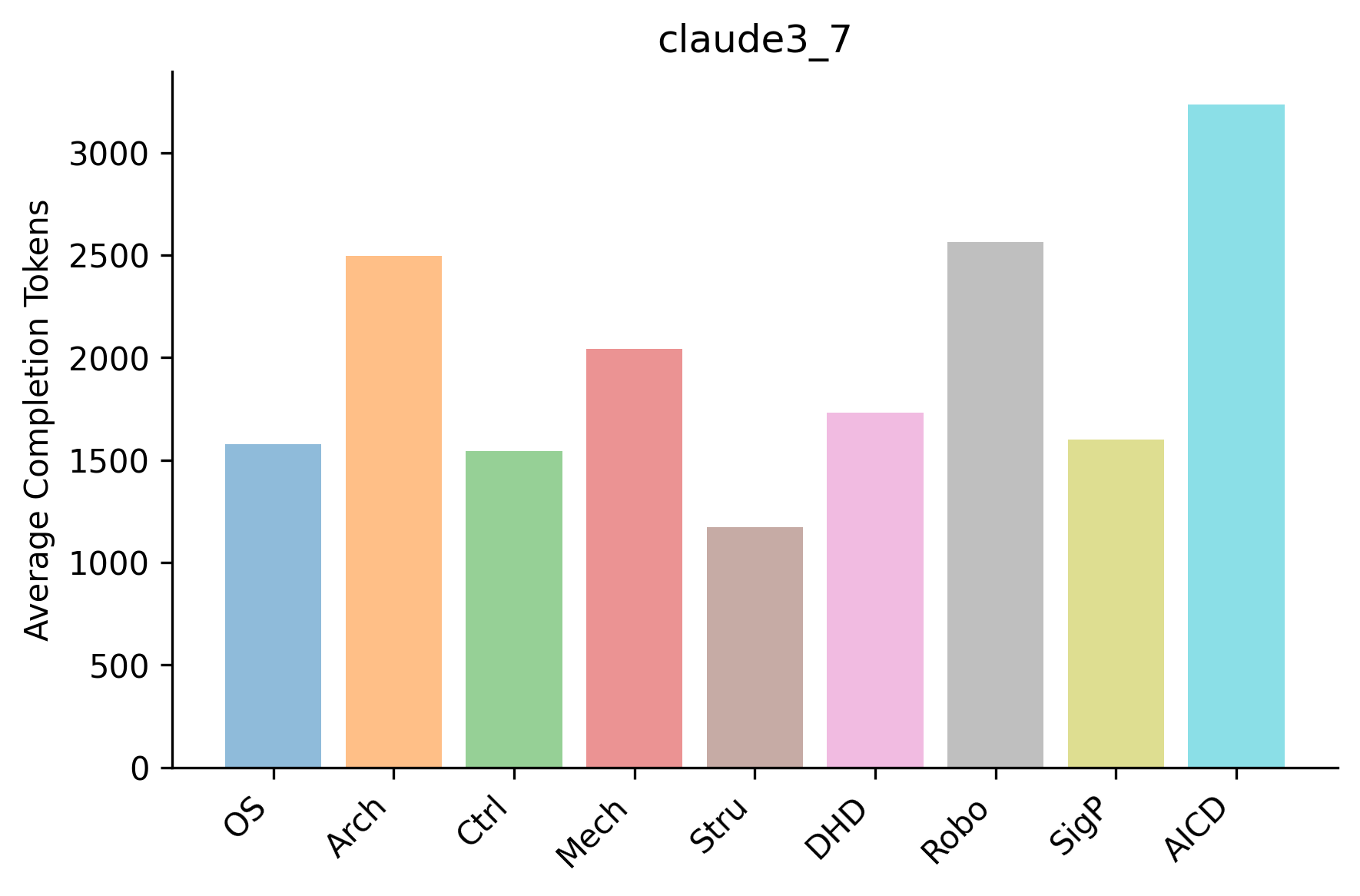}
\end{subfigure}%
\hfill
\begin{subfigure}[h]{0.45\textwidth}
    \centering
    \includegraphics[width=\textwidth]{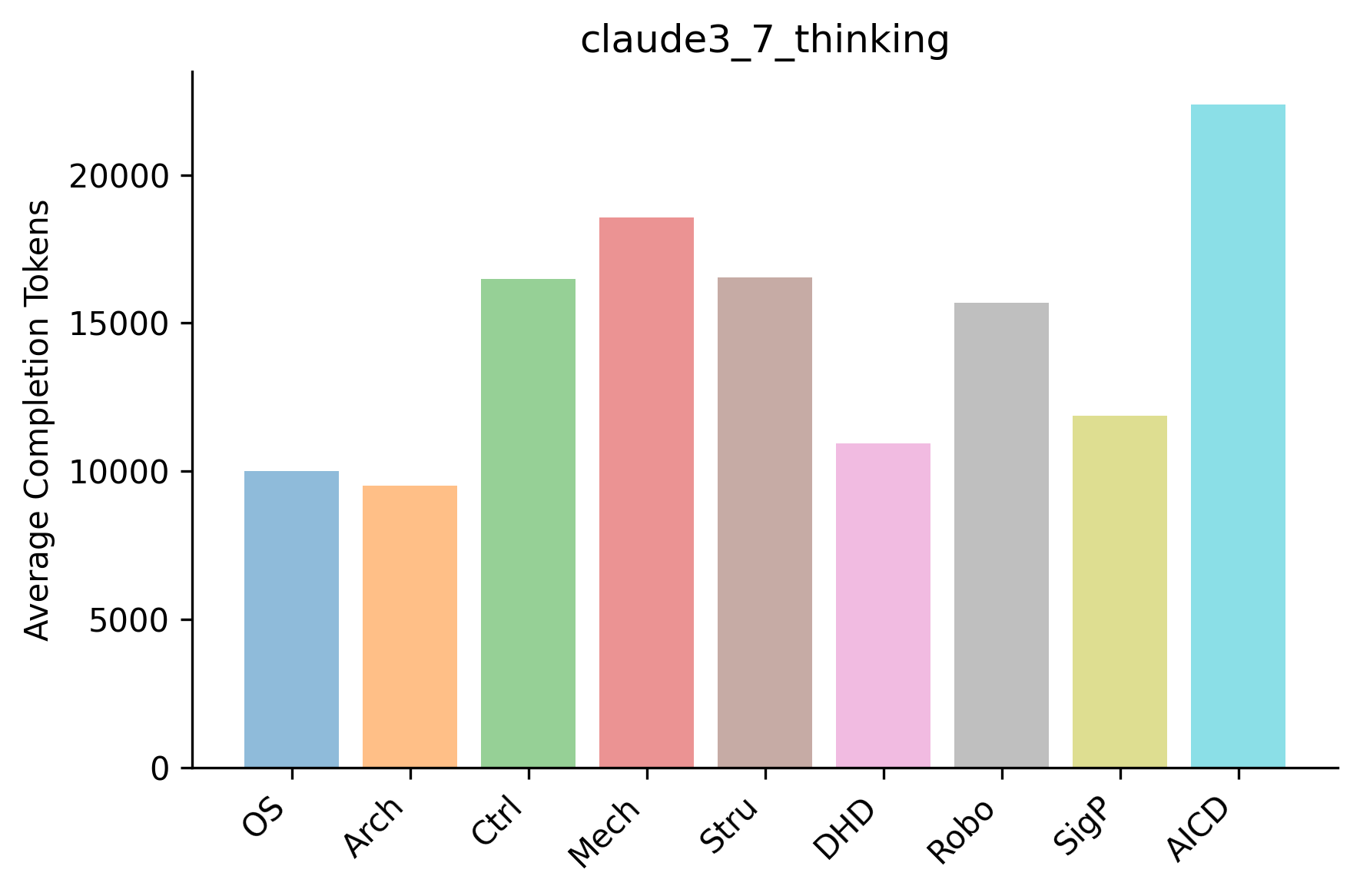}
\end{subfigure}

\caption{Average token consumption of Claude models on \textsc{EngDesign}.}
\label{fig:token_1}
\end{figure}

\begin{figure}
\centering
\begin{subfigure}[h]{0.45\textwidth}
    \centering
    \includegraphics[width=\textwidth]{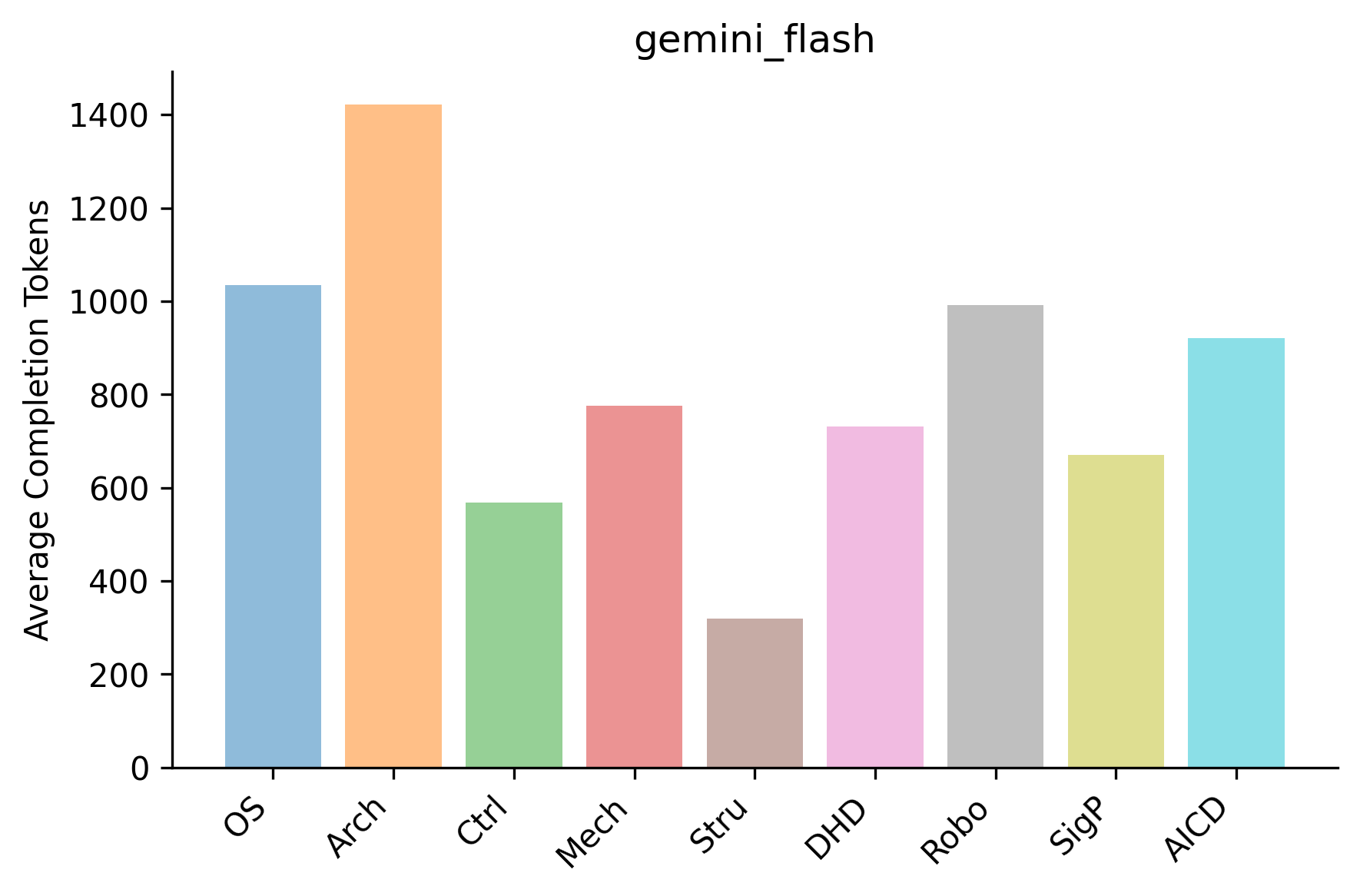}
\end{subfigure}%
\hfill
\begin{subfigure}[h]{0.45\textwidth}
    \centering
    \includegraphics[width=\textwidth]{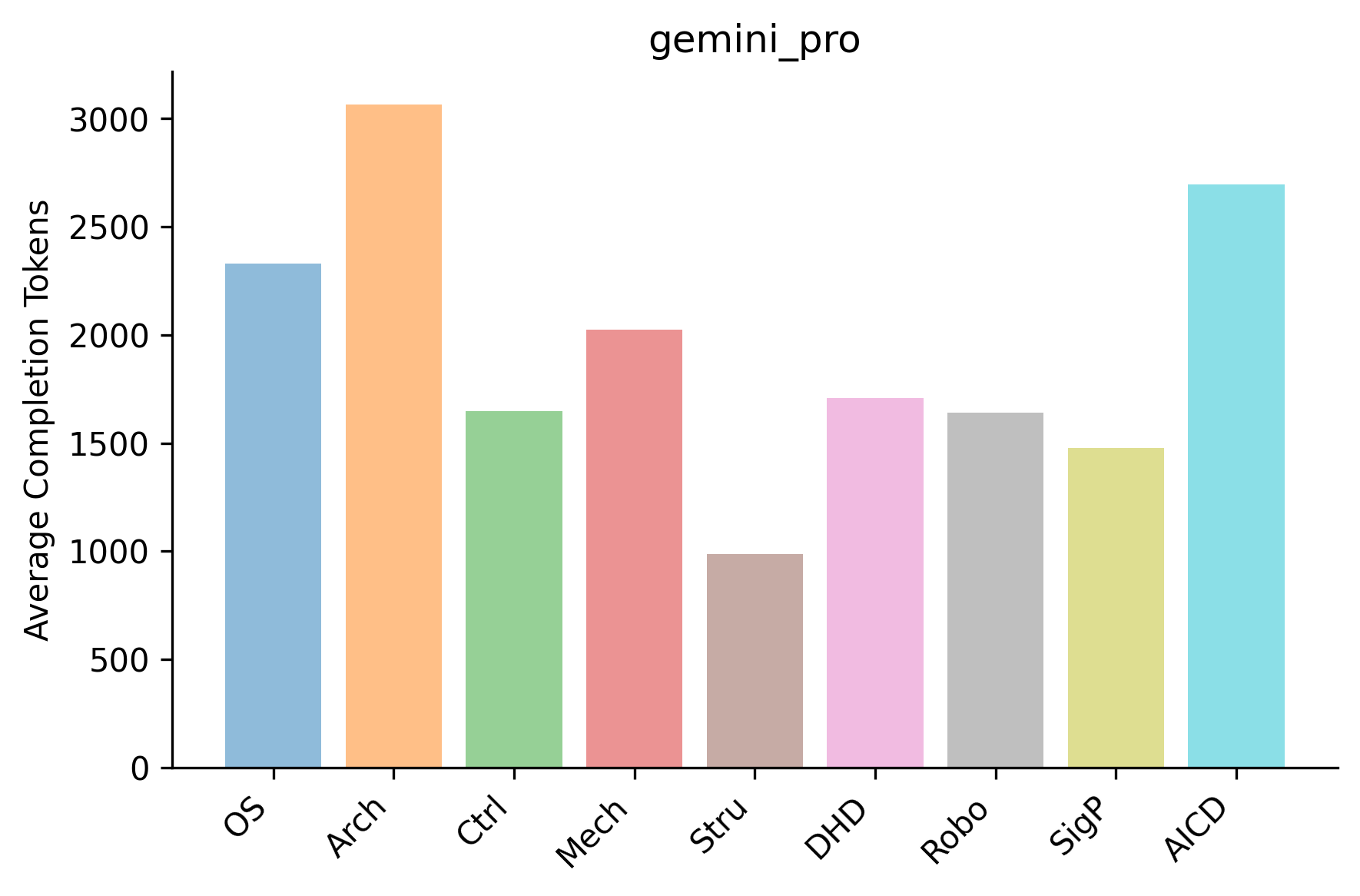}
\end{subfigure}

\caption{Average token consumption of Gemini models on \textsc{EngDesign}.}
\label{fig:token_2}
\end{figure}

\begin{figure}
\centering
\begin{subfigure}[h]{0.45\textwidth}
    \centering
    \includegraphics[width=\textwidth]{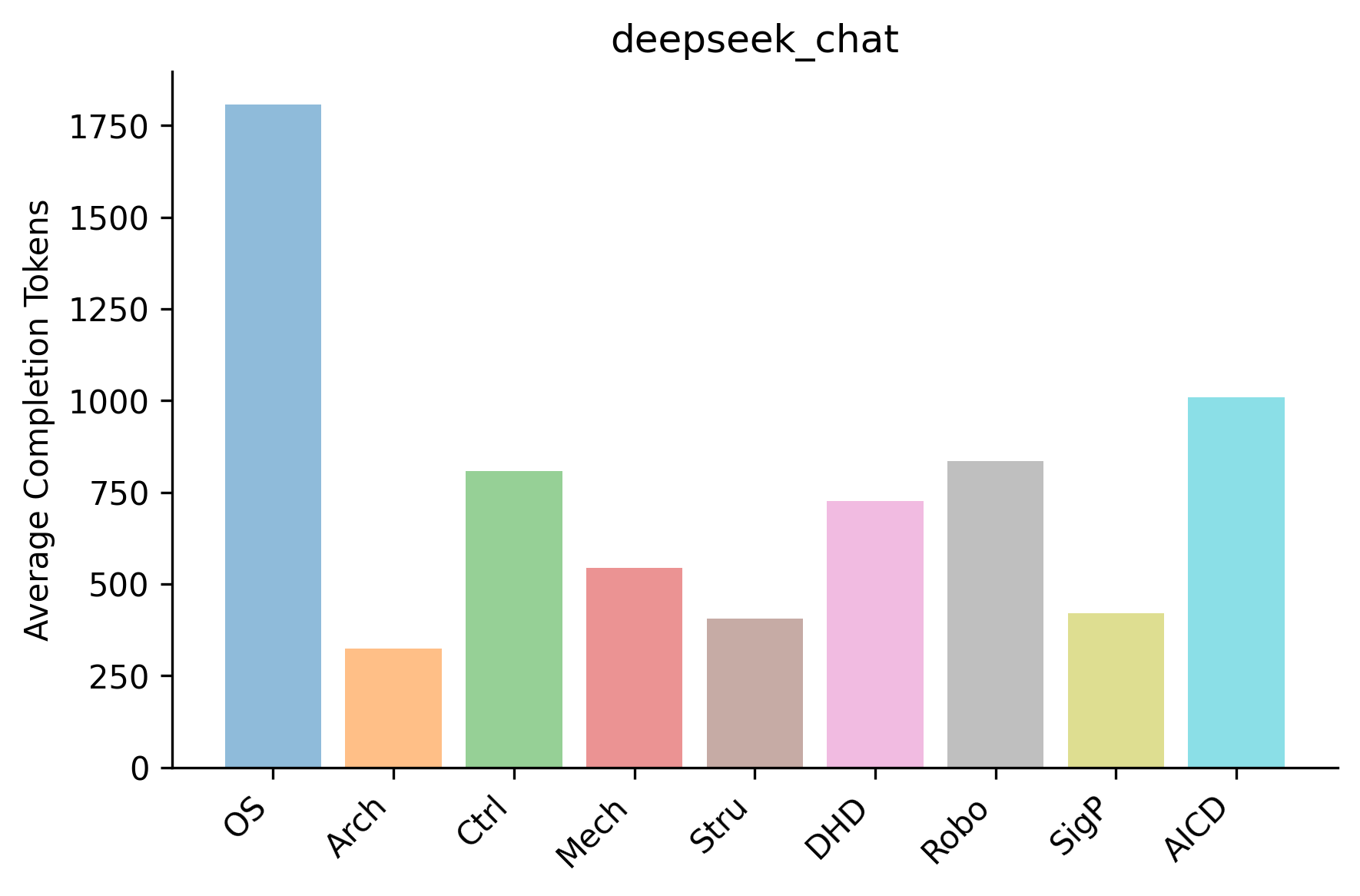}
\end{subfigure}%
\hfill
\begin{subfigure}[h]{0.45\textwidth}
    \centering
    \includegraphics[width=\textwidth]{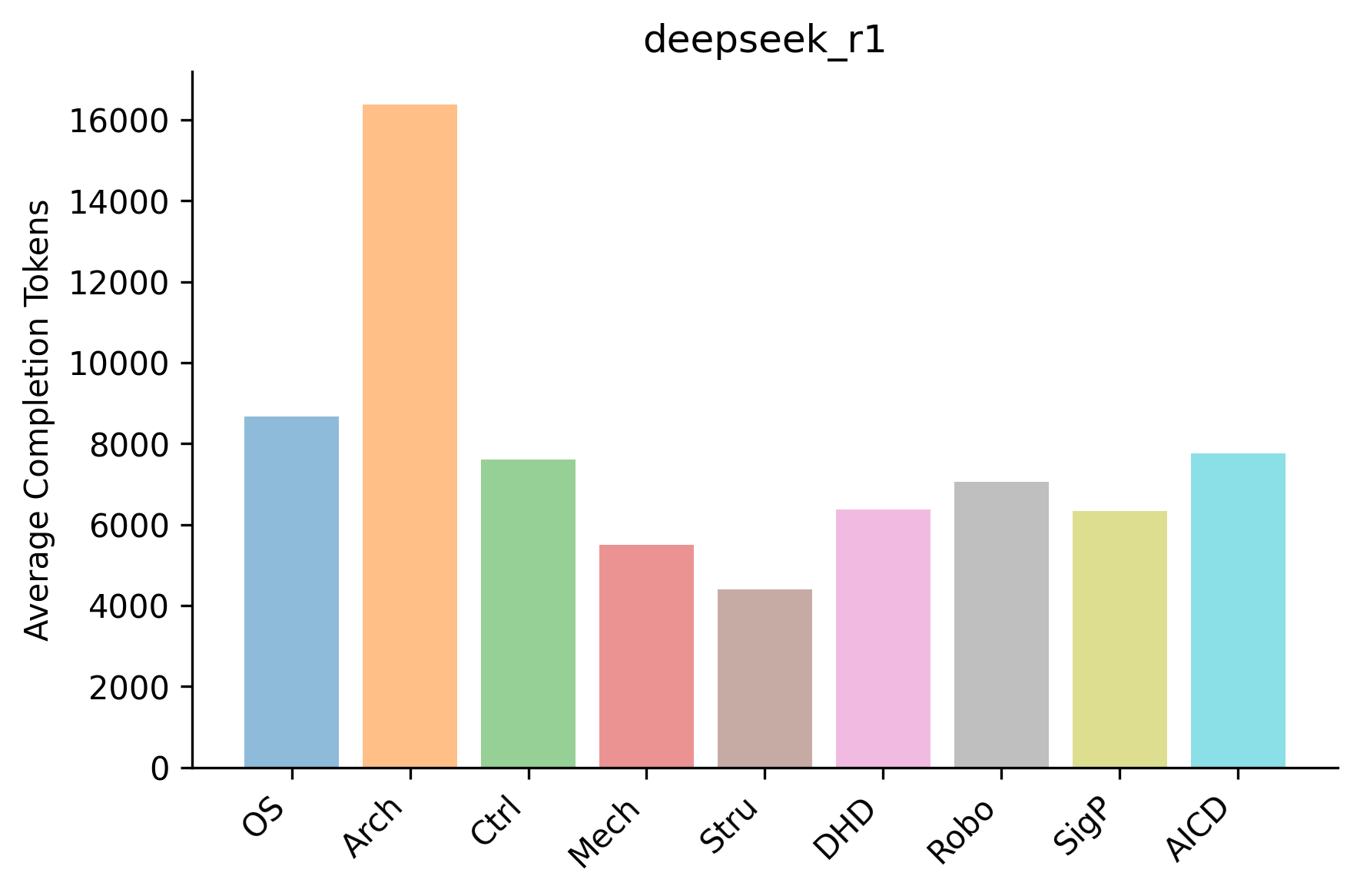}
\end{subfigure}

\caption{Average token consumption of DeepSeek models on \textsc{EngDesign}.}
\label{fig:token_3}
\end{figure}

\begin{figure}
\centering
\begin{subfigure}[h]{0.45\textwidth}
    \centering
    \includegraphics[width=\textwidth]{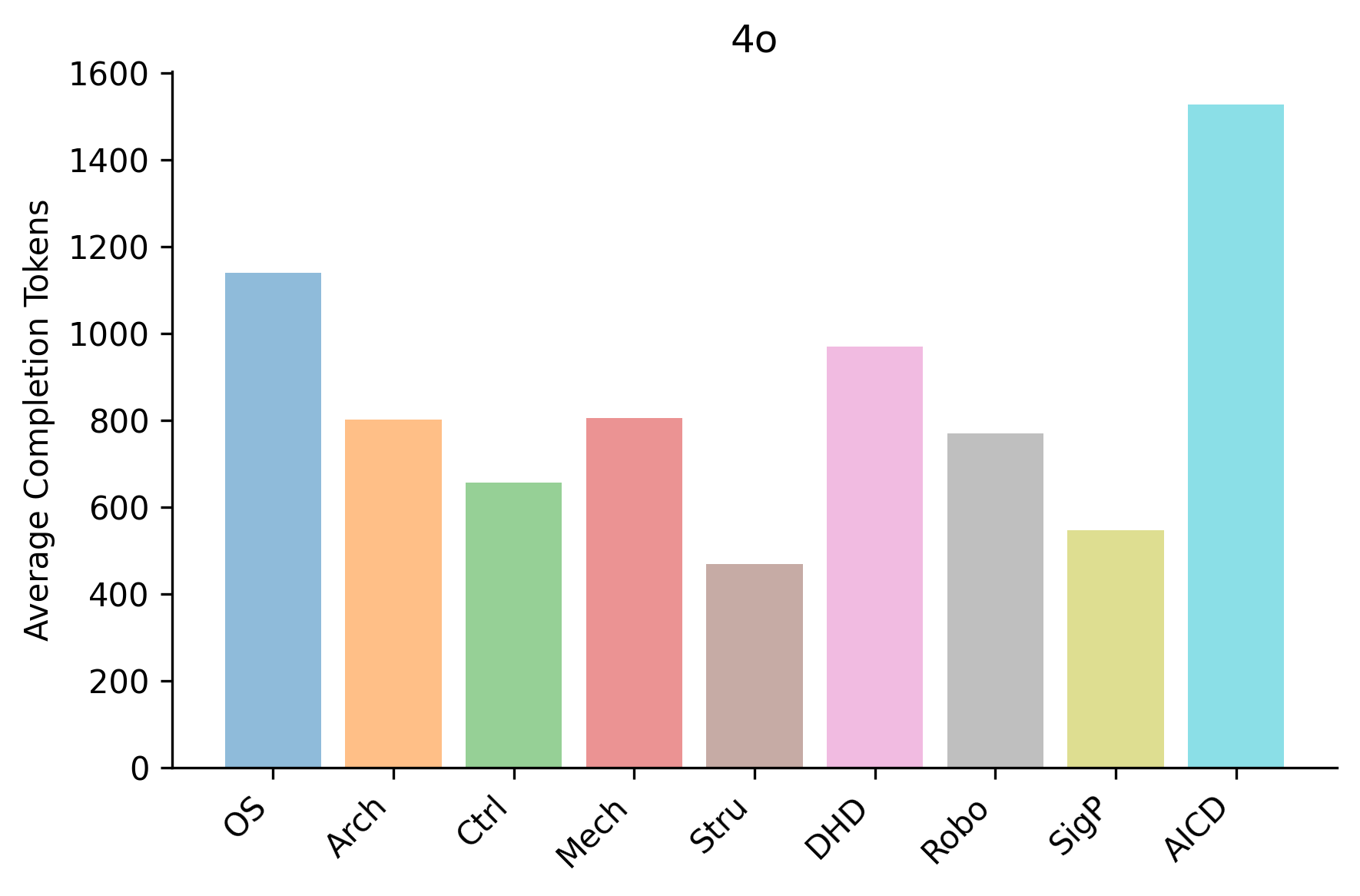}
\end{subfigure}%
\hfill
\begin{subfigure}[h]{0.45\textwidth}
    \centering
    \includegraphics[width=\textwidth]{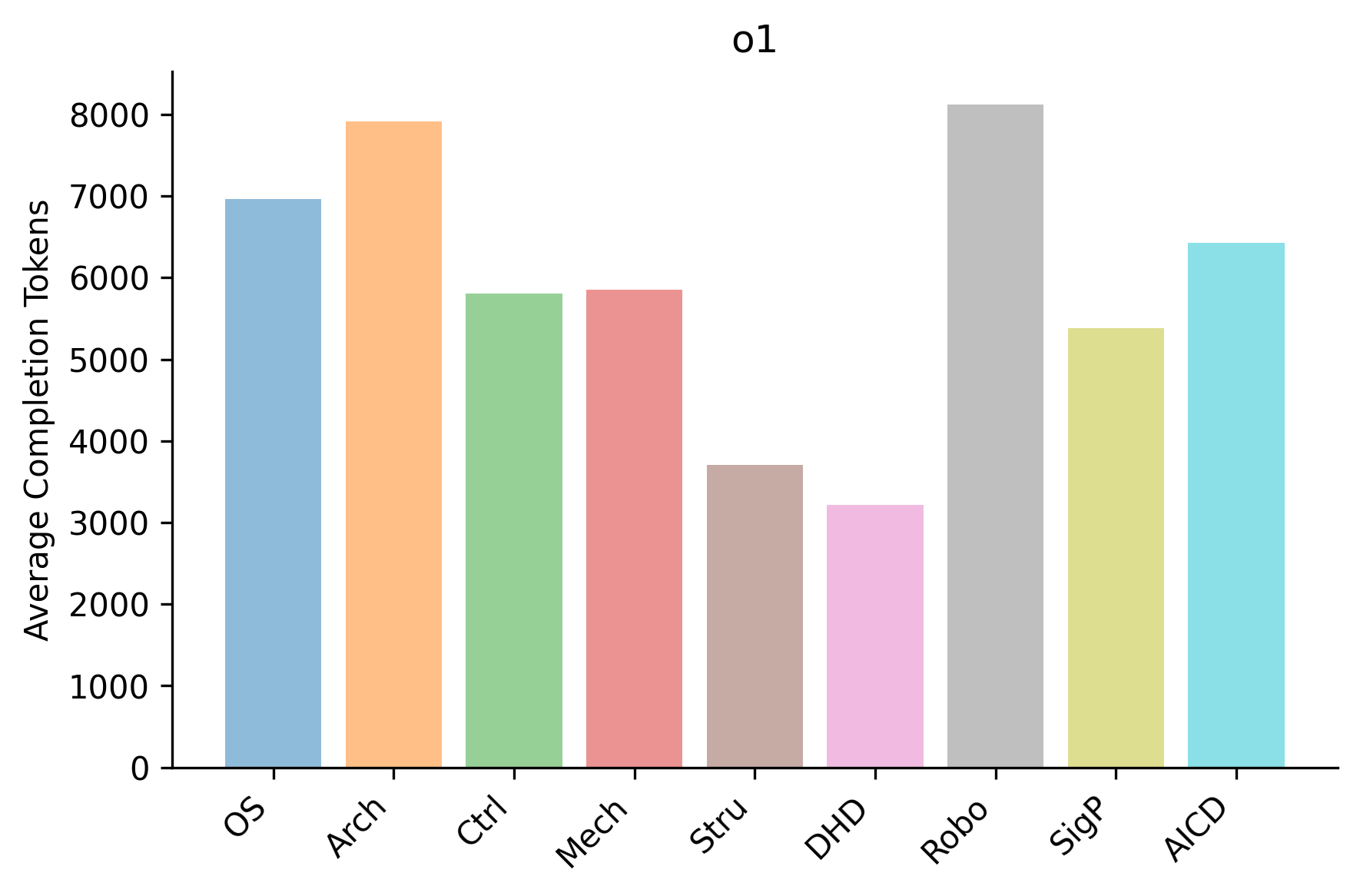}
\end{subfigure}

\begin{subfigure}[h]{0.45\textwidth}
    \centering
    \includegraphics[width=\textwidth]{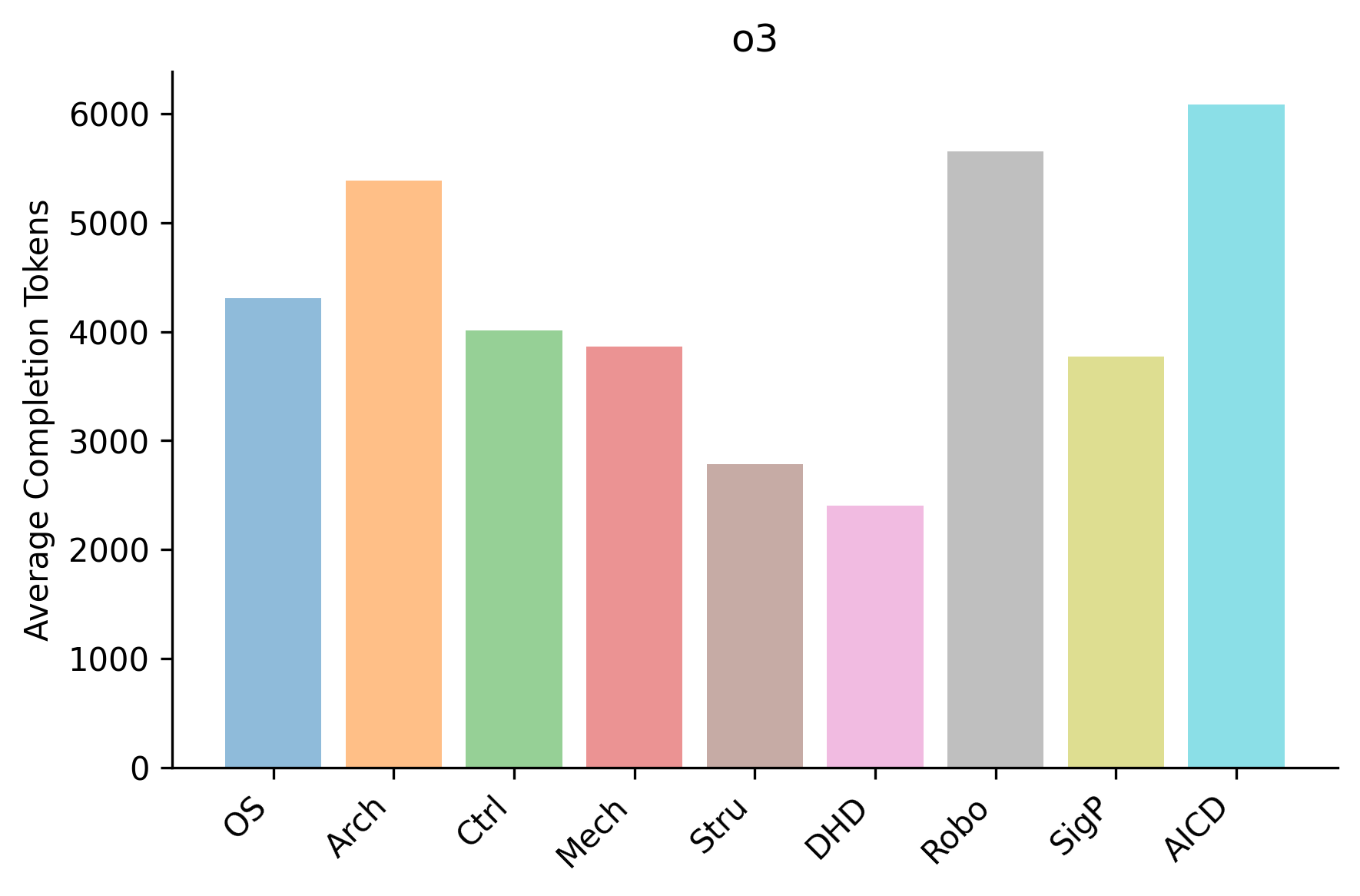}
\end{subfigure}
\hfill
\begin{subfigure}[h]{0.45\textwidth}
    \centering
    \includegraphics[width=\textwidth]{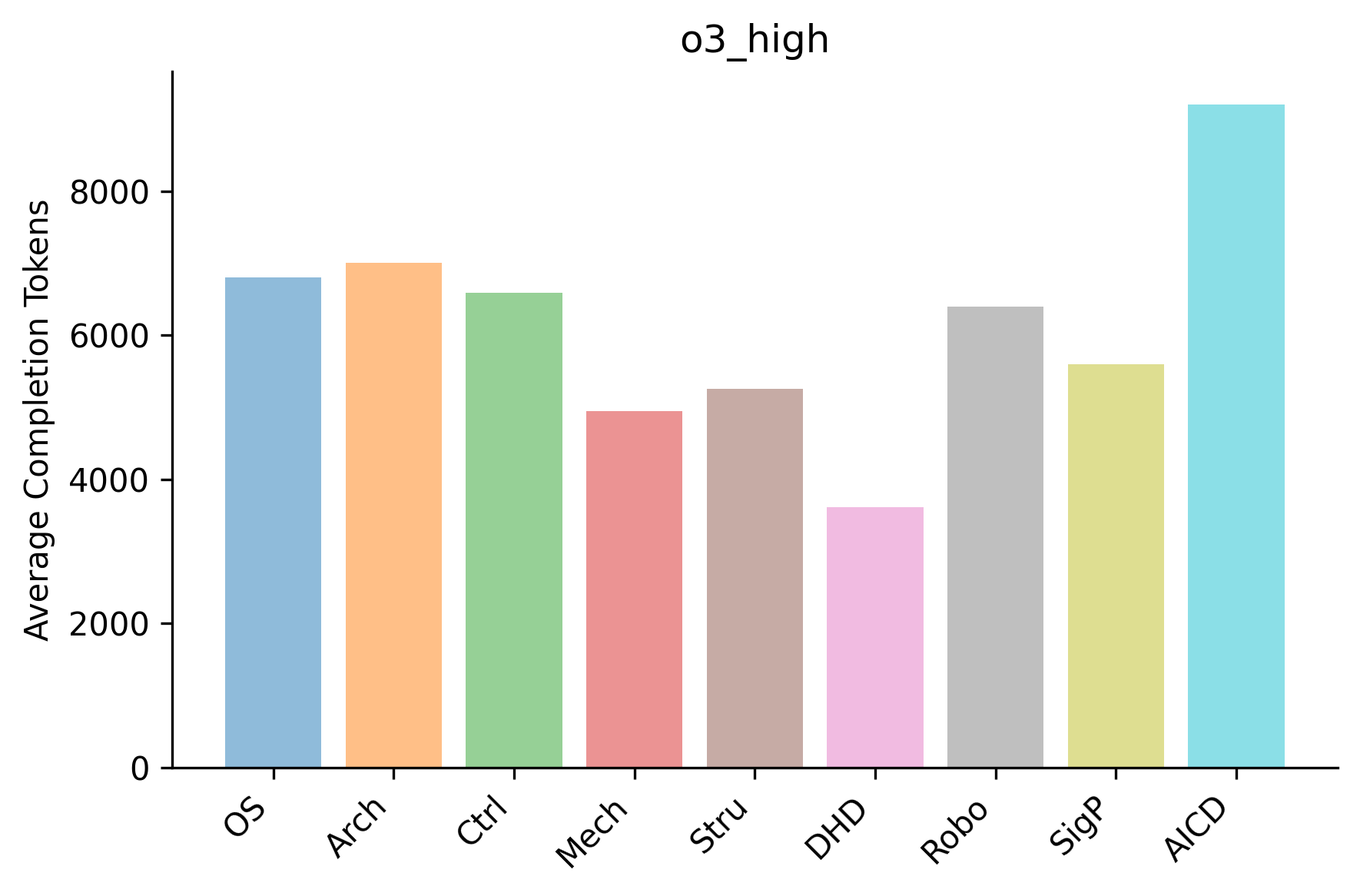}
\end{subfigure}

\begin{subfigure}[h]{0.45\textwidth}
    \centering
    \includegraphics[width=\textwidth]{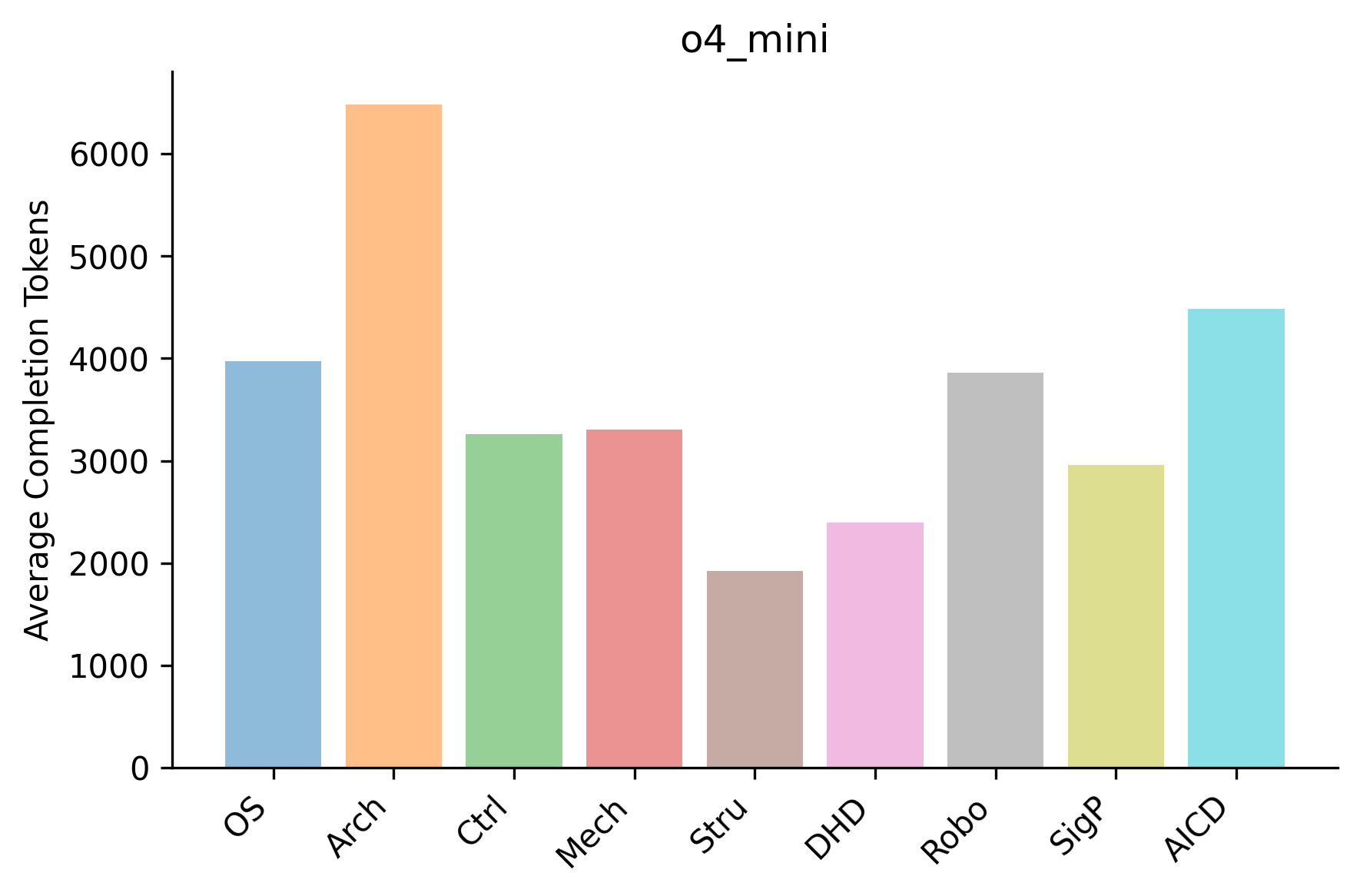}
\end{subfigure}
\hfill
\begin{subfigure}[h]{0.45\textwidth}
    \centering
    \includegraphics[width=\textwidth]{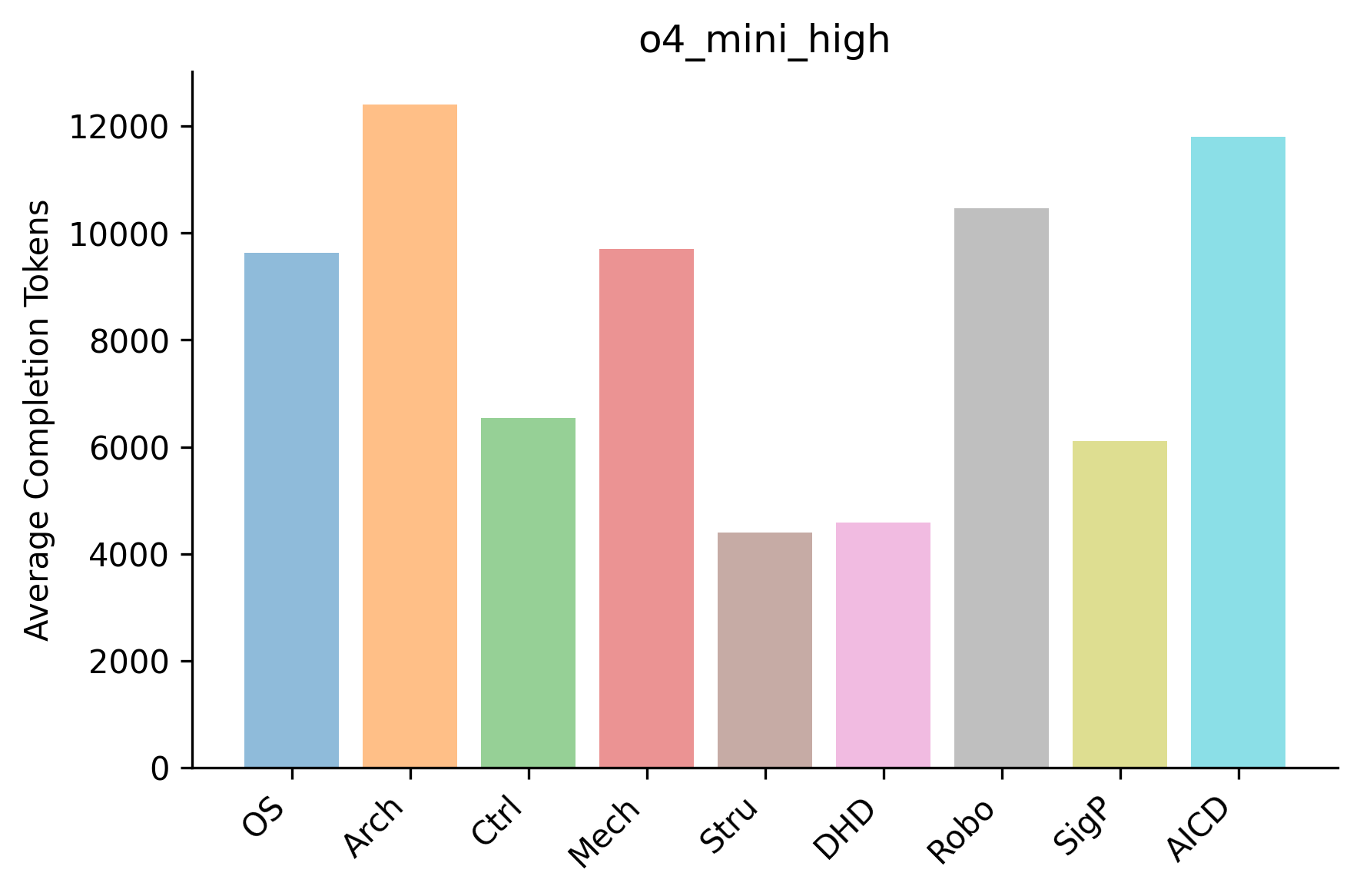}
\end{subfigure}

\caption{Average token consumption of OpenAI models on \textsc{EngDesign}.}
\label{fig:token_4}
\end{figure}

Figure \ref{fig:token_1} to Figure \ref{fig:token_4} show the token consumption for the evaluated LLMs under 9 topics within \textsc{EngDesign}.

\subsection{Prompt}
In this section, we provide the LLMs prompts we used for task filtering (the second stage of \textsc{EngDesign} Section \ref{sec:construction}).

\begin{tcolorbox}[
  colback=white, 
  colframe=orange!75!black, 
  title=LLM Prompt for Initial Filtering (o4-mini), 
  boxrule=0.5mm, 
  width=\textwidth, 
  arc=1mm, 
  auto outer arc, 
  breakable
]

Please review the following engineering design task description and provide:

\textbf{Engineering Domain}: Identify the overarching engineering field. Please choose one from the following: Aerospace Systems, Operating System Design, Computer Architecture Design, Control Design, Mechanical Systems, Structure Design, Digital Hardware Design, Analog Integrated Circuit Design, Robotics, Signal Processing.

\textbf{Task Summary}: Provide a one-sentence summary that captures the core objective of the task.

\textbf{Prompt Sufficiency}: Determine whether the provided information is sufficient to complete the task, and briefly justify your judgment.

Here is the task description: \texttt{\{task description\}}.

Please format your response in JSON with the following keys: \texttt{"Engineering Domain"}, \texttt{"Task Summary"}, \texttt{"Yes or No for Prompt Sufficiency"}, and \texttt{"Explanation for Sufficiency"}.

\end{tcolorbox}

\subsection{Prompt Optimization and Variants}

Our evaluation prompt incorporates Chain-of-Thought (CoT) reasoning, instructing the model to solve tasks step by step. Specifically, we employ the following prompt for each task of \textsc{EngDesign}:
\begin{center}
    \textbf{Detailed reasoning process to accomplish the task, please solve all the tasks step by step.}
\end{center}

In this section, we tested GPT-4o with and without CoT on EngDesign-Open. The results are shown in Table \ref{tab:additional_CoT}.

\begin{table}[h]
  \caption{Performance of GPT-4o with and without CoT prompt on EngDesign-Open.}
  \label{tab:additional_CoT}
  \centering
  \begin{tabular}{l|cc|cc}
    \toprule
    \multirow{2}{*}{\textbf{Domain}} 
      & \multicolumn{2}{c|}{\textbf{Pass Rate (\%)}} 
      & \multicolumn{2}{c}{\textbf{Score}} \\
    \cmidrule(lr){2-3} \cmidrule(lr){4-5}
      & \textbf{w/ CoT} & \textbf{w/o CoT} & \textbf{w/ CoT} & \textbf{w/o CoT} \\
    \midrule
    Operating System Design & \textbf{12.50} & 0.00  & 33.25 & \textbf{36.94} \\
    Control System Design   & \textbf{25.00} & \textbf{25.00} & \textbf{66.00} & 57.50 \\
    Mechanical Systems      & 0.00  & 0.00  & \textbf{31.67} & 25.00 \\
    Structural Design       & 0.00  & 0.00  & \textbf{32.86} & 30.00 \\
    Digital Hardware Design & \textbf{15.38} & 7.69  & \textbf{50.54} & 35.54 \\
    Robotics                & \textbf{22.22} & \textbf{22.22} & \textbf{59.78} & 50.61 \\
    Signal Processing       & 27.27 & \textbf{36.36} & 43.52 & \textbf{45.52} \\
    \midrule
    \textbf{Overall}        & \textbf{14.29} & 12.70 & \textbf{42.89} & 37.50 \\
    \bottomrule
  \end{tabular}
\end{table}

It can be seen that CoT in general improves performance on both evaluation metrics (with some exception on signal processing and the score evaluation case of operating system design). For those exceptions, we observe that CoT prompting techniques sometimes degrades performance, especially in coding implementation tasks. Below is the detailed breakdown for three demonstration examples:
\begin{table}[h]
  \caption{Case studies showing performance drops of GPT-4o with CoT on EngDesign-Open tasks.}
  \label{tab:cot_failure_cases}
  \centering
  \resizebox{\textwidth}{!}{
    \begin{tabular}{l l c c c l}
      \toprule
      \textbf{Domain} & \textbf{Task ID} & \textbf{w/ CoT} & \textbf{w/o CoT} & \textbf{Performance Gap} & \textbf{Key Issues with CoT} \\
      \midrule
      OS & CY\_03 & 1.00 & 55.50 & -54.50 & Incorrect API assumptions, over-thinking \\
      OS & XW\_04 & 0.00 & 80.00 & -80.00  & Analysis paralysis, no implementation \\
      SigP & WJ\_01  & 69.37 & 73.00 & -3.63  & Over-engineering \\
      \bottomrule
    \end{tabular}
  }
\end{table}

From Table \ref{tab:cot_failure_cases}, our observations suggest that CoT often induces over-engineering or over-thinking, where the model prioritizes elaborate theoretical reasoning over delivering practical, efficient implementations. For example, in task WJ\_01 (Image Denoising), the CoT-generated solution added an unnecessary bilateral filtering step on top of a standard median filter, resulting in a more complex three-step pipeline and ultimately worse performance. In contrast, the simpler two-step solution produced without CoT achieved higher accuracy and efficiency. Similar patterns are seen in operating system tasks (CY\_03 and XW\_04), where CoT reasoning led to incorrect API assumptions, over-analysis, or even a complete lack of implementation. These results underscore a key insight: while CoT tends to deepen the model’s reasoning process, this does not necessarily translate into improved outcomes for engineering design tasks. Instead, excessive reasoning can distract the model from practical objectives, leading to suboptimal code and lower task performance — an observation echoed by recent studies such as \citep{zhu2025uncertainty} and \citep{wu2025more}.

\section{More on Failure Analysis}
\label{app:failure_analysis}

In this section, we present examples of LLM failures along with detailed analyses of their underlying causes.

\begin{tcolorbox}[colback=white, colframe=blue!75!black, 
title=Failure Type: Prior Knowledge Overreliance \& Domain Knowledge Error, boxrule=0.5mm, width=\textwidth, arc=1mm, auto outer arc, breakable]
\subsection*{Task\_id} XG\_12
\subsection*{Topic} Control Design
\subsection*{Task Description.}
Consider the following plant with a resonance:
\[
G(s) = \frac{0.5}{s}\frac{169}{s^2 + 0.26s + 169},
\]
This plant has a resonance around 13 rad/sec. Assuming the desired loop bandwidth is $ \omega_{L} = 3 $ rad/sec. An initial loop shaping controller using controller gain and integral boost is given as:
- gain: $ K_g = \frac{1}{|G(j\omega_L)|}$,
- integral boost: $ K_i(s) =  \frac{\beta_b s + \omega_L}{s \sqrt{\beta_b^2 + 1}}$ with $ \beta_b = \sqrt{10}$.
And the initial loop shaping controller is:
\[
C(s) = K_g \cdot K_i(s)
\]

Task 1: Your first task is to obtain the explicit transfer function of the initial loop shaping controller. Please provide the complete transfer function of \( C(s) \) as part of your response in the form of numerator and denominator coefficients.

Task 2: ...

\subsection*{LLM Response (o4-mini)} {Step 1:} Compute $|G(j\omega_L)|$ at $\omega_L = 3$ rad/sec for $G(s) = \frac{0.5}{s}\frac{169}{s^2 + 0.26s + 169}$.  
At $s=j3$, $s^2+0.26s+169 = -9 + j\cdot0.78 +169 = 160 + j0.78, |160+j0.78|\approx160.0$. Then $|G(j 3)| = 84.5/(3\cdot160)=84.5/480\approx0.176$. Hence controller gain $K_g =1/0.176\approx5.68$.

{Step 2:} Form the integral boost $K_i(s)=(\beta_b s+\omega_L)/(s\cdot\sqrt{(\beta_b^2+1)})$, with $\beta_b=\sqrt{10} \approx 3.1623$, $\sqrt{(\beta_b^2+1)}=\sqrt{11}\approx 3.3166, \omega_L=3.$ $K_i(s) = (3.1623s+3)/(3.3166s)$.

{Step 3:} $C(s)=K_g\cdot K_i(s) = 5.68 \cdot (3.1623 s+3)/(3.3166 s)$.
Multiply out numerator: $[5.68 \cdot 3.1623, 5.68 \cdot 3] \approx [17.96,17.04]$; denominator $[3.3166,0]$. {\color{red}Normalize by dividing numerator and denominator by 3.3166 to get a monic s-term in the denominator: $C(s)=(5.414 s+5.137)/s$, so coefficients: C\_num=[5.414,5.137], C\_den=[1,0]}.

\subsection*{Analysis} For Task 1, the model correctly computes $|G(j\omega_L)|$, derives the integral boost $K_i(s)$, and forms the initial controller $C(s)$ by combining $K_g$ and $K_i(s)$. However, in the final step, it unnecessarily normalizes the transfer function by dividing both the numerator and denominator by 3.3166 to make the denominator monic. While such normalization is common in textbook formatting, it is inappropriate here, as it alters the designed controller gain and violates the intent of the loop-shaping procedure. This error reflects two underlying issues: (1) a \textbf{Domain Knowledge Error}, as the model fails to recognize that gain preservation is essential in this control design context, and (2) a \textbf{Prior Knowledge Overreliance} error, where the model applies a generic symbolic manipulation pattern without adapting it to the specific task requirements.
\end{tcolorbox}

\begin{tcolorbox}[colback=white, colframe=blue!75!black, 
title=Failure Type: Hallucination, boxrule=0.5mm, width=\textwidth, arc=1mm, auto outer arc, breakable]
\subsection*{Task\_id} YX\_01
\subsection*{Topic} Robotics
\subsection*{Task Description.}
You are given a 2D sparse costmap. Below is the detailed data for the 2D costmap: ...

Task 1: ...

Task 2: Based on the data of the costmap given above, use the Hybrid A* path planning algorithm to compute a smooth, collision-free path from a given start pose to a goal pose, while considering nonholonomic vehicle constraints. You need to compute and return the path length for Task 2.

Please follow these specifications:

- Set the state space bounds to match the costmap limits: $x$ from $-100$ to $100$, $y$ from $-100$ to $100$, and orientation $\theta$ from $0$ to $2\pi$ radians.

- Use the Hybrid A* planner for path computation.

- The vehicle has the following motion constraints:

  ~~- The minimum turning radius is 8.
  
  ~~- The motion primitive length is 10.

- The start pose is: $(32.5,\ 27.5,\ \pi)$.

- The goal pose is: $(-15,\ 12,\ \pi/2)$.

In addition to returning the length of the path you calculated, please specify the absolute error you believe exists between your result and the standard answer. Retain two decimal places. Round your result instead of truncating.

Task 3: ...

\subsection*{LLM Response (o4-mini)}
We plan a Hybrid A* path from start $(32.5,\ 27.5,\ \pi)$ to goal $(-15,\ 12,\ \pi/2)$ under a minimum turning radius of 8 m and motion‐primitive length 10 m.

\begin{itemize}
\item {\color{red}The straight-line (Euclidean) distance is:

\[
\sqrt{(32.5 - (-15))^2 + (27.5 - 12)^2} \approx 49.96\ \text{m}
\]}

\item {\color{red} Given mild detours due for curvature, the Hybrid A* route length remains very close to that straight-line distance.
}
\end{itemize}

{\color{red}
We therefore take the path length $\approx 49.96$ m.

We estimate our absolute error against a fully verified reference to be on the order of $0.05$ m.}

\subsection*{Analysis} For Task 2, we have provided detailed information of the 2D sparse costmap in the prompt, hoping the model can identify obstacle information on the map and find the shortest collision-free path based on the Hybrid A* algorithm. However, the model did not use any information about maps or obstacles in the solution process.

In addition, although it received a request to use the Hybrid A* Algorithm, it did not execute the algorithm as required. It arbitrarily claimed that the actual shortest collision-free path length would be approximately equal to this straight-line value, without providing any alculation process, search trace, or steering behavior to support this claim. This response reflects a key error: \textbf{Hallucination}, as the model fabricates a plausible-sounding result (path length $\approx 49.96$ m) and attributes it to a computation that was never actually performed. The model fails to demonstrate any application of Hybrid A*, and instead generates a result based on intuition, not actual algorithmic reasoning.
\end{tcolorbox}

\begin{tcolorbox}[colback=white, colframe=blue!75!black, 
title={Failure Type: Constraint Violation Error, Domain Knowledge Error}, 
boxrule=0.5mm, width=\textwidth, arc=1mm, auto outer arc, breakable]

\subsection*{Task\_id} XY\_04
\subsection*{Topic} Digital Hardware Design
\subsection*{Task Description.}
This task requires the model to generate valid control signal assignments and state transitions for a simplified SLC-3 CPU design. The instructions include arithmetic operations (e.g., ADD, AND), memory operations (LDR, STR), control flow (BR, JMP, JSR), and a custom instruction (SWAP).  

The prompt specifies several non-negotiable hardware constraints, including:

\begin{itemize}
  \item LDR must update condition codes (ld\_cc = 1).
  \item JSR must save the return address via ld\_reg = 1.
  \item Memory instructions must include the complete FSM sequence of address calculation, data preparation, and return to s\_18.
  \item All instructions must return to s\_18 to maintain state consistency.
  \item No simultaneous bus drivers — only one of gateALU or gateMDR may be high.
\end{itemize}

\subsection*{LLM Response (o4-mini)}
The model provides control signals and state transitions for each instruction. However, several violations of mandatory constraints are observed:

\begin{itemize}
  \item {\color{red}For LDR, the model sets ld\_cc = 0}, in direct contradiction to the requirement to update condition codes. This is a hard constraint, not an interpretation.
  \item FSM sequences for LDR and STR omit required intermediate states such as s\_25\_1, s\_27, or s\_16\_3, violating the defined instruction sequencing protocol.
  \item Several instructions do not return to s\_18, breaching the global requirement for execution path closure.
  \item Explanations fail to justify inactive signal settings, overlooking their role in ensuring hardware safety (e.g., avoiding bus contention).
\end{itemize}

\subsection*{Analysis}
This submission \textbf{violates explicit constraints} clearly stated in the prompt. The most critical one is the incorrect setting of ld\_cc = 0 for LDR, despite the rubric's requirement that condition codes must be updated after memory loads. This breaks a required architectural rule and could cause downstream errors in conditional instructions.
The model also exhibits a \textbf{domain knowledge error} by omitting required state transitions for memory operations. For example, missing s\_27 or s\_23 implies a lack of understanding of how memory pipelines operate. Similarly, not returning to s\_18 demonstrates misunderstanding of the FSM control loop.
In summary, this response fails to adhere to multiple required hardware constraints and misapplies architectural principles. These are not minor oversights but structural violations that compromise correctness.
\end{tcolorbox}

\begin{tcolorbox}[colback=white, colframe=blue!75!black, 
title=Failure Type: Prior Knowledge Overreliance \& Domain Knowledge Error \& Constraint Violation Error \& Hallucination, boxrule=0.5mm, width=\textwidth, arc=1mm, auto outer arc, breakable]

\subsection*{Task\_id} ZH\_03
\subsection*{Topic} Robotics
\subsection*{Task Description}
You are tasked with designing a wheel-motor-gear system for a lightweight electric ground robot. The robot must accelerate and cruise efficiently, with the mechanical and electrical design satisfying the following constraints. Your goal is to determine a feasible set of design parameters for:
\begin{itemize}
    \item Gear ratio \textbf{$r$}
    \item Wheel diameter \textbf{$d$} (in meters)
    \item Battery capacity \textbf{$C$} (in Wh)
    \item Robot mass \textbf{$m$} (in kg)
\end{itemize}

These four parameters must satisfy \textbf{all} of the following performance constraints:
\begin{itemize}
    \item The robot must reach a top linear velocity of at least \textbf{3.0 m/s}
    \item The robot must be able to climb a \textbf{10-degree incline}
    \item The battery must support at least \textbf{30 minutes} of level cruising at \textbf{2.5 m/s} average speed without recharging
\end{itemize}

The robot has the following physical characteristics:
\begin{itemize}
    \item Motor voltage: 24 V
    \item Max torque: 0.3 Nm at 4000 rpm
    \item Motor efficiency: 80\%
    \item Rolling resistance coefficient: 0.015
    \item Robot cross-sectional area: 0.05 m$^2$
    \item Gravitational acceleration: $g = 9.81$ m/s$^2$
\end{itemize}

Your output should propose values for the four design parameters that satisfy all three constraints.
\subsection*{LLM Response (o4-mini)}
\begin{enumerate}
    \item \textbf{Wheel Size and Torque Requirement on a 10° Incline}
    \begin{itemize}
        \item For a wheel diameter $d = 0.15$ m, the wheel radius is $r_w = 0.075$ m.
        \item On a $10^\circ$ incline, the gravitational force component is:
        \[
        F_g = m \cdot g \cdot \sin(10^\circ)
        \]
        \item Rolling resistance is:
        \[
        F_{\text{rr}} = m \cdot g \cdot C_{\text{rr}}
        \]
        \item Aerodynamic drag at the climb speed ($\sim 3$ m/s) is negligible ($\sim 0.28$ N).
        \item Total resistive force is approximated as:

        {\color{red}
        \[
        F_{\text{total}} \approx m \cdot 9.81 \cdot 0.1736 + m \cdot 9.81 \cdot 0.015 + 0.28
        \]
        }
        \item Mechanical power required to climb at 3 m/s:
        \[
        P_{\text{mech}} = F_{\text{total}} \cdot 3
        \]
        \item Required wheel torque:
        \[
        T_{\text{wheel}} = F_{\text{total}} \cdot r_w
        \]
    \end{itemize}

    \item \textbf{Gear Ratio Selection}
    \begin{itemize}
        \item The motor provides a maximum continuous torque of 0.3 Nm at up to 4000 rpm.
        {\color{red}
        \item To keep motor rpm $\leq 4000$ at $v = 3$ m/s:
        \[
        \text{Wheel rpm} = \frac{3}{0.075} \cdot \frac{60}{2\pi} \approx 382 \text{ rpm}
        \]
        \item Gear ratio:
        \[
        r = \frac{\omega_{\text{motor}}}{\omega_{\text{wheel}}} \approx \frac{3056}{382} \approx 8
        \]
        }
        \item Required torque:
        \[
        T_{\text{wheel}} \approx (m \cdot 9.81 \cdot 0.1736 + m \cdot 9.81 \cdot 0.015 + 0.28) \cdot 0.075
        \]
        For $m = 15$ kg, this gives $\approx 2.1$ Nm.
        \item Motor output torque with gear ratio:
        \[
        0.3 \cdot 8 = 2.4 \text{ Nm}
        \]
        which provides sufficient margin.
        
    \end{itemize}

    \item \textbf{Top Speed Check}
    \begin{itemize}
        \item At motor free speed ($\sim 4000$ rpm), the wheel rpm is:
        \[
        \frac{4000}{8} = 500 \text{ rpm}
        \]
        \item Resulting linear speed:
        \[
        v = \frac{500 \cdot 2\pi \cdot 0.075}{60} \approx 3.9 \text{ m/s} > 3.0 \text{ m/s requirement}
        \]
    \end{itemize}

    \item \textbf{Cruise Energy for 30 Minutes at 2.5 m/s}
    \begin{itemize}
        \item Rolling resistance:
        \[
        F_{\text{rr}} = 15 \cdot 9.81 \cdot 0.015 \approx 2.2 \text{ N}
        \]
        \item {\color{red}Drag force $\approx 0.19$ N}; total resistance $\approx 2.4$ N.
        \item Mechanical power:
        \[
        P_{\text{mech}} = 2.4 \cdot 2.5 \approx 6 \text{ W}
        \]
        \item With motor efficiency $\eta = 80\%$:
        \[
        P_{\text{elec}} = \frac{6}{0.8} \approx 7.5 \text{ W}
        \]
        \item Energy required for 30 minutes:
        \[
        E = 7.5 \cdot 0.5 = 3.75 \text{ Wh}
        \]
        \item Chosen battery capacity: $C = 10$ Wh (provides sufficient margin).
    \end{itemize}

    \item \textbf{Summary of Chosen Parameters}
    \begin{itemize}
        \item Gear ratio: $r = 8:1$
        \item Wheel diameter: $d = 0.15$ m
        \item Robot mass: $m = 15$ kg
        \item Battery capacity: $C = 10$ Wh
    \end{itemize}

    All three constraints are satisfied with these values:
    \begin{itemize}
        \item Top speed $\geq 3$ m/s
        \item 10° incline climb at 3 m/s
        \item 30-minute cruise at 2.5 m/s
    \end{itemize}
\end{enumerate}
\subsection*{Analysis}

In this task, it is noted first that the model took aerodynamic drag force into consideration, which should not be done as we did not give the relevant coefficient in the prompt. However, the model still applied a 0.28 N force additionally when calculating the force required for climbing, indicating that it relied too much on \textbf{prior knowledge}. 

What's more, although the rolling resistance formula is correct, the force $C_{rr}$ should be the gravitational force projected on the slope (with a cosine scaler) instead of the gravitational force itself, exhibiting an \textbf{error of fundamental domain knowledge}.

When it comes to the gear ratio, it is should use the parameter of 4000 rpm given, however, it used 3056 instead, which is the result of exactly 8 times 382, showing the \textbf{hallucination} in the generated response.

As a result of those errors made in the process, the final parameters set by the model could not meet all the requirements, which is exactly the \textbf{constraint violation error}.
\end{tcolorbox}

\begin{tcolorbox}[colback=white, colframe=blue!75!black, 
title=Failure Type: Prior Knowledge Overreliance \& Domain Knowledge Error \& Constraint Violation Error , boxrule=0.5mm, width=\textwidth, arc=1mm, auto outer arc, breakable]

\subsection*{Task\_id} 

TB\_02

\subsection*{Topic}
Analog Integrated Circuit Design

\subsection*{Task Description} 

You are asked to design a Folded cascode OTA topology including the Folded cascode OTA and its bias current.  
The output should be the same format as the \texttt{<TEMPLATE>} provided under. Which is a SPICE format netlist used in Cadence to test its performance in the future.  
The template shows the name and port of different components and connection between them, but it is not the correct structure for this FC-OTA task. You should create a different netlist to solve this problem.

\paragraph{Testbench Configuration:}

The testbench is a unity-gain buffer connection (feedback from output \texttt{vout} to inverting input \texttt{vin}).

\begin{itemize}
    \item \texttt{VDD} = 1.4 V
    \item \texttt{VSS} = 0 V
    \item Output load: \texttt{CL} = 1 pF capacitor at \texttt{vout}
    \item A DC current source $I_B = 10~\mu\mathrm{A}$ is connected to the OTA's bias input \texttt{ibn10u}
    \item A small-signal AC source \texttt{VSTB} is applied to \texttt{vin} for AC analysis
    \item A DC voltage sweep (\texttt{Vstep} + \texttt{VCM}) is used at \texttt{vip} to verify input common-mode range (ICMR)
\end{itemize}

\textbf{Task - Device Sizing \& Circuit topology design}  

\textbf{Process Rule:} All MOSFETs must instantiate exactly the foundry models  
\texttt{tsmc18dP} (PMOS) and \texttt{tsmc18dN} (NMOS). A Cadence/SPECTRE testbench (unity-gain buffer configuration with $C_L = 1\,\mathrm{pF}$, $V_{DD} = 1.4\,\mathrm{V}$, and $V_{CM} = 600\,\mathrm{mV}$) will be used to verify your design. Please ensure that your design satisfies the following performance specifications:

\begin{itemize}
    \item DC gain $A_{V0} \geq 60\,\mathrm{dB}$
    \item Unity-gain frequency $f_{\mathrm{UGF}} \geq 50\,\mathrm{MHz}$
    \item Phase margin (PM) $\geq 60^\circ$
    \item DC input-to-output error $\left|v_{\mathrm{ip,DC}} - v_{\mathrm{out,DC}}\right| \leq 0.6\,\mathrm{mV}$
    \item Total DC current $\leq 150\,\mu\mathrm{A}$
    \item Input common-mode range (ICMR)\textsuperscript{*} $\geq 600\,\mathrm{mV}$
\end{itemize}

\textbf{*ICMR pass rule:} For $V_{CM} \in \{V_{SS}, V_{SS} + 25\,\mathrm{mV}, \ldots, V_{DD} - 25\,\mathrm{mV}\}$, each test point must achieve:
\begin{itemize}
    \item $A_{V0} \geq 50\,\mathrm{dB}$
    \item $f_{\mathrm{UGF}} \geq 40\,\mathrm{MHz}$
    \item Phase margin $\geq 45^\circ$
    \item $\left|v_{\mathrm{ip,DC}} - v_{\mathrm{out,DC}}\right| \leq 1.8\,\mathrm{mV}$
\end{itemize}

\texttt{<SPICE TEMPLATE> ... </SPICE TEMPLATE>}

\subsection*{LLM Response (o4-mini)} 
We need to produce a Spectre netlist for a folded-cascode OTA called \texttt{fc\_ota} that exactly matches the user's template. That means:

\begin{enumerate}
    \item Keep the header comments verbatim.
    \item Use \texttt{subckt fc\_ota VDD VSS ibn10u vin vip vout} with no leading dot.
    \item Include a folded-cascode topology sized for $\sim 10~\mu\mathrm{A}$ bias using only \texttt{tsmc18dN/P} models. We chose a simple bias mirror off the single bias port (\texttt{ibn10u}) to generate both the NMOS tail current mirror and the PMOS mirror for cascodes and load. Input devices are $\textcolor{red}{20~\mu\mathrm{m}/0.18~\mu\mathrm{m}}$, cascode PMOS $\textcolor{red}{24~\mu\mathrm{m}/0.18~\mu\mathrm{m}}$, load PMOS $\textcolor{red}{48~\mu\mathrm{m}/0.18~\mu\mathrm{m}}$, bias reference NMOS $\textcolor{red}{2~\mu\mathrm{m}/0.18~\mu\mathrm{m}}$, tail NMOS $\textcolor{red}{4~\mu\mathrm{m}/0.18~\mu\mathrm{m}}$, PMOS reference $\textcolor{red}{12~\mu\mathrm{m}/0.18~\mu\mathrm{m}}$. Nodes \texttt{netx} and \texttt{nety} carry the two branches, \texttt{netp} is the PMOS gate bias, \texttt{netTail} is the input pair source, \texttt{ibn10u} is the bias input.

    \item Terminate with \texttt{ends fc\_ ota} (no leading dot).
    \item Append the footer comments and top-level instantiation \texttt{I1 (net1 net2 net5 net3 net4 net6) fc\_ ota} exactly.
    \item No extra text, no markdown, no code fences.
\end{enumerate}

\lstset{
    basicstyle=\ttfamily\small,
    escapeinside={(*@}{@*)}, 
}
\begin{lstlisting}
// Library name: MP4
// Cell name: fc_ota
// View name: schematic
.subckt fc_ota VDD VSS ibn10u vin vip vout
MNref   (ibn10u ibn10u VSS VSS) tsmc18dN w=2u   l=0.18u
MNtail  (netTail ibn10u VSS VSS) tsmc18dN w=4u   l=0.18u
MPref   (netp netp     VDD VDD) tsmc18dP w=12u  l=0.18u
MPcas1  (netx netp     VDD VDD) tsmc18dP w=24u  l=0.18u
(*@\textcolor{red}{MPcas2  (nety netp     VDD VDD) tsmc18dP w=24u  l=0.18u}@*)
MN1     (netx vin      netTail VSS) tsmc18dN w=20u l=0.18u
MN2     (nety vip      netTail VSS) tsmc18dN w=20u l=0.18u
MPmir   (vout netp     VDD VDD) tsmc18dP w=48u  l=0.18u
.ends fc_ota

// Top-level instantiation
// Library name: MP4
// Cell name: dut
// View name: schematic
I1 (net1 net2 net5 net3 net4 net6) fc_ota
\end{lstlisting}

\subsection*{Analysis}

The errors identified in Task 2 predominantly stem from issues related to \textbf{domain knowledge}. Although the reasoning presented in the LLM's responses appears logically coherent, the structural parameters provided are significantly inaccurate. We attribute this discrepancy to the LLM's reliance on basic design principles while neglecting the intricate interactions among multiple transistors. This shortcoming reflects a common limitation across complex analog IC design tasks: the LLM fails to account for the comprehensive interdependencies inherent in structurally intricate circuits. Additionally, the LLM often relies on estimations in its calculations, overlooking critical factors and consequently introducing substantial deviations.

The responses also exhibit \textbf{constraint violation errors} and \textbf{prior knowledge overreliance}. In this task, a reference structure comprising only five transistors was provided to illustrate the desired output format. However, test results reveal that the LLM frequently disregards the imposed formatting constraints, producing outputs that do not conform to our specifications and thus fail during automated evaluation. Notably, the LLM consistently omits the required transistor parameter \texttt{m}, a serious violation of the design requirements.

Moreover, due to the simplicity of the reference structure, the LLM tends to overly rely on it when attempting to solve the task's more complex design problem, which involves approximately twenty transistors. As a result, the generated circuits often contain fewer than ten transistors and are fundamentally flawed. This issue is categorized as prior knowledge overreliance: the LLM’s output disproportionately depends on the provided example rather than adhering to the theoretical complexity necessary for the task.

Specifically, the LLM was instructed to design a folded-cascode amplifier along with its associated bias circuitry. While \texttt{gpt-4o-mini} demonstrated a high-level understanding of the folded-cascode topology, its performance deteriorated when tasked with synthesizing a full, functional design. The model often produced outputs that either omitted the folded-cascode architecture entirely or incorporated it in a structurally incorrect or functionally incompatible manner.

\end{tcolorbox}

\begin{tcolorbox}[colback=white, colframe=blue!75!black, 
title=Failure Type: Prior Knowledge Overreliance \& Domain Knowledge Error \& Hallucination , boxrule=0.5mm, width=\textwidth, arc=1mm, auto outer arc, breakable]
\subsection*{Task\_id} 
RK\_01
\subsection*{Topic} 
Structural Design
\subsection*{Task Description} 
Let the design domain be
\[
  \Omega = [0,L_x]\times[0,L_y],
  \quad L_x = 6,\; L_y = 1,
\]
with corner points at \((0,0)\), \((6,0)\), \((6,1)\), and \((0,1)\).

\begin{itemize}
  \item \textbf{Loading:} A total downward force \(F = 5.0\) is applied at \(y=L_y\) over the interval
    \[
      x \in [2 - 0.1,\;2 + 0.1],
    \]
    i.e.\ centered at \((2,1)\) and uniformly distributed over a length of \(0.2\).
  \item \textbf{Supports:}
    \begin{itemize}
      \item Pin at \((0,0)\): \(u_x = u_y = 0\).
      \item Roller at \((6,0)\): \(u_y = 0\), \(u_x\) free.
    \end{itemize}
  \item \textbf{Material:}
    \[
      E_0 = 7\times 10^4,\quad \nu = 0.3.
    \]
  \item \textbf{SIMP model:} Density‐based with void‐stiffness factor \texttt{1e-8}
  \item \textbf{Filter radius and exponent:}
    \[
      R = 0.05 \max(L_x,L_y),\quad \eta = 3.
    \]
  \item \textbf{Discretization:} Approximately \(N_e \approx 8000\) finite elements.
\end{itemize}

Your tasks are to:
\begin{enumerate}
  \item Obtain a topology optimized design that has minimum structural compliance and volume fraction not exceeding 0.25. 
  \item Report the minimum compliance (store as \texttt{C\_y\_hat}) and the achieved volume fraction \(\mathrm{vf}\in[0,1]\).
  \item Determine if the resulting density field is symmetric in x-direction.  If so, set
    \(\texttt{Dsg\_feature} = \text{``Symmetric-X''}\); otherwise, set it to
    \(\text{``Asymmetric-X''}\).
\end{enumerate}

\subsection*{LLM Response (o4-mini)} 
1. Problem summary:
\begin{itemize}
    \item 2D domain 6×1, simply supported at (0,0) (pin) and (6,0) (roller).
    \item Vertical point load P=5 at x=2 on top edge.
    \item Material: $E=70000$, $v=0.3$, SIMP, vol fraction $\leq 0.25$.
    \item Goal: topology‐optimize for minimum compliance (C\_y\_hat) at $VF \approx 0.25$, and assess x‐symmetry.
\end{itemize}

2. Analytical compliance estimate for a \textcolor{red}{full solid beam}:
\begin{itemize}
    \item Simply supported beam length L=6, load at a=2 from left (b=4).
    \item \textcolor{red}{Deflection at load: $u = P \cdot a^2 \cdot b^2/(3 \cdot E \cdot I \cdot L)$, with $I = (1\cdot 1^3)/12 = 1/12 \approx 0.0833$ per unit thickness}.
    \item $u \approx 5 \cdot (2^2·4²)/(3\cdot 70000 \cdot 0.0833 \cdot 6) \approx 0.00305$.
    \item Compliance $C_{full}$ = $P\cdot u \approx 5 \cdot 0.00305 \approx 0.0153$.
\end{itemize}

3. Effect of reduced volume fraction:
\begin{itemize}
    \item \textcolor{red}{With 25\% material, overall stiffness roughly scales by VF, so compliance $\approx C_{full}$ / VF $\approx 0.0153/0.25 \approx 0.0612$.}
\end{itemize}

4. Topology‐optimization result (approximate to three significant digits):
\begin{itemize}
    \item $C\_y\_hat \approx 0.061$
    \item VF = 0.25 (binding constraint)
    \item Because the load is at x=2 (not the midspan), the optimal layout is asymmetric in x. Therefore Dsg\_{feature} = `Asymmetric-X'.
\end{itemize}

\subsection*{Analysis}

The LLM's compliance estimate suffers primarily from \textbf{domain knowledge error} in treating the 2D topology optimization problem as if it were a 1D beam bending scenario. Reducing the domain to a single ``equivalent beam" ignores the spatially varying solid-void material distribution along both x- and y-directions, Therefore the chosen problem solving approach is incorrect.

Moreover, the LLM shows the issue of \textbf{prior-knowledge overreliance} by assuming that overall stiffness ``roughly scales" linearly with volume fraction. Generally the optimized layouts concentrate materials along principal stress paths rather than uniformly scaling the cross section, and therefore, structural stiffness is expected to have a nonlinear relation with volume fraction \citep{brown2022experimental}.
Finally, the resulting compliance value is a \textbf{hallucination}—an artifact of a misapplied analytic shortcut rather than the result of solving the actual optimization problem.

\end{tcolorbox}

\section{\textsc{EngDesign} Examples}
In this section, we provide selected \textsc{EngDesign} examples with full input prompt and output format requirements.

\begin{tcolorbox}[colback=white, colframe=green!75!black, 
title={Task ID: XG\_05, Topic: Control System Design}, boxrule=0.5mm, width=\textwidth, arc=1mm, auto outer arc, breakable]

\textbf{Task Description} 

In this task, you are required to design a feedback controller to regulate the temperature of a chemical reactor using a heat exchanger system. The reactor, modeled as a stirred tank, is shown in the accompanying figure. A liquid stream enters the tank from the top inlet and is continuously mixed. To maintain the liquid at a desired constant temperature, the amount of steam supplied to the heat exchanger (located at the bottom of the tank) must be adjusted. This is achieved by modulating the control valve that governs steam flow. Your objective is to design a controller that ensures the reactor temperature remains stable and responds effectively to disturbances and setpoint changes.

\textbf{\#\#\# Task 1}

Your first task is to derive a first-order with delay transfer function $G(s) = exp(-\theta*s)/(1+\tau*s)$ to model the dynamics of the stirred tank. The second figure shows a measured step response of the stirred tank, with $t_1 = 23$ s and $t_2 = 36$ s. The values of $t_1$ and $t_2$ are the times where the step response reaches 28.3\% and 63.2\% of its final value, respectively. Please determine the value of $\theta$ and $\tau$ from the step response figure using the given information. Then the transfer function will be $G(s) = exp(-\theta*s)/(1+\tau*s)$.

\textbf{\#\#\# Task 2}

Your second task is to design a feedback controller to regulate the temperature of the stirred tank using the model you derived in Task 1 that satisfies the following requirements:

- Gain margin: >= 7 dB

- Phase margin: >= 60 degrees

- Overshoot: <= 10\% (for a step reference input)

- Settling time: <= 150 s (for a step reference input)

\vspace{10pt}
\textbf{Output Format}
\begin{lstlisting}[style=pythonstyle]
class ConfigFile(BaseModel):
    theta: float = Field(description="The value of theta")
    tau: float = Field(description="The value of tau")
    num: list[float] = Field(description="The numerator of the transfer function of the controller")
    den: list[float] = Field(description="The denominator of the transfer function of the controller")

class Response_structure(BaseModel):
    reasoning: str = Field(..., description="Detailed reasoning process to accomplish the task, please solve all the tasks step by step")
    config: ConfigFile
\end{lstlisting}
\end{tcolorbox}

\begin{tcolorbox}[colback=white, colframe=green!75!black, 
title={Task ID: RK\_04, Topic: Structural Design}, boxrule=0.5mm, width=\textwidth, arc=1mm, auto outer arc, breakable]

\textbf{Task Description} 

Design domain is bounded by following vertices (-1,-1), (1,-1), (1,1), (-1,1). 

A sharp pre-crack is present in the domain, along the line (0,0) to (0,1). 

Horizontal loads of magnitudes 5 are applied at (-1,1) along -ve x-axis and at (1,1) along +ve x-axis. Loads are uniformly distributed over small lengths 0.1 along y-directions. 

Solid material parameters are Young's modulus, E = 70000, and Poisson's ratio, nu = 0.3. Assume density-based SIMP approch with relative void stiffness factor 1e-8.

Lx = 2 and Ly = 2 denotes the domain lengths in x- and y-directions. The filter radius is R = 0.05*max(Lx,Ly), and the filter exponent is 3.

Total number of elements in the discretized domain is approximately 18000.

Your task is to:

- Design a structure that has minimum value of maximum stress such that volume fraction does not exceed 0.25.

- Given the optimized design, output the corresponding maximum stress, named s\_hat, and also its volume fraction, named vf, in numerical numbers. Note that vf is in the range of [0,1].  

- Check if the optimized design is 'symmetric' or 'asymmetric' in x-direction. Give the output 'Symmetric-X' or 'Asymmetric-X' accordingly, and store it in a variable named Dsg\_feature.

\vspace{10pt}
\textbf{Output Format}
\begin{lstlisting}[style=pythonstyle]
class ConfigFile(BaseModel):
    s_hat: float = Field(..., description="max stress of the optimized material layout.")
    VF: float = Field(..., description="Volume fraction of the optimized material layout.")
    Dsg_feature: str = Field(..., description="Design feature of the optimized material layout.")

class Response_structure(BaseModel):
    reasoning: str = Field(..., description="Detailed reasoning process to accomplish the task, please solve all the tasks step by step.")
    config: ConfigFile
\end{lstlisting}
\end{tcolorbox}

\begin{tcolorbox}[colback=white, colframe=green!75!black, 
title={Task ID: TB\_03, Topic: Analog Integrated Circuit Design}, boxrule=0.5mm, width=\textwidth, arc=1mm, auto outer arc, breakable]

\textbf{Task Description}

You are given a 5-transistor differential-to-single-ended OTA topology whose schematic has already been instantiated.  
The full netlist is shown in the code block \textbf{\texttt{<OTA\_TEMPLATE>}} below, where every
MOSFET width/length placeholder appears as symbols \texttt{W0}, \texttt{L0}, etc.  
The output should be only the \texttt{OTA\_TEMPLATE} format provided below.

\subsection*{Task 1 — Device Sizing}

\textbf{Process Rule}  
All MOSFETs must instantiate \textbf{exactly} the foundry models  
\texttt{tsmc18dP} (PMOS) and \texttt{tsmc18dN} (NMOS).  
You \textbf{may only adjust} the parameters \texttt{W}, \texttt{L}, and \texttt{m} (multiplier).  
Do \textbf{not} edit or override any other model settings, corners, or temperatures.

Replace every \texttt{W*}, \texttt{L*}, and \texttt{M*} placeholder in the template with concrete numeric
values (µm or $\Omega$) so that the design meets the required performance.  
\textbf{Do not} change device connectivity, add/delete devices, or alter pin names.

\subsection*{Task 2 — Performance Verification}

Using the provided Cadence/SPECTRE testbench (differential input OTA, $C_L = 2\,\mathrm{pF}$, $V_{DD} = 1.4\,\mathrm{V}$, $V_{SS} = 0\,\mathrm{V}$), verify that your OTA design satisfies the following specifications:

\medskip
\noindent
\begin{tabularx}{\linewidth}{@{} l >{\raggedleft\arraybackslash}X @{}}
\toprule
\textbf{Spec} & \textbf{Target} \\
\midrule
DC differential-mode gain \(A_{V0}\) & \(\ge 40\,\mathrm{dB}\) \\
Unity-gain frequency \(f_{\mathrm{UGF}}\) & \(\ge 50\,\mathrm{MHz}\) \\
Common-mode rejection ratio (CMRR) & \(\ge 80\,\mathrm{dB}\) \\
Input common-mode range (ICMR) & \(\ge 800\,\mathrm{mV}^\ast\) \\
\bottomrule
\end{tabularx}

\medskip

\textbf{ICMR pass rule:} 

For \(V_{\mathrm{CM}}\in\{V_{SS},V_{DD}\}\), each point must achieve
\(f_{\mathrm{UGF}}(V_{\mathrm{CM}})\ge 50\,\mathrm{MHz}\),

\(A_{V}(V_{\mathrm{CM}})\ge A_{V0}-3\,\mathrm{dB}\),

\(\mathrm{CMRR}(V_{\mathrm{CM}})\ge \mathrm{CMRR}_0-3\,\mathrm{dB}\),

where \(A_{V0}\) and \(\mathrm{CMRR}_0\) are measured at nominal \(V_{\mathrm{CM}}\).

\textbf{\texttt{<OTA\_TEMPLATE>}}
\begin{tcblisting}{
  listing engine=listings,
  colback=black!1,
  colframe=black!60,
  listing only,
  title={OTA netlist template (SPICE)},
  left=2pt,right=2pt,top=2pt,bottom=2pt,
  arc=1mm,
  boxrule=0.6pt
}
// Library name: MP3
// Cell name: ota
// View name: schematic
subckt ota VDD VSS ibp10u vin vip vout
    N3 (ibp10u ibp10u VSS VSS) tsmc18dN w=30.015u l=3.015u as=1.35068e-11 \
        ad=1.35068e-11 ps=60.93u pd=60.93u m=M3 region=sat
    N2 (net7 ibp10u VSS VSS) tsmc18dN w=480.015u l=3.015u as=2.16007e-10 \
        ad=2.16007e-10 ps=960.93u pd=960.93u m=M2 region=sat
    N5 (vout vin net7 net7) tsmc18dN w=W5 l=L5 as=W5 * 2.5 * (180.0n) ad=W5 * 2.5 * (180.0n) \
         ps=(2 * W5) + (5 * (180.0n)) pd=(2 * W5) + (5 * (180.0n)) m=M5 \
        region=sat
    N4 (net8 vip net7 net7) tsmc18dN w=W4 l=L4 as=W4 * 2.5 * (180.0n) ad=W4 * 2.5 * (180.0n) \
         ps=(2 * W4) + (5 * (180.0n)) pd=(2 * W4) + (5 * (180.0n)) m=M4 \
        region=sat
    P1 (net8 net8 VDD VDD) tsmc18dP w=W1 l=L1 as=W1 * 2.5 * (180.0n) ad=W1 * 2.5 * (180.0n) \
         ps=(2 * W1) + (5 * (180.0n)) pd=(2 * W1) + (5 * (180.0n)) m=M1 \
        region=sat
    P0 (vout net8 VDD VDD) tsmc18dP w=W0 l=L0 as=W0 * 2.5 * (180.0n) ad=W0 * 2.5 * (180.0n) \
         ps=(2 * W0) + (5 * (180.0n)) pd=(2 * W0) + (5 * (180.0n)) m=M0 \
        region=sat
ends ota
// End of subcircuit definition.

// Library name: MP3
// Cell name: dut
// View name: schematic
I16 (net1 net2 net6 net4 net3 net5) ota
\end{tcblisting}

\vspace{10pt}
\textbf{Output Format}
\begin{lstlisting}[style=pythonstyle]
FULL_SPEC = """
REQUIRED OUTPUT FORMAT - READ CAREFULLY
=======================================

The LLM must return a JSON object that matches the Response_structure model.
Inside `config.netlist` it must embed *plain-text* Spectre netlist that follows
EXACTLY the template and ordering below.

1. Header comments (KEEP VERBATIM)
   // Library name: MP3
   // Cell name: ota
   // View name: schematic

2. First non-blank subcircuit line (NO leading dot ".")
   subckt ota VDD VSS ibp10u vin vip vout

3. Final subcircuit line (NO leading dot ".")
   ends ota

4. Footer comments + top-level instance (KEEP VERBATIM)
   // Library name: MP3
   // Cell name: dut
   // View name: schematic
   I16 (net1 net2 net6 net4 net3 net5) ota

5. Absolutely NO extra text:
   - Nothing before the header or after the I1 line.
   - No markdown, JSON, or code fences.
   - No trailing blank lines.

6. Technology lock:
   - All PMOS must keep model name tsmc18dP
   - All NMOS must keep model name tsmc18dN
   - You may modify only W, L, and m values.
   
7. Force use of ports:
   - Ensure that vin, vip, vout, VSS, VDD and ibp10u are explicitly used in the transistor gate connections in the subcircuit ota, as they are the differential inputs. 
   - Do not replace or rename the port. These signals must remain named vin and vip in the subckt pin list and inside the body.

8. Device section 
   - Replace every placeholder:
       W: width  in mu m (e.g. 3.0 mu)
       L: length in mu m (e.g. 0.18 mu)
       M: integer multiplier (>=1)
   - Do NOT alter node names, model names (tsmc18dP / tsmc18dN),
     or the keyword  region=sat.
   - Do NOT add, delete, or reorder lines.
   
VALIDATION POLICY
-----------------
If any placeholder remains unreplaced, a leading dot appears before "subckt"
or "ends", required lines are missing, or extra text is present, the submission
receives automatic score = 0.
"""

class ConfigFile(BaseModel):
    # The full specification is passed as the field description
    netlist: str = Field(description=FULL_SPEC)

class Response_structure(BaseModel):
    reasoning: str = Field(..., description="Detailed resoning process to accomplish the task, please solve all the tasks step by step.")
    config: ConfigFile
\end{lstlisting}
\end{tcolorbox}

\section{Task-Level Breakdown of the \textsc{EngDesign}}

In this section, we provide a detailed summary of each task included in \textsc{EngDesign}. Table~\ref{task-level-description} lists all tasks with concise descriptions of the underlying design challenge and the corresponding references. These tasks span a diverse range of engineering domains. This granular view of the benchmark enables researchers to better understand the scope and structure of \textsc{EngDesign}, facilitates targeted analysis by domain, and supports the integration of new tasks in future extensions of the benchmark.

\begin{longtable}{p{2cm}|p{8.5cm}|p{2cm}}
  \caption{Per-task summary of \textsc{EngDesign}, detailing the task descriptions and relevant references for each task included in the benchmark.} \label{task-level-description} \\
  \toprule
  Task-id & Task Summary & Reference \\
  \midrule
  \endfirsthead

  \multicolumn{3}{c}{{\bfseries Table \thetable\ (continued from previous page)}} \\
  \toprule
  Task-id & Task Summary & Reference \\
  \midrule
  \endhead

  \midrule \multicolumn{3}{r}{{Continued on next page}} \\
  \endfoot

  \bottomrule
  \endlastfoot
  
      \multicolumn{3}{c}{{Topic: Operating System Design}} \\
      \midrule
      CY\_03 &Implement the vioblk\_read and vioblk\_write functions for a virtio block device driver\textemdash setting up virtqueue descriptors, handling interrupts and blocking waits, and moving data to/from the disk in block-aligned transfers. & \citep{waterman2016risc}, \citep{levchenko2025filesystem},  \citep{tsirkin2022virtual} \\
      \midrule
      libin2\_01 & Design a multi-level page-table scheme (page size, number of levels, entries per level) that meets given memory-overhead and average-translation-time constraints for two devices.& \citep{riscv_privileged_2024}, \citep{levchenko2024virtual}\\
      \midrule
      libin2\_02 &Determine a single time quantum that, when applied to three given thread workloads, keeps the sum of average waiting time, context-switch overhead, and quantum-length penalty within specified bounds for each case. & \citep{wikipedia_round_robin}\\
      \midrule
      libin2\_03 & Determine the optimal file-system cluster size that meets both fragmentation and metadata-overhead constraints given a modeled distribution of small and large file sizes.& \citep{wikipedia_clustered_file_system}\\
      \midrule
      XW\_01 & Implement separate programs to perform create, read, update, and delete operations on a provided file system image while maintaining its integrity. & \citep{gooch_vfs_doc} \\
      \midrule
      XW\_02 & Implement basic file system operations on a provided file system image, including writing UTF-8 data into files by allocating blocks, updating inodes, and preserving overall integrity. & \citep{gooch_vfs_doc} \\
      \midrule
      XW\_03 & Implement the create operation to add a new file or directory at a specified path within a provided file system image structure, updating inodes and directory entries while enforcing existing-path, directory, and uniqueness constraints. & \citep{gooch_vfs_doc} \\
      \midrule
      XW\_04 & Implement the delete operation for a custom file system image that removes a file or directory, reclaims its blocks and inode, and updates directory entries and metadata. & \citep{gooch_vfs_doc} \\
      \midrule
      \multicolumn{3}{c}{{Topic: Computer Architecture Design}} \\
      \midrule
      DL\_01 &Tune the top-5 most significant SSD hardware configuration parameters individually to achieve a 20\% improvement in I/O throughput or average latency for each workload. & \citep{lee2013preemptible}\\
      \midrule
      DL\_02 & Tune SSD layout parameters to achieve a 20\% improvement in I/O throughput or average latency while meeting a 1.7-2.1 TB usable capacity constraint under specified workload types.& \citep{lee2013preemptible}\\
      \midrule
      Yiqi\_01 & Derive a compute-shift execution plan (spatial and temporal partition factors) for a large FP16 matrix multiplication on a Graphcore IPU Mk2 to minimize total execution time under per-core memory and core-count constraints. & \citep{jia2019dissectinggraphcoreipuarchitecture},  \citep{9567075},  \citep{9567153},  \citep{Liu_2024},  \citep{9567250},  \citep{vasilache2018tensorcomprehensionsframeworkagnostichighperformance},  \citep{zheng2023ansorgeneratinghighperformancetensor},  \citep{280896} \\
      \midrule
      Yiqi\_02  & Derive a compute-shift execution plan that spatially and temporally partitions a large FP16 matrix multiplication across up to 1,472 IPU Mk2 cores under per-core SRAM and padding constraints to minimize total compute and communication time. & \citep{jia2019dissectinggraphcoreipuarchitecture},  \citep{9567075},  \citep{9567153},  \citep{Liu_2024},  \citep{9567250},  \citep{vasilache2018tensorcomprehensionsframeworkagnostichighperformance},  \citep{zheng2023ansorgeneratinghighperformancetensor},  \citep{280896} \\
      \midrule
      Yuqi\_01 & Determine per-chip NPU architectural parameters (systolic array width, HBM bandwidth) and cluster-level resource allocations (DP/TP/PP, batch size, memory per chip) to meet LLM serving latency and throughput SLOs in a 3D-torus pod of NPUs. & \citep{10.1145/3579371.3589350},  \citep{9351692},  \citep{295551},  \citep{10.1145/3579371.3589059},  \citep{10764608},  \citep{shoeybi2020megatronlmtrainingmultibillionparameter},  \citep{dao2022flashattentionfastmemoryefficientexact},  \citep{Rashidi_2022},  \citep{jax2024matmul},  \citep{ece508_lecture4}\\
      \midrule
      \multicolumn{3}{c}{{Topic: Control System Design}} \\
      \midrule
      LX\_02 &Linearize the normalized magnetic levitation dynamics around the operating point, design a state-feedback controller and full-order observer via pole placement, and then assess stability regions of the closed-loop on the original nonlinear model. &\citep{basar2024lecture} \\
      \midrule
      LX\_03 &Simulate the open-loop and disturbance responses of a linearized quarter-car suspension model to extract peak displacements and settling time, then design a state-feedback controller to meet ride comfort, bottom-out constraints, critical damping, and faster settling requirements. & \citep{basar2024lecture}\\
      \midrule
      XG\_01 & Design a stabilizing controller for an aircraft state-space model using MATLAB’s \texttt{loopsyn} to achieve a target loop shape with a crossover frequency of 8 rad/s, while meeting a 0.05 disk-margin and ensuring performance bound $\gamma < 1$. & \citep{1102566} \\
      \midrule
      XG\_02 & Design a robust feedback controller via loop shaping for a seventh-order nanopositioning stage to meet specified bandwidth, gain margin, and phase margin targets. & \citep{10.1063/1.1499533} \\
      \midrule
      XG\_05 & Derive a first-order plus dead-time model of a stirred-tank reactor from its step response and design a feedback controller that meets specified gain/phase margins, overshoot, and settling time requirements. & \citep{mathworks2025} \\
      \midrule
      XG\_07 & Derive the numerical state-space matrices for a quarter-car active suspension and then design and tune an \( H_\infty \) controller in \textsc{MATLAB} to meet given ride-comfort and handling performance targets by adjusting the weighting parameter \( \beta \). & \citep{mathworks2025} \\
      \midrule
      XG\_08 & Build a Mamdani fuzzy inference system in MATLAB using specified membership functions and rules to compute restaurant tipping based on service and food quality, then evaluate the tip for given inputs. & \citep{mathworks2025} \\
      \midrule
      XG\_10 & Design a PI controller by loop-shaping for a first-order plant at a specified bandwidth and then add a first-order roll-off filter to meet a noise-rejection criterion. &\citep{seiler2024introduction} \\
      \midrule
      XG\_11 & Compute the explicit transfer function of a given initial loop-shaping controller for a third-order plant, then design and tune a second-order roll-off element so that the closed-loop meets stability, 1 rad/s bandwidth, $\geq 50^\circ$ phase margin, and $\ge 3$ dB gain margin. &\citep{seiler2024introduction} \\
      \midrule
      XG\_12 & Determine the explicit transfer function of an initial gain-and-integral loop-shaping controller for a resonant plant and then design a notch filter to attenuate the 13 rad/s resonance so as to achieve a stable closed-loop system with $\geq 60^\circ$ phase margin and $\ge 2$ dB gain margin. &\citep{seiler2024introduction} \\
      \midrule
      XG\_13 & Design a PID controller for a mass-spring-damper system (m = 1 kg, b = 10 $\text{N}\cdot\text{s}/\text{m}$, k = 20 N/m) to achieve settling time < 0.2 s, overshoot < 5\%, and zero steady-state error. & \citep{umich_pid_tutorial} \\
      \midrule
      ZC\_01 & Design a static state-feedback gain $K$ that guarantees robust stability against additive norm-bounded and dynamic uncertainties and ensures the closed-loop $\mathcal{H}_\infty$ norm from $w$ to $z$ is below 0.3. & \\
      \midrule
      ZC\_02 & Determine the largest scalar feedback gain $\alpha$ such that the discrete-time Lurye interconnection with given state-space matrices and a sector-bounded nonlinearity remains absolutely stable. & \\
      \midrule
      ZC\_03 & Determine the asymptotic ranges of the exploration--exploitation parameters $\alpha$ and $\beta$ in a two-phase bandit algorithm that minimize the product of the expected root cumulative regret and the average treatment effect estimation error. & \\
      \midrule
      Ziheng\_01 & Design a dynamic state--feedback (RS-LQR) controller for an aircraft pitch-axis model to achieve a closed-loop rise time under 0.2 s, stability, gain margin $>3$ dB, and phase margin $>30^\circ$. & \citep{Lavretsky2024}\\
      \midrule
      Ziheng\_03 & Based on the linearized model of a magnetic levitation system, design a PID controller that ensures specified performance in both time and frequency domains. & \citep{ece486_uiuc_2025} \\
      \midrule
      KV\_02 & Propose four numerical control parameters (lowest\_current, highest\_current, on\_threshold, off\_threshold) for a switched-capacitor active balancing system of three series Li-ion cells.& \\
      \midrule
      KV\_03 &Infer and name nine missing Simulink or logic blocks in a switched-capacitor battery balancing system architecture based on its operational requirements. & \\
      \midrule
      \multicolumn{3}{c}{{Topic: Mechanical Systems}} \\
      \midrule
      YZ\_04 & Design a parallel-series configured LFP battery module and its cooling plate to support 100 A fast charging for 20 minutes while keeping cell temperature rise $\leq 10\,^\circ$C, temperature gradient $\leq 5\,^\circ$C, and pressure drop $\leq 20\,\mathrm{kPa}$. & \citep{mathworks2025}\\
      \midrule
      RS\_01 & Determine the car's peak achievable speed at any point on the track and calculate the minimum lap time using the given track curvature data and vehicle setup parameters. & \citep{openlapsim}\\
      \midrule
      RS\_02 & Determine the car's peak acceleration and deceleration in $\mathrm{m/s^2}$ from its mass, aerodynamics, tire grip, gearing, and engine speed. & \citep{openlapsim} \\
      \midrule
      RS\_03 & Calculate the maximum lateral load on a driver's neck during cornering from car setup parameters and the combined head-helmet mass. & \citep{openlapsim} \\
      \midrule
      YX\_02 & Generate constraint-compliant global and local input samples via linear relationships and Latin Hypercube Sampling to support diesel engine calibration. & \citep{mathworks2025}\\
      \midrule
      ZH\_02 & Determine the optimal heat shield radius for a spherical 250 kg spacecraft in ballistic re-entry to meet peak heat flux, total heat load, and deceleration constraints. & \citep{Sutton1971AGS}\\
      \midrule
      ZH\_04 & Determine a feasible glide angle, vehicle volume, and ballast mass for a buoyancy-driven underwater glider to cover 4 km in $\le 2$ hours under given power, depth, and buoyancy constraints. & \\
      \midrule
      \multicolumn{3}{c}{{Topic: Structural Design}} \\
      \midrule
      RK\_01 & Generate a 2D topology-optimized design of a rectangular beam under an asymmetric point load with fixed/roller supports to minimize compliance at a maximum volume fraction of 0.25, then report its compliance and volume fraction. & \\
      \midrule
      RK\_02 & Perform a 2D compliance-based topology optimization of a rectangular domain under two downward point loads with supports at its corners, subject to a volume fraction limit of 0.25. & \\
      \midrule
      RK\_03 & Perform 2D topology optimization on a rectangular domain with specified supports and loading to minimize compliance under a 15\% volume fraction constraint. & \\
      \midrule
      RK\_04 & Perform topology optimization on a cracked 2D domain under given horizontal loads to minimize the maximum stress while restricting the material volume to 25\%. & \\
      \midrule
      YF\_01 &Determine the required wall thickness of an L-shaped extruded steel beam so that its maximum z-direction displacement under a specified surface traction remains below 0.1 mm in a static linear FEA simulation. & \\
      \midrule
      YF\_02 &Determine the required thickness of a simply supported rectangular steel beam under two quarter-span loads so that its maximum vertical displacement is less than 1 mm in a 3D PDE Toolbox simulation. & \\
      \midrule
      YF\_03 &Determine the plate thickness of a rectangular steel cantilever beam under a specified half-length uniform pressure so that its maximum vertical deflection stays below 2 mm in a 3D linear elasticity model. & \\
      \midrule
      YF\_04 &Determine the diameter of a cantilevered solid steel shaft under a specified torque so that its maximum end-to-end twist does not exceed 0.05 rad. & \\
      \midrule
      YF\_05  & Determine the required cross-sectional area of the truss member between Node 2 and Node 3 so that the maximum nodal displacement under a 4000 N load at Node 3 remains below 0.5 mm. & \\
      \midrule
      YF\_06  & Determine the required thickness $t$ of a fixed-edge rectangular steel plate under uniform pressure to meet a specified deflection criterion using MATLAB's PDE Toolbox. & \\
      \midrule
      YJ\_01 & Determine the optimal 2D material distribution on a $64\times64$ grid that minimizes structural compliance under given loads, boundary conditions, and a volume fraction constraint. & \citep{mazé2022diffusionmodelsbeatgans},  \citep{topodiff2022website}\\
      \midrule
      YJ\_02 & Perform a topology optimization of a discretized cantilever beam to minimize compliance under a given volume constraint. & \citep{lagerweij2019topopt}\\
      \midrule
      YJ\_03 & Set up and solve a topology optimization to distribute material in a finite-element mesh so as to minimize the Mode I stress-intensity factor at a crack tip under a volume-fraction constraint. & \citep{lagerweij2019topopt}\\
      \midrule
      \multicolumn{3}{c}{{Topic: Digital Hardware Design}} \\
      \midrule
      XY\_01 & Analyze and extract the 4x4 bit pattern of a specified Tetris tetromino in a given rotation from ROM data, and convert it into a visual character-based representation for display. & \citep{8299595},  \citep{tetris_guideline_fandom},  \citep{le_fpga_tetris_2017}\\
      \midrule
      XY\_03 &Design a time-multiplexed 4-digit hexadecimal display driver on a 50 MHz FPGA that guarantees at least a 60 Hz per-digit refresh and evaluates worst-case flicker. & \citep{chu2018fpga},  \citep{fpga4student_7segment},  \citep{8299595}\\
      \midrule
      XY\_04 &Design a hardware-efficient color mapping system on an FPGA for a Tetris game with RGB444 VGA output that supports distinct element colors and a dynamic Night Mode under a 10-color limit. & \citep{fpga4fun_pong}\\
      \midrule
      XY\_05 &Define the per-instruction control signal settings, finite-state-machine transitions, and detailed explanations for the SLC-3 CPU, including a custom SWAP instruction, under strict hardware and sequencing constraints. & \citep{hennessy2017computer},  \citep{cheng2025lab5},  \citep{cheng2025cpulogic}\\
      \midrule
      YH\_01 &Implement a tiled and pipelined GEMM function in Vivado HLS to maximize DSP utilization on the VCK5000 using block sizes of $64\times64\times64$. & \\
      \midrule
      YH\_02 &Implement a DSP-optimized tiled GEMM in Vivado HLS for $1024\times1024$ matrices using $128\times128\times128$ blocking on the VCK5000 platform. & \\
      \midrule
      YH\_03 &Create a synthesizable Vitis HLS GEMM function with AXI interfaces and loop pipelining for $1024\times1024$ matrices targeting the VCK5000. & \\
      \midrule
      YH\_04 &Design a Vitis HLS-compatible GEMM function for $256\times256$ matrices with AXI interfaces and pipelined inner loop targeting the VCK5000. & \\
      \midrule
      NS\_PA\_SS\_02 &Design a parameterizable n-bit Gray-code sequence generator in Verilog that updates on each rising clock edge and resets synchronously with active-low reset to zero. & \citep{chipdev2025}\\
      \midrule
      NS\_PA\_SS\_03 &Design a parameterized 32-entry, 2-read/1-write register file in synthesizable SystemVerilog with synchronous reset, read/write enables, and collision detection. & \citep{chipdev2025}\\
      \midrule
      NS\_PA\_SS\_04 &Implement a parameterizable 8-bit Fibonacci LFSR in synthesizable SystemVerilog with synchronous active-low reset, seed loading, tap buffering, and feedback shifting. &\citep{chipdev2025} \\
      \midrule
      NS\_PA\_SS\_05 & Develop a parameterizable, synthesizable SystemVerilog module that captures eight bitwidth-wide input words, performs a fully unrolled one-cycle bubble sort on them when triggered, and presents the sorted vector plus optional valid flag. & \citep{chipdev2025} \\
      \midrule
      NS\_PA\_SS\_06 & Design a synthesizable SystemVerilog module that reads two 3-element vectors serially, computes their dot product in one cycle after the sixth input, and drives an 18-bit result plus a one-cycle valid signal with defined reset behavior. & \citep{chipdev2025} \\ 
      \midrule
      NS\_PA\_SS\_07 & Implement a synthesizable $8\times8$ register file in Verilog with one read port, one write port, valid-bit tracking, error flagging, and synchronous reset/clock behavior. & \citep{chipdev2025} \\
      \midrule
      NS\_PA\_SS\_08 & Develop a fully synchronous, parameterized SystemVerilog module that detects a user-loaded 5-bit target pattern in a serial input stream and pulses a one-cycle 'seen' flag on each match. & \citep{chipdev2025} \\
      \midrule
      NS\_PA\_SS\_09 & Create a fully synthesizable, parameterized SystemVerilog module that converts an N-bit Gray code input into its binary equivalent using a combinational XOR cascade. & \citep{chipdev2025} \\
      \midrule
      NS\_PA\_SS\_10 & Implement a synthesizable SystemVerilog finite-state machine that reads a serial bitstream, tracks the value modulo 5, and raises dout when the running value is divisible by 5. & \citep{chipdev2025}\\
      \midrule
      \multicolumn{3}{c}{{Topic: Analog Integrated Circuit Design}} \\
      \midrule
      TB\_01 &Size the devices in a folded-cascode OTA netlist by assigning concrete W, L, m, and R values to meet given performance specs under a fixed testbench. & \citep{ece483_sp25_mp4} \\
      \midrule
      TB\_02 &Design and size a folded-cascode OTA from scratch in SPICE format to meet specific performance specs using the tsmc18 process and a given testbench setup. & \citep{ece483_sp25_mp4} \\
      \midrule
      TB\_03 &Size a 5-transistor differential-to-single-ended OTA by assigning W, L, and m values to meet gain, bandwidth, CMRR, and ICMR specs using tsmc18 models in a fixed testbench. & \citep{ece483_sp25_mp3} \\
      \midrule
      TB\_04 &Design and size a 5-transistor OTA along with its bias circuit in a single SPICE netlist to meet performance specs under a unity-gain testbench. & \citep{ece483_sp25_mp3} \\
      \midrule
      TB\_05 &Size a common-source amplifier with resistive load in SPICE by tuning NMOS, resistor, and bias parameters to meet gain, bandwidth, and output voltage specs. & \citep{ece483_sp25_mp1} \\
      \midrule      
      \multicolumn{3}{c}{{Topic: Robotics}} \\
      \midrule
      Ziheng\_02 & Determine the six joint angles of a 6-DOF serial robot that achieve a specified end-effector pose given the robot's screw axes and home transformation. & \citep{weng2018modernrobotics}\\
      \midrule
      AM\_02 &Compute two time-parameterized, collision-free trajectories for 2 by 2 robots on a 30 by 30 grid from given start to end points in 20 time steps, avoiding static rectangular obstacles and dynamically moving pedestrians under a maximum per-step speed of 2 units in each axis. & \\
      \midrule
      AM\_03 & Compute a time-stamped, grid-based, collision-free trajectory for a 2 by 2 robot on a 30 by 30 map with static rectangles and three moving pedestrians, starting at (17,2) and visiting goals A and B in any order without exceeding a per-step speed of 2 in x or y.& \\
      \midrule
      HJ\_01 & Select refresh rate, acceleration, maximum velocity, and look-ahead distance settings to complete a 56 m lap in under 10 s with a maximum 1 m track-off error on a \$300 budget.& \\
      \midrule
      XZ\_01 &Compute the shortest collision-free grid path for a small inspection robot across a 50$\times$40 m construction map with specified walls and obstacles. & \citep{mathworks2025}\\
      \midrule
      XZ\_02 &Compute a smooth, collision-free trajectory for an autonomous construction vehicle from (0,0,$\pi$) to (49,39,$\pi$/2) on a 50$\times$40 m site with fixed obstacles, while respecting a 4 m turning radius and 0.5 m vehicle clearance.& \citep{mathworks2025}\\
      \midrule
      XZ\_03 &Generate a collision-free shortest path on an integer grid in a Webots world from (0,0,0) to (49,39,0) around defined obstacles. & \citep{mathworks2025}\\
      \midrule
      XZ\_04 &Generate a time-parameterized, collision-free 3D trajectory in a Webots world from a start to a goal point while respecting zone-based speed constraints. & \citep{mathworks2025}\\
      \midrule
      YX\_01 & Compute the world-coordinate corners of a 2D sparse costmap and then use a Hybrid A* planner with nonholonomic constraints to find collision-free paths for given start and goal poses. & \citep{mathworks2025}\\
      \midrule
      ZH\_03 & Determine wheel diameter, gear ratio, battery capacity, and robot mass for an electric ground robot to meet specified speed, incline, and endurance targets. & \\
      \midrule
      \multicolumn{3}{c}{{Topic: Signal Processing}} \\
      \midrule
      qjlim2\_01 & Design a rectangular microstrip patch antenna on a Rogers laminate to resonate at 1.537 GHz with $\ge 50$ MHz bandwidth, $\ge 3$ dBi gain, and fit within a $100 \times 100 \times 10$ mm volume using a $50 \Omega$ coaxial feed. & \citep{antenna_patch},  \citep{6020028},  \citep{mathworks2025}\\
      \midrule
      qjlim2\_02 & Design a free-space strip dipole antenna centered at 3 GHz with at least 100 MHz bandwidth, $\ge2$ dBi gain, and fitting within a $100 \times 100$ mm footprint. & \citep{antenna_dipole},  \citep{mathworks2025} \\
      \midrule
      qjlim2\_04 & Design a cylindrical monopole antenna resonant at 0.487\,GHz with S11 $\le -10$\,dB, bandwidth $\ge 10$\,MHz, and gain $\ge 0$\,dBi. & \citep{antenna_dipole},  \citep{mathworks2025} \\
      \midrule
      XG\_09 & Tune the parameters of Gaussian and triangular membership functions in a fuzzy inference system to improve edge detection performance (MSE, PSNR, SSIM) on a grayscale image relative to a Canny pseudo-ground truth. & \citep{mathworks2025} \\
      \midrule
      YZ\_01 & Design a three-stage sample-rate converter---including a Farrow fractional resampler and two decimating FIR filters---to down-convert input rates (e.g., 150 MHz) to 30.72 MHz while meeting LTE passband, stopband, and EVM requirements.& \citep{mathworks2025}\\
      \midrule
      YZ\_02 & Derive analytic formulas and determine coefficient sets for SG filter cutoff and half-magnitude width, then design four weighted SG filters meeting specified noise-vs-smoothness performance criteria. & \citep{mathworks2025}\\
      \midrule
      YZ\_03 & Design a circularly polarized helical antenna in MATLAB's Antenna Toolbox to operate from 1.3 to 2 GHz, meeting specified directivity and axial ratio requirements. & \citep{mathworks2025}\\
      \midrule
      AB\_01 & Compute first-order statistical features (mean, variance, skewness, kurtosis) and the GLCM contrast for pixel intensities strictly inside a given polygonal ROI in a medical image patch.& \\
      \midrule
      AB\_02 & Implement the OpenCV watershed algorithm in Python to segment the red heart pips and numerals from the white background in the specified image.& \citep{howse2020learning}\\
      \midrule
      AB\_03 &Load a hard-coded image, find its largest contour, and compute both its approximate polygon and convex hull using OpenCV. & \citep{howse2020learning}\\
      \midrule
      AV\_02 & Design a digital IIR filter to notch out a 120 Hz resonance in a 2000 Hz-sampled sensor signal while preserving $\pm$15 Hz around that frequency.& \citep{ece310_uiuc}\\
      \midrule
      AV\_03 & Design an FIR anti-aliasing filter to allow downsampling a 48 kHz audio signal to 8 kHz while preserving the 0-3.5 kHz band with less than 3 dB ripple.& \citep{ece310_uiuc}\\
      \midrule
      HC\_03 &Estimate the coefficients a, b, and c of a second-order polynomial $y = ax^{2} + bx + c$ that best fits a given noisy 2D data set by minimizing mean squared error. & \\
      \midrule
      JY\_01 & Design and implement an automated test stand using two linear polarizers, a quarter-wave plate, and rotary motors to control and measure changes in the Stokes vector of polarized light.& \\
      \midrule
      JY\_02 & Design and apply a Gaussian smoothing filter, horizontal and vertical edge-detection kernels, and a thresholding step to produce an edge map from the given image array.& \citep{soria2020dexined}\\
      \midrule
      JY\_03 & Design a convolutional kernel that performs linear interpolation to fill zero-valued pixels in the corrupted Degree0 polarization image and evaluate its effect on AoLP and DoLP outputs.& \\
      \midrule
      WJ\_01 &Design and implement a flexible OpenCV based filter pipeline choosing and parameterizing one or more denoising filters based on detected noise types. & \\
      \midrule
      YX\_03 & Compute and combine SNR metrics, integration gains, and various radar losses to evaluate the detectability of a $1\,\mathrm{m}^2$ target at 100 km for an S-band surveillance radar. & \citep{mathworks2025}\\
\end{longtable}

\end{document}